\documentclass{aa}  
\usepackage{graphicx}
\usepackage{txfonts}
\begin{document} 
\title{An imaging spectroscopic survey of the planetary nebula NGC~7009 with 
MUSE\thanks{Based on observations collected at the European Organisation for 
Astronomical Research in the Southern Hemisphere, Chile in Science Verification 
(SV) observing proposal 60.A-9347(A).}
\fnmsep
\thanks{FITS files corresponding to the images in Figs. 2, 3, 4, 5, 6, 7, 8, 9, 10, 
11, 12, 13, 14, 15, 16, 17, 18, 19, 20, 21 and 22 are available at the CDS via 
anonymous ftp to cdsarc.u-strasbg.fr (130.79.128.5) or via 
http://cdsweb.u-strasbg.fr/cgi-bin/qcat?J/A+A/.}
}
\titlerunning{MUSE survey of NGC~7009}

\author{J. R. Walsh\inst{1}
        \and
        A. Monreal-Ibero\inst{2,3}   
        \and
        M. J. Barlow\inst{4}
        \and
        T. Ueta\inst{5} 
        \and
        R. Wesson\inst{6} 
        \and
        A. A. Zijlstra\inst{7,8}
        \and 
        S. Kimeswenger\inst{9,10}
        \and
        M. L. Leal-Ferreira\inst{11, 12}
        \and
        M. Otsuka\inst{13}
}
 \institute{European Southern Observatory. Karl-Schwarzschild Strasse 2,  D-85748 Garching, Germany \\
            \email{jwalsh@eso.org}
            \and      
            Instituto de Astrof\'{i}sica de Canarias (IAC), E-38025 La Laguna, Tenerife \\
            \email{amonreal@iac.es}
            \and
            Universidad de La Laguna, Dpto. Astrof\'{i}sica, E-38206 La Laguna, Tenerife, Spain
            \and
            Dept. of Physics and Astronomy, University College London, Gower Street, London WC1E 6BT, United Kingdom \\
            \email{mjb@star.ucl.ac.uk}
            \and
            Department of Physics and Astronomy, University of Denver, 2112 E. Wesley Ave., Denver, CO, 80210, USA \\
            \email{Toshiya.Ueta@du.edu}                          
            \and
            Dept. of Physics and Astronomy, University College London, Gower Street, London WC1E 6BT, United Kingdom \\
            \email{rw@nebulousresearch.org}          
            \and
            Jodrell Bank Centre for Astrophysics, Alan Turing Building, University of Manchester, 
            Manchester M13 9PL, UK
            \and
            Laboratory for Space Research, University of Hong Kong, Pokfulam Road, Hong Kong \\ 
            \email{albert.zijlstra@manchester.ac.uk}
            \and
            Instituto de Astronom{\'i}a, Universidad Cat{\'o}lica del Norte, Av. Angamos 0610, Antofagasta, Chile
            \and
            Institut f{\"u}r Astro- und Teilchenphysik, Leopold--Franzens Universit{\"a}t Innsbruck, Technikerstrasse 25, 6020 Innsbruck, Austria \\
            \email{skimeswenger@ucn.cl}
            \and
            Leiden Observatory, University of Leiden, Leiden, the Netherlands
            \and
            Oberkasseler Stra{\ss}e 130, D-40545 D\"{u}sseldorf, Germany \\
            \email{mllferreira@gmail.com}
            \and
            Okayama Observatory, Kyoto University Honjo, Kamogata, Asakuchi, Okayama, 719-0232, Japan \\
            \email{otsuka@kusastro.kyoto-u.ac.jp}
}

\date{Received: 17 May 2018; accepted: 14 September 2018 }
\authorrunning{Walsh, J. R., et al.}

\abstract
{}
{The spatial structure of the emission lines and continuum over the 
50$''$ extent of the nearby, O-rich, PN NGC~7009 (Saturn Nebula) have been 
observed with the MUSE integral field spectrograph on the ESO Very Large 
Telescope. This study concentrates on maps of line emission and their 
interpretation in terms of physical conditions.
}
{MUSE Science Verification data, in $<$0.6$''$ seeing, have 
been reduced and analysed as maps of emission lines and continuum over the
wavelength range 4750~--~9350\,\AA. The dust extinction, the electron densities 
and temperatures of various phases of the ionized gas, abundances of species 
from low to high ionization and some total abundances are determined 
using standard techniques.
}
{Emission line maps over the bright shells are presented, from neutral to 
the highest ionization available (\ion{He}{II} and [\ion{Mn}{V}]). For
collisionally excited lines (CELs), maps of electron temperature 
($T_{\rm e}$ from [\ion{N}{II}] and [\ion{S}{III}]) and density 
($N_{\rm e}$ from [\ion{S}{II}] and [\ion{Cl}{III}]) are 
available and for optical recombination lines (ORLs) temperature (from the 
Paschen jump and ratio of \ion{He}{I} lines) and density (from high Paschen 
lines). These estimates are compared: for the first time, maps of the 
differences in CEL and ORL $T_{\rm e}$'s have been derived, and 
correspondingly a map of $t^{2}$ between a CEL and ORL temperature,
showing considerable detail. Total abundances of only He and O were 
formed, the latter using three 
ionization correction factors. However the map of He/H is not 
flat, departing by $\sim$2\% from a constant value, with remnants 
corresponding to ionization structures. An integrated spectrum 
over an area of 2340 arcsec$^{2}$ was also formed and compared to 
1D photoionization models.
}
{The spatial variation of a range of nebular parameters illustrates the 
complexity of the ionized media in NGC~7009. These MUSE data 
are very rich with detections of hundreds of lines over areas of 
hundreds of arcsec$^{2}$ and follow-on studies are outlined.
}

\maketitle
%

\section{Introduction}
\label{Intro}

Planetary nebulae (PNe) provide self-contained laboratories of a wide range 
of low density ionization conditions and thus form a corner-stone of studies 
of the interstellar medium in
general, from the diffuse interstellar medium, to H\,II and 
He\,II regions surrounding hot stars of all evolutionary stages to the 
highest density regions in stellar coronae and active galactic nuclei. 
However this advantage also proves a challenge, as the range of ionization 
conditions occur within a single object and all but a few nearby Galactic 
PNe are compact or slightly extended targets. Observation with a small 
aperture or slit is bound to sample a range of conditions,
framing high and low ionization, low to higher density (typical range 
$10^{1}$~--~$10^{6}$ cm$^{-3}$), low to moderate electron temperature (range 
a few $\times 10^{3}$ to $\sim$ 25\, 000 K), including X-ray emitting 
gas with temperatures to $\sim$ 10$^{6}$K (for example 
\citet{Guerrero2000} and the results from the ChanPlaNS consortium: 
\citet{Kastner2012, Freeman2014, Montez2015}). 
The resulting spectrum contains a summation of these conditions, the exact 
mixture depending on the nebular morphology, the temperature of the central 
star and the velocity and mass loss rate of its stellar wind, the age of 
the PN (younger PN generally being of higher density), the homogeneity of the 
circumstellar gas, the metal abundance and the content of dust. Spectroscopy 
with 2D coverage leads to increased understanding of the nebular structure 
and the role of the local physical conditions on the emitted spectrum, since
PNe generally show a gradient in ionization from close to the central star 
to lower values in the outer regions.

Slit spectroscopy has until recently been the primary tool for PN spectroscopy,
where the typical slit can sample the central higher ionization zone to the low
ionization or neutral outer regions, depending on whether the nebula is density 
or ionization bounded. For moderately extended PNe (sizes 10~--~100's $''$), the 
selection of the slit is all-important. Often the placement is made based on 
shape, extension matching the slit, features of interest such as knots, 
filaments/jets, bipolar axis, etc, which in terms of sampling the widest possible 
range of conditions may not be representative. A good example is the seminal 
study of \object{NGC~7009} by X.-H. Liu and X. Fang and 
co-workers \citep{Liu1993, Liu1995, FangLiu2011, FangLiu2013}, which has presented 
the deepest and most comprehensive optical (3000~--~10\,000\,\AA)
spectroscopy to date. Their study is primarily based on spectra derived from 
a single long slit aligned along the major axis and including the horns 
(ansae) and the outer low ionization knots. To step from this particular
spatial sampling of the conditions 
in the ionized and neutral gas to a comprehensive description 
of the whole nebula is clearly an extrapolation of unknown magnitude, unless a 
wider range of conditions are sampled (in this case as next obvious choice a 
slit along the minor axis). In addition comparison of nebular properties based 
on single slit spectra between nebulae may not be strictly valid since various 
differing choices may have been applied in choosing the slit position and 
orientation.

Studies devoted to imaging of selected emission lines, chosen to sample the
range of conditions in typical PNe, have been undertaken, often in 
co-ordination with slit spectroscopy, the latter targeting regions with 
particular conditions. Typical examples, which have focussed on NGC~7009, are 
\citet{Bohigas1994} and \citet{Lame1996}. With imaging in bright permitted 
and forbidden lines from \ion{He}{II} to [\ion{O}{I}], a wide span of ionization 
conditions can be sampled and narrow filters tuned to line pairs sensitive to 
electron density and temperature provide maps of the diagnostics, such as 
[\ion{S}{II}]6716.4 and 6730.8\,\AA, and [\ion{O}{III}]5006.9 and 4363.2\,\AA. 
Of course these images have to be 
corrected for the continuum collected within the filter bandwidth, whose origin
is bound-free, free-free from ionized H and He and the H$^{+}$ and He$^{++}$
two photon emission, as well as direct or scattered starlight from the 
central star (or any field stars along the line of sight). In the infrared, 
\citet{Persi1999} used the circular variable filter (CVF) on ISOCAM for 
wavelength stepped monochromatic imaging of six PNe in the 5~--~16.5 $\mu$m 
range.

The HST imaging study of NGC~7009 using WF/PC2 
\citep{Rubin2002}, unmatched in photometric fidelity and spatial resolution,
was conducted in a search for spatial evidence of temperature variations based 
on the [\ion{O}{III}] $T_{\rm e}$ sensitive ratio. Clearly such studies must carefully
select lines to observe and match them, as closely as possible, to the 
interference filter parameters in order to avoid significant contamination 
from adjacent emission lines. Even with very narrow filters, some strong line 
close pairs, such as H$\alpha$ and the [\ion{N}{II}] doublet at 6548.0, 6583.5\,\AA,
fall within the same filter passband and accurate separation can be 
problematic. Inevitably given the density of lines in PNe spectra, there is 
often a non-optimal match of the off-line filter for subtraction of the 
continuum contribution to the passband. The passbands and throughputs of 
interference filters are also notoriously difficult to calibrate depending 
on input beam, ambient temperature and age, so the correction for the 
continuum and adjacent line contribution can be kept generally low, but 
can rarely be optimal.

Multiple slit observations provide the next step towards full spectroscopic
coverage of the surface variations of PNe. These studies are usually limited
to high surface brightness PNe, and even then necessitate extensive 
observing periods for good coverage of moderately extended (e.g., 
$^{>}_{\sim}$ 20$''$) nebulae, with the added risk of changing observing 
conditions within or across different slit positionings (thus the stability 
of space-based spectroscopy, such as with STIS on HST, e.g., 
\citet{Dufour2015}, are preferred). The early study \citep{Meaburn1981}
of the [\ion{S}{II}] electron density across NGC~7009 with multiple slit 
positions is an example. The above mentioned problems of deriving line ratio
diagnostics from emission line maps can be partially overcome by combining
them with aperture or long slit spectra in order to check, or refine, the 
maps by comparison to sampled regions from spectroscopy, as indeed done 
by \citet{Rubin2002}.

Integral field spectroscopy in the optical has been applied to PNe beginning
with Fabry-Perot studies using the TAURUS wide field imaging Fabry-Perot 
interferometer \citep{Taylor1980}, such as for studies of 
\object{NGC~2392} and \object{NGC~5189} \citep{Reay1983a,
Reay1984}. Whilst these observations targeted 
small wavelength ranges at relatively high spectral resolution for radial 
velocity studies, lower spectral resolution using the Potsdam Multi Aperture 
Spectrometer (PMAS) has targeted the faint halos of several nebulae, such
as \object{NGC~6826} and \object{NGC~7662} \citep{Sandin2008}. The VIMOS 
IFU on the VLT has also been used for spectroscopy of the faint halos of 
\object{NGC~3242} and \object{NGC~4361} \citep{MonrealIbero2005}; see also 
\citep{MonrealIbero2006}. NGC~3242 was observed with the 
VIMOS IFU in its wide field mode (field 
$54'' \times 54''$ mode with 0.67$''$ spaxels) covering the bright rings, 
from which electron densities, temperatures and chemical abundances of 
various morphological features were measured \citep{Monteiro2013}. 
\citet{Tsamis2008} obtained IFU spectra with the VLT FLAMES
IFU, Argus, of small regions of several PNe, including NGC~7009. The maximum
field of view of the FLAMES IFU ($22 \times 12 ''$) only sampled parts of 
these nebulae but at excellent spectral resolution (in the sense that 
the instrument line spread function full width at half maximum (FWHM) 
matches the typical intrinsic width of the emission lines of 
$\sim 10 - 20$ km s$^{-1}$).

Most recently a number of southern hemisphere extended PNe have been 
targetted using the Wide Field Spectrograph (WiFeS) on the 2.3-m ANU 
telescope at Siding Spring Observatory. This IFU has a field of 25$''$ 
by 38$''$ with either $1 \times 1$ or $1 \times 0.5$ arcsecond spaxels 
and spectral resolving powers from 3000~--~7000 over the optical (3400~--~
7000\,\AA) range. Results of emission line ratio maps, physical
conditions and abundances in either the integrated nebulae or 
for selected regions in Galactic PNe have been published in a
series of papers (e.g., \citet{Ali2015, Ali2016, Ali2017}). Imaging 
Fourier transform spectrometry continues to be applied to PNe, as 
exemplified by the SITELLE study of \object{NGC~6720} by 
\citet{Martin2016}.

In the near to mid-infrared, IFU techniques have also been applied
to a few PNe. \citet{Matsuura2007} mapped the $H_2$ emission in some of 
the compact knots of the very extended Helix Nebula (\object{NGC~7293}). 
Herschel also had imaging spectroscopy facilities and the SPIRE FTS 
has been used for mapping CO and OH$^{+}$ in the Helix Nebula 
\citep{Etxaluze2014}; OH$^{+}$ was also detected in three other PNe 
(\object{NGC~6445} , NGC~6720, and \object{NGC~6781}) by 
\citet{Aleman2014}. The HerPlans programme has mapped 11 PNe with 
PACS/SPIRE, and IFU maps of NGC~6781 over the wavelength range 
51~--~220 $\mu$m and aperture maps over 194~--~672 $\mu$m are 
featured in \citet{Ueta2014}.

The central theme of this paper is that integral field spectroscopy 
(IFS) provides an ideal approach to the spatial study of line and 
continuum spectroscopy of extended Galactic PNe. 
The most advanced optical IFS to date is the Multi-Unit Spectroscopic 
Explorer (MUSE; \citet{Bacon2010}) mounted on VLT UT4 (Yepun). Its 
$60 \times 60''$ field in Wide Field Mode, is ideally 
suited to the dimensions of a broad swathe of extended Galactic PNe,
fulfilling the aim of sampling the full range of ionic conditions 
across the PNe surface. The spaxel size in Wide Field Mode (0.2$''$) 
advantageously samples the best ground-based seeing. Targets larger 
than the MUSE field can of course be fully sampled by mosaicing, such
as performed for the Orion Nebula by \citet{Weilbacher2015}. However 
the wavelength range of MUSE (4750~--~9300\,\AA) is not suited for 
following-up the rich near-UV range (from the atmospheric 
cut-off to $\sim$4500\,\AA), extensively studied by ground-based
optical spectroscopy. It is shown in this paper 
that, except for the $T_{\rm e}$-sensitive [\ion{O}{III}] 
4363.2\AA\ line and the oxygen M1 multiplet recombination lines (which are 
to some extent available in the MUSE extended wavelength mode, reaching 
to below 4650\,\AA), many emission lines of interest over the full range of 
ionization conditions (except molecular species) are observable with MUSE.

NGC~7009 (PNG 037.7-34.5) was chosen for this demonstration of MUSE PN 
imaging spectroscopy based on several considerations:
\begin{itemize}
\item high surface brightness (2.4$\times 10^{-13}$ ergs 
cm$^{-2}$ s$^{-1}$ arcsec$^{-2}$ based on a diameter of 30$''$ and 
observed H$\beta$ flux of $1.66 \times 10^{-10}$ ergs 
cm$^{-2}$ s$^{-1}$ \citep{Peimbert1971}; 
\item full extent of about 55$''$ excluding the faint extensions 
of the halo \citep{Moreno-Corral1998};
\item moderate excitation \citep[excitation class 5;][]{Dopita1990}), 
thus including \ion{He}{II};
\item excellent complementary optical spectra covering wavelengths
bluer than the MUSE cut-off and at higher spectral resolution;
\item extensive data to shorter wavelengths (UV and X-ray, the
latter from ChanPlans) and mid- and far-infrared imaging and
spectroscopy, such as \citet{Phillips2010} and HerPlans.
\end{itemize}
NGC~7009 can be considered as an excellent 'average' Galactic PN,
sampling the O-rich nebulae since NGC~7009 has C/O = 0.45 
\citep{KingBarlow1994}. The results of mapping the extinction
across NGC~7009 from MUSE observations were previously presented in 
\citet{Walsh2016}, hereafter Paper I. An overview of 
observations and results on NGC~7009 is presented in the introduction 
to Paper I.

Details of the MUSE observations are presented in Sect. 2, followed by
an overview of the reduction and analysis of the spectra in Sect 3. 
Then flux maps in a range of emission lines are presented in Sect. 4 and
the electron temperature and density maps in Sect. 5. The method of
Voronoi tesselation is applied to some of the emission line
data to examine the properties of the faint outer shell in Sect. 6.
Ionization maps are presented in Sect. 7 and abundance maps in Sect. 8.
Integrated spectra of the whole nebula and selected long slits
are briefly discussed in Sect. 9, where 1D photoionization modelling 
is presented. Section 10 closes with a discussion of
some aspects of the data. Further analyses are being developed and 
will be presented in future work. A summary of the conclusions is 
presented in Sect. 11.


\section{MUSE observations}
\label{MUSEobs}

During the MUSE Science Verification in June and again in August 2014, 
NGC~7009 was observed in service mode with the MUSE Wide Field Mode (WFM-N) 
spatial setting WFM (0.2$''$ spaxels) with the standard (blue filter) 
wavelength range (4750 - 9300\,\AA, 1.25\,\AA\ per pixel). Table 
\ref{tab:observations} presents a complete log of the observations; CS 
refers to the pointing centred on the central star. The position 
angle (PA $\sim$ 33$^{\circ}$) of the detector was chosen to place the 
long axis of the nebula, from the east to the west ansae, along 
the diagonal of the detector in order to ensure coverage of the full extent 
of the nebula within the one arcminute square MUSE field of view (but 
excluding the very extended, faint halo). A second PA, rotated 
by 90$^{\circ}$ from the first, was included in order to assist in smoothing 
out the pattern of slicers and channels, as recommended in the MUSE Manual 
\citep{Richard2017}. For each sequence of observations at a given exposure 
time, a four-cornered dither around the first exposure at the central 
pointing, with offsets ($\Delta \alpha$ = 0.6$''$, $\Delta \delta$ = 0.6$''$), 
was made to improve the spatial sampling.

The Differential Image Motion Monitor (DIMM) seeing (specified at 5000\,\AA) 
during the time of the observations is listed in Tab. \ref{tab:observations}. 
The requested maximum seeing was 0.70$''$ FWHM. The repeat observations at 
another instrument position angle (180$^{\circ}$ displaced from the first 
sequence at PA 33$^{\circ}$) in August 2014 indicated a much worse DIMM 
seeing value, but the measured FWHM on the images from the reduced cubes 
was only $\sim$0.75$''$. Since the final image quality of the second 
set was significantly worse than for the June observations, these cubes 
were neglected for the subsequent analysis.

\begin{table*}
\caption{Log of MUSE observations of NGC~7009}
\centering
\begin{tabular}{lllrrll}
\hline\hline
Pointing & Date         & UT      & Exposure & PA       & Airmass & DIMM seeing \\
         & (YYYY-MM-DD) & (h:m)   &   (s)    & ($^{\circ}$) & range   &    ($''$)   \\
\hline
NGC~7009CS & 2014-06-22 & 09:23 &  10 &  33.0   & 1.115 - 1.128 & 0.64 \\
NGC~7009CS & 2014-06-22 & 09:13 &  60 &  33.0   & 1.093 - 1.110 & 0.67 \\
NGC~7009CS & 2014-06-22 & 09:35 & 120 &  33.0   & 1.131 - 1.164 & 0.57 \\
Offset sky & 2014-06-22 & 09:45 & 120 &  33.0   & 1.173          & 0.53 \\
           &            &       &     &                &      &  \\
NGC~7009CS & 2014-06-24 & 07:26 &  10 & 123.0   & 1.029 - 1.028 & 0.63 \\
NGC~7009CS & 2014-06-24 & 07:15 &  60 & 123.0   & 1.034 - 1.030 & 0.67 \\
NGC~7009CS & 2014-06-24 & 07:37 & 120 & 123.0   & 1.028 - 1.029 & 0.60 \\
Offset sky & 2014-06-24 & 07:48 & 120 & 123.0   & 1.029          & 0.68 \\
           &            &       &     &                &      &  \\
NGC~7009CS & 2014-08-20 & 01:54 &  10 & 303.0     & 1.176 - 1.159 & 1.16 \\
NGC~7009CS & 2014-08-20 & 01:44 &  60 & 303.0     & 1.212 - 1.183 & 1.07 \\
NGC~7009CS & 2014-08-20 & 02:07 & 120 & 303.0     & 1.155 - 1.124 & 1.18 \\
Offset sky & 2014-08-20 & 02:17 & 120 & 303.0     & 1.118          & 0.98 \\
           &            &       &     &                &      &  \\
NGC~7009CS & 2014-08-20 & 02:37 &  10 & 303.0     & 1.107 - 1.090 & 0.93 \\
NGC~7009CS & 2014-08-20 & 02:27 &  60 & 303.0     & 1.212 - 1.183 & 0.89 \\
NGC~7009CS & 2014-08-20 & 02:49 & 120 & 303.0     & 1.075 - 1.058 & 1.24 \\
Offset sky & 2014-08-20 & 02:59 & 120 & 303.0     & 1.055          & 1.27 \\
           &            &       &     &         &                &      \\
\hline
\end{tabular}
\tablefoot{The coordinates of the central star (CS) are: $21^{h} 04^{m} 10.^{s}8$, 
$-11^{\circ} 21' 48.''3$ (J2000).
}
\label{tab:observations}
\end{table*}

An offset sky position was only observed once per sequence and was located
at ($\Delta \alpha$ = $+$90.0$''$, $\Delta \delta$ = $+$120.5$''$) ensuring 
that the field was well beyond the full extent of the halo of NGC~7009. 
This field turned out to be ideal with only a few dozen stars, none bright 
and with no detection of a halo (real or telescope-induced) from NGC~7009.

\section{Reduction and analysis of MUSE cubes}
\label{Reduction}

\subsection{Basic reduction}
The MUSE data sets were reduced with the instrument pipeline version 1.0 
\citep{Weilbacher2014}. For each night of observation, a bias frame, master 
flat and wavelength calibration table were produced using the MUSE pipeline 
tasks from the closest available bias, flat and arc lamp exposures 
respectively, generally taken on the same night. The mean spectral 
resolving power quoted by the fit of the $muse\_wavecal$ task to the arc 
lamp exposures was 3025$\pm$75. The line spread function derived on each 
night (using $muse\_lsf$) was employed for extraction of each slice as a 
function of wavelength. The default (V1.0) pipeline geometry tables, 
giving the relative location of 
the slices in the field of view, the flux calibration for the WFM-N mode 
together with the standard Paranal atmospheric extinction curve were used.
Comparison of the datasets from the nights 2014 June 22 and 24 showed
excellent agreement in terms of wavelength repeatability and flux 
stability, in confirmation of the monitoring of the instrument and ambient 
conditions.

\begin{figure}
\resizebox{\hsize}{!}{
\includegraphics[width=0.97\textwidth,angle=0,clip]{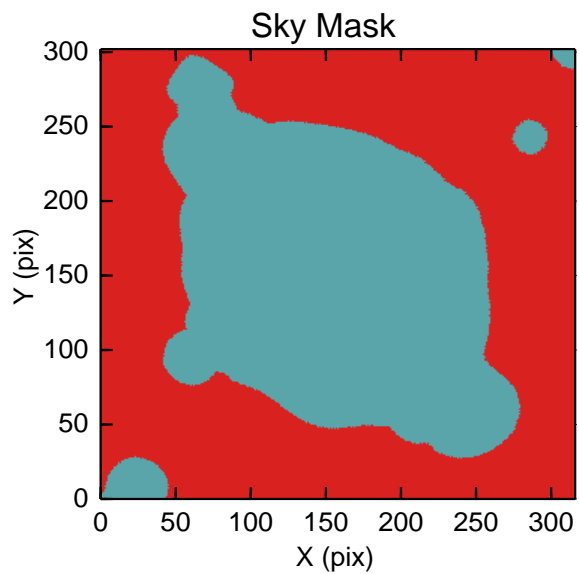}
}
\caption{Sky mask for the PA 33$^{\circ}$ exposures. The dark areas (green) 
are masked as non-sky (nebula and stars).
}
\label{fig:skymask}
\end{figure}

Following the basic reduction of each cube in the dithered sequence at the two
PA's of 33 and 123$^{\circ}$, all ten cubes were combined with the 
task $sci\_post$ to produce the wavelength calibrated, atmospheric extinction
and refraction corrected, fluxed combined cube \citep[see][]{Weilbacher2014}. 
The offset sky exposure was processed (with $muse\_create\_sky$) to form the 
sky spectrum. A sky mask was produced to ensure that genuine line-free regions 
around the nebula in the field of view were used for sky subtraction based on 
the sky spectrum for the offset sky position. Figure \ref{fig:skymask} shows 
the mask for the PA 33$^{\circ}$ exposures; the only nebular emission line 
present in the sky region is faint [\ion{O}{III}]5006.9\,\AA. Several stars in 
this background area were also included in the mask (see Fig. \ref{fig:skymask}). 
The final sky-subtracted cubes have dimensions 426 ($\alpha$) by 433 ($\delta$) 
by 3640 ($\lambda$) pixels (4750~--~9300\,\AA\ at the default binning of 1.25\,\AA).

Particular care was taken to ensure the accurate alignment of all the cubes so that
line ratios with high spatial fidelity across the wavelength range of each cube and
between cubes of different exposure times were preserved. Offsets between each cube, 
based on the position of the NGC~7009 central star, were measured in the projected 
images (both in the dithered images and those at the 90$^\circ$ offset PA) 
averaged over the wavelength range 5800-6200\,\AA. Final registration between 
the three cubes for each exposure time, to better than 0.10 pixels was achieved. 
Table \ref{tab:imagequality} lists the derived Gaussian FWHM determined with the 
IRAF\footnote{IRAF is distributed by the National Optical Astronomy Observatories, 
which are operated by the Association of Universities for Research in Astronomy, Inc., 
under cooperative agreement with the National Science Foundation.} $imexam$ task for 
three wavelengths bands, 4880~--~4920\,\AA, 6800~--~6840\,\AA\ and 9180~--~9220\,\AA, 
chosen to be free of all but very weak emission lines. The smaller FWHM of 
star images at redder wavelengths compared to 4900\,\AA\ (c.f., Tab.
\ref{tab:imagequality}) shows the expected 
inverse weak dependence of atmospheric seeing on wavelength ($\lambda^{-1/5}$ 
\citep{Fried1966}, although the FWHM falls off slightly faster than 
$\lambda^{-1/5}$) for the MUSE data cubes.

At each exposure level (10, 60 and 120s) a single cube was produced combining
all the dithers at the two PA's from the data on 22 and 24 June 2014.
Thus the total exposure per cube is 10 times the listed single dither exposure
(one central position, four offset dithers and two PA's). 

\begin{table}
\caption{Image quality with wavelength for final combined cubes (June 
2014 data only)}
\centering
\begin{tabular}{rcrr}
\hline\hline
Exposure & $\lambda$, $\Delta \lambda$ & DIMM FWHM & Image FWHM \\
(s)      &  (\AA)         & ($''$)    & ($''$)     \\
\hline
10       & 4900, 40 & 0.59 & 0.52 \\
10       & 6820, 40 & 0.58 & 0.48 \\
10       & 9200, 40 & 0.49 & 0.42 \\
         &      &      &    \\
60       & 4900, 40 & 0.61 & 0.61 \\
60       & 6820, 40 & 0.55 & 0.56 \\
60       & 9200, 40 & 0.49 & 0.48 \\
         &      &      &   \\
120      & 4900, 40 & 0.57 & 0.56 \\
120      & 6820, 40 & 0.54 & 0.53 \\
120      & 9200, 40 & 0.46 & 0.45 \\
\hline
\end{tabular}
\label{tab:imagequality}
\end{table}


\subsection{Line emission measurement}
\label{Linemis}

The emission lines in each of the three final exposure level cubes were measured 
by fitting Gaussians to the 1D spectrum in each spaxel, using a line list derived
from the tabulation of \citet{FangLiu2013}, taking account of line blending 
arising from the lower spectral resolution of MUSE.  Not all lines in this 
extensive list were used: those above about 0.002 $\times$ F(H$\beta$) were 
searched and if above a given signal-to-noise (2.5) per spaxel, then they 
were fitted. Lines within a separation $\leq$ 10\,\AA\ (8 pixels) were fitted 
by a double Gaussian, up to a set of eight consecutive close lines. A cubic 
spline fit to the line-free continuum (of bound-free,
free-free and 2-photon origin, and at the position of the central star, stellar 
continuum) provided the estimate of the continuum under the emission lines. The 
statistical errors for each voxel ($\Delta \alpha$, $\Delta \delta$, $\Delta
\lambda$ element) delivered by the MUSE pipeline were 
considered to provide errors on the Gaussian fits from the covariance matrix 
of the $\chi^{2}$ minimization (the MINUIT code \citep{James1975} is employed). 
The tables of lines fits per spaxel were then processed into emission line maps 
with corresponding error maps. The mean of the ratio of the value to the 
propagated error (referred to as signal-to-error) is reported for the various 
quantities derived from the maps, such as line ratios, electron temperatures and 
densities, in the caption of the relevant figure.


\section{Presentation of emission line maps}
\label{Emismaps}

\begin{figure*}
\centering
\resizebox{\hsize}{!}{
\includegraphics[width=0.97\textwidth,angle=0,clip]{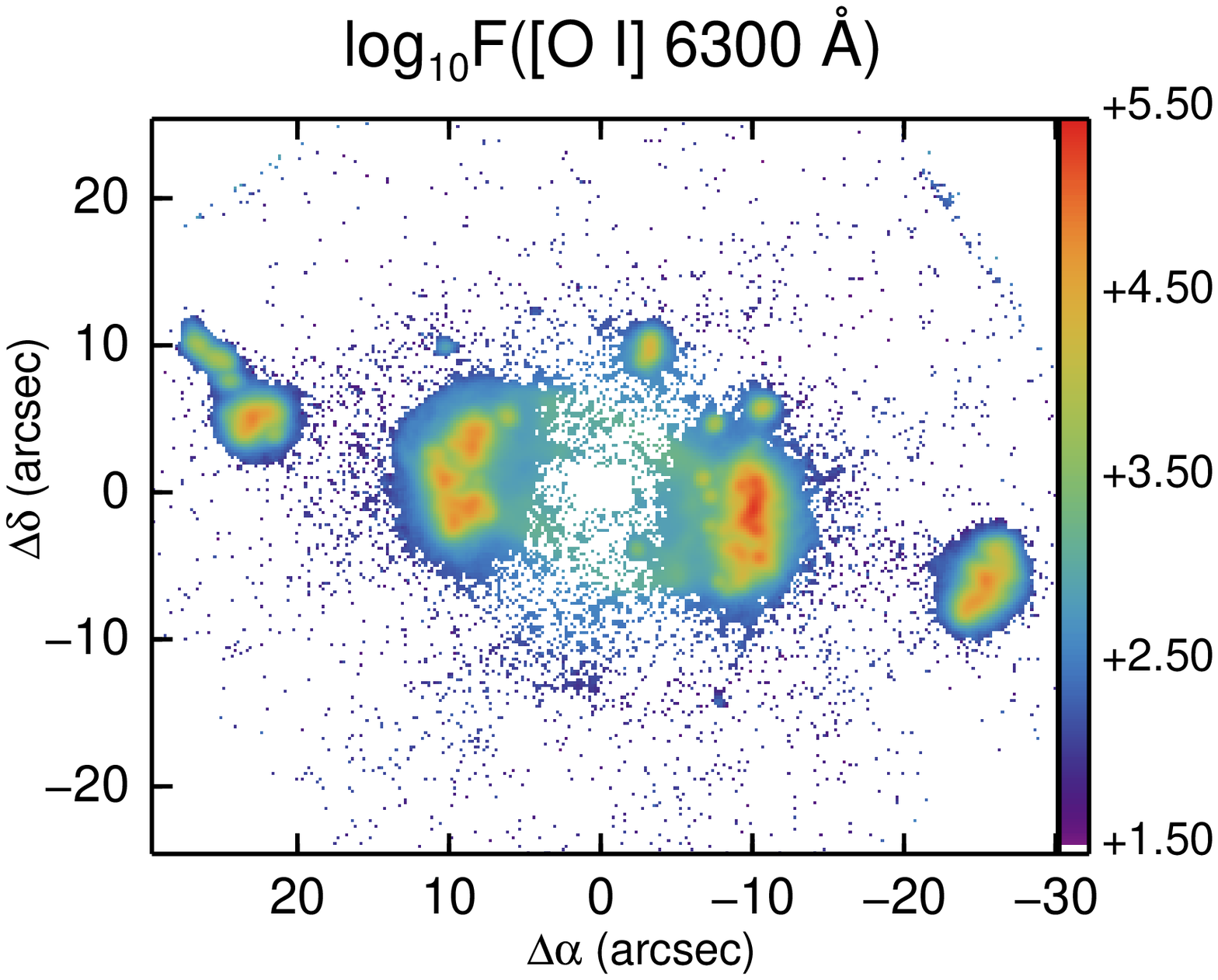}
\hspace{0.2truecm}
\includegraphics[width=0.97\textwidth,angle=0,clip]{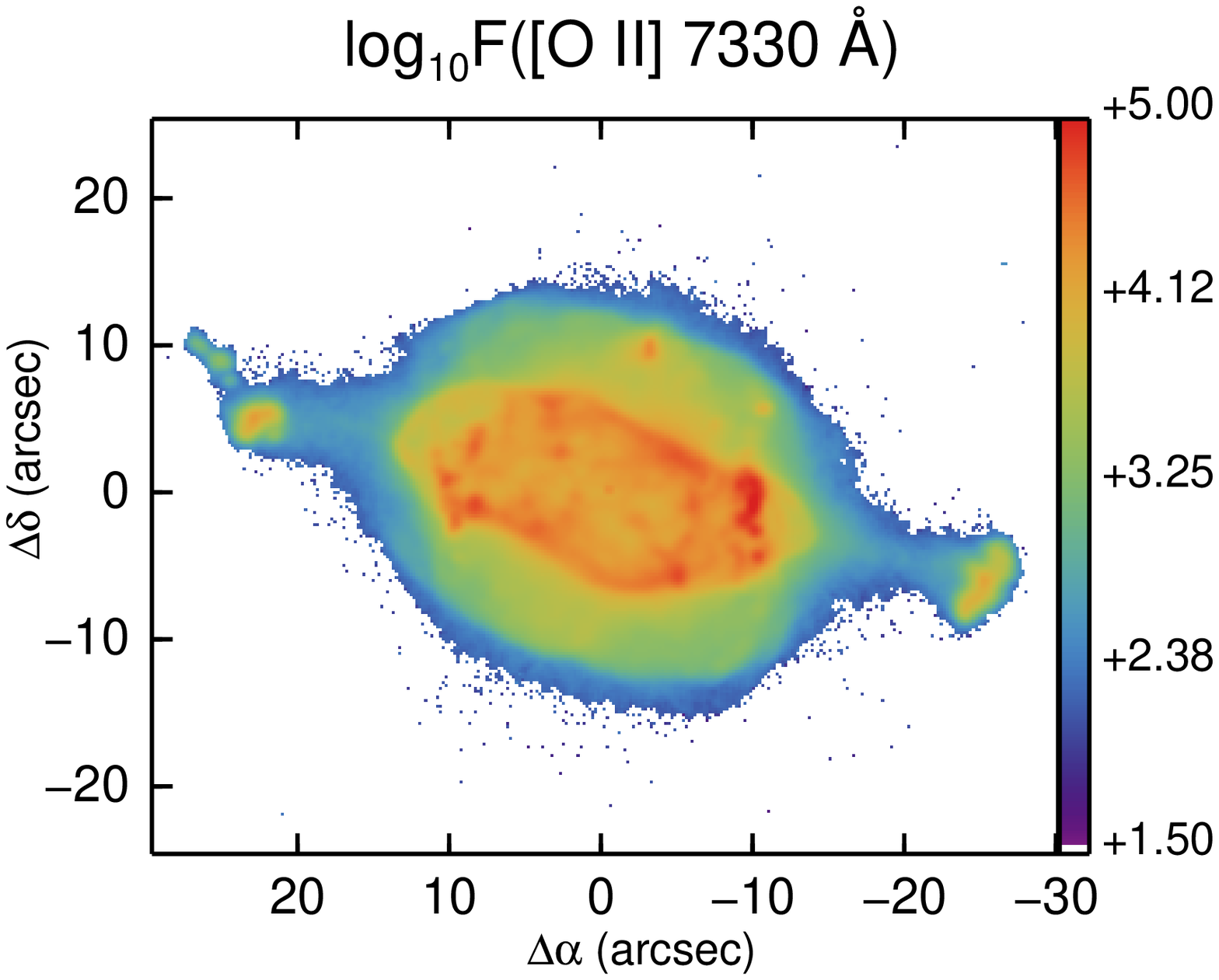}
\hspace{0.2truecm}
\includegraphics[width=0.97\textwidth,angle=0,clip]{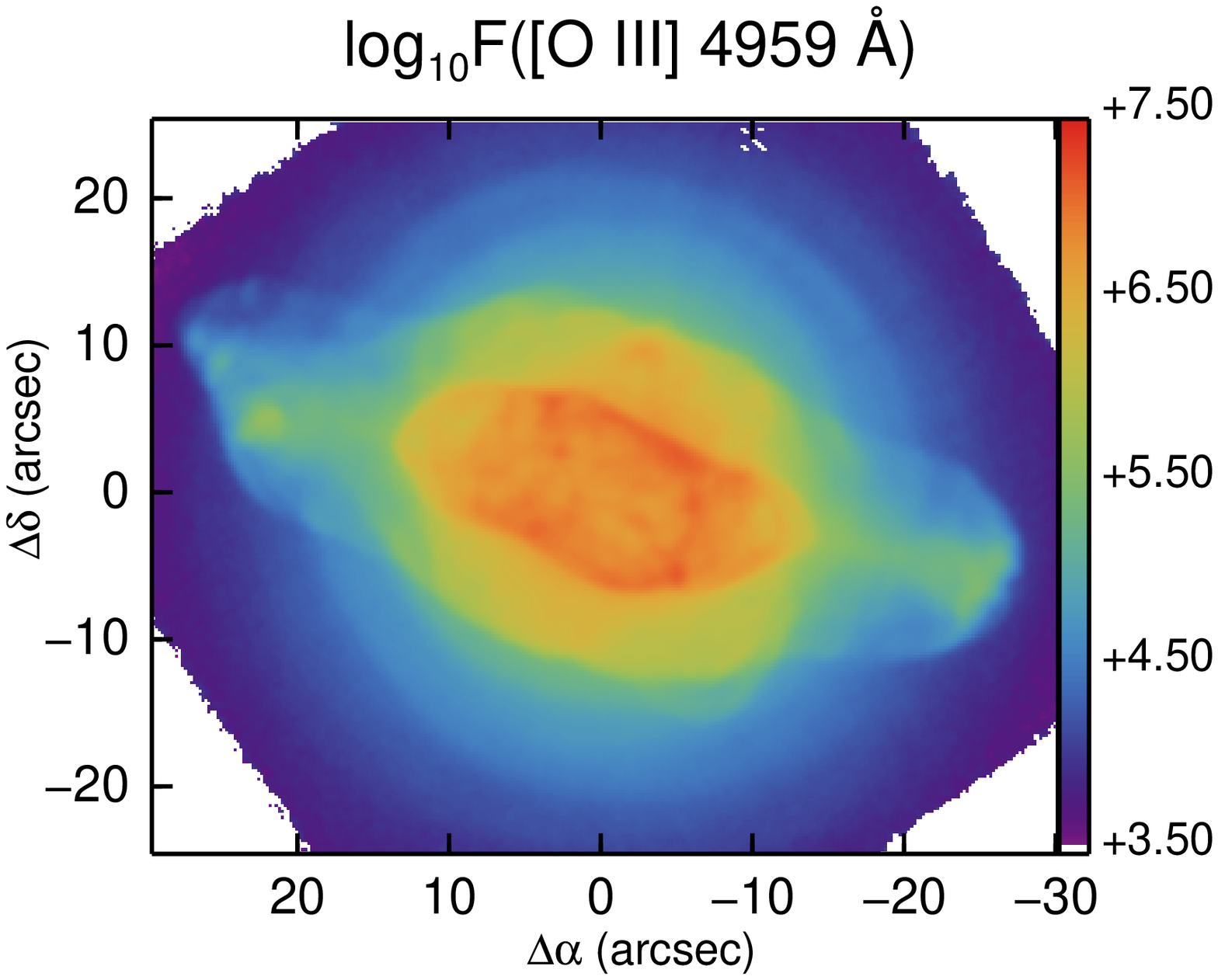}
}
\caption{Images of NGC~7009 in the O emission lines: O$^{0}$ (from the 
[\ion{O}{I}]6300.3\,\AA\ line, 120s cube), O$^{+}$ (from the 
[\ion{O}{II}]7330.2\,\AA\ line, 
120s cube) and O$^{++}$ (from the [\ion{O}{III}]4958.9\,\AA\ 10s cube).
}
\label{fig:omaps}
\end{figure*}

\begin{figure*}
\centering
\resizebox{\hsize}{!}{
\includegraphics[width=0.97\textwidth,angle=0, clip]{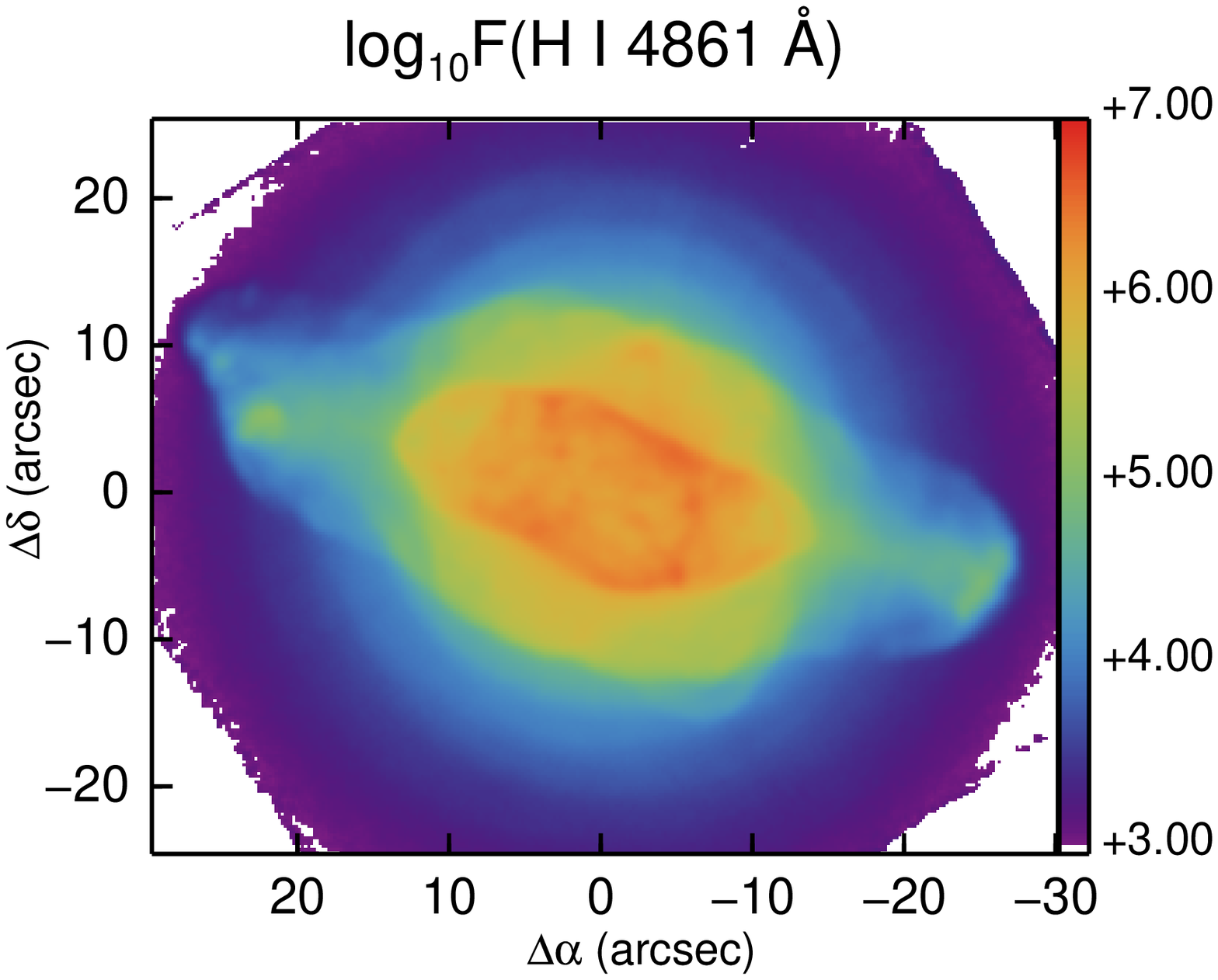}
\hspace{0.2truecm}
\includegraphics[width=0.97\textwidth,angle=0,clip]{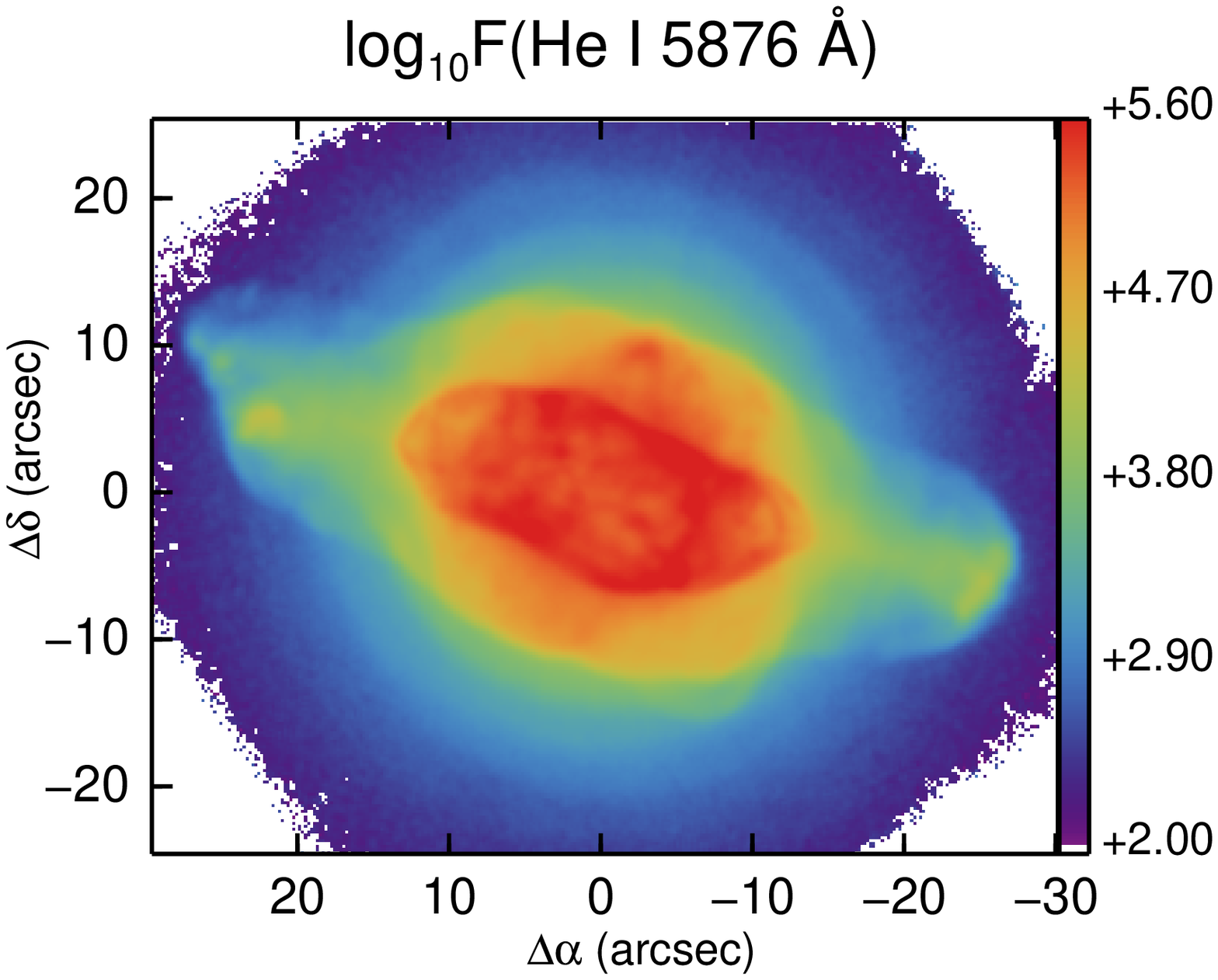}
\hspace{0.2truecm}
\includegraphics[width=0.97\textwidth,angle=0,clip]{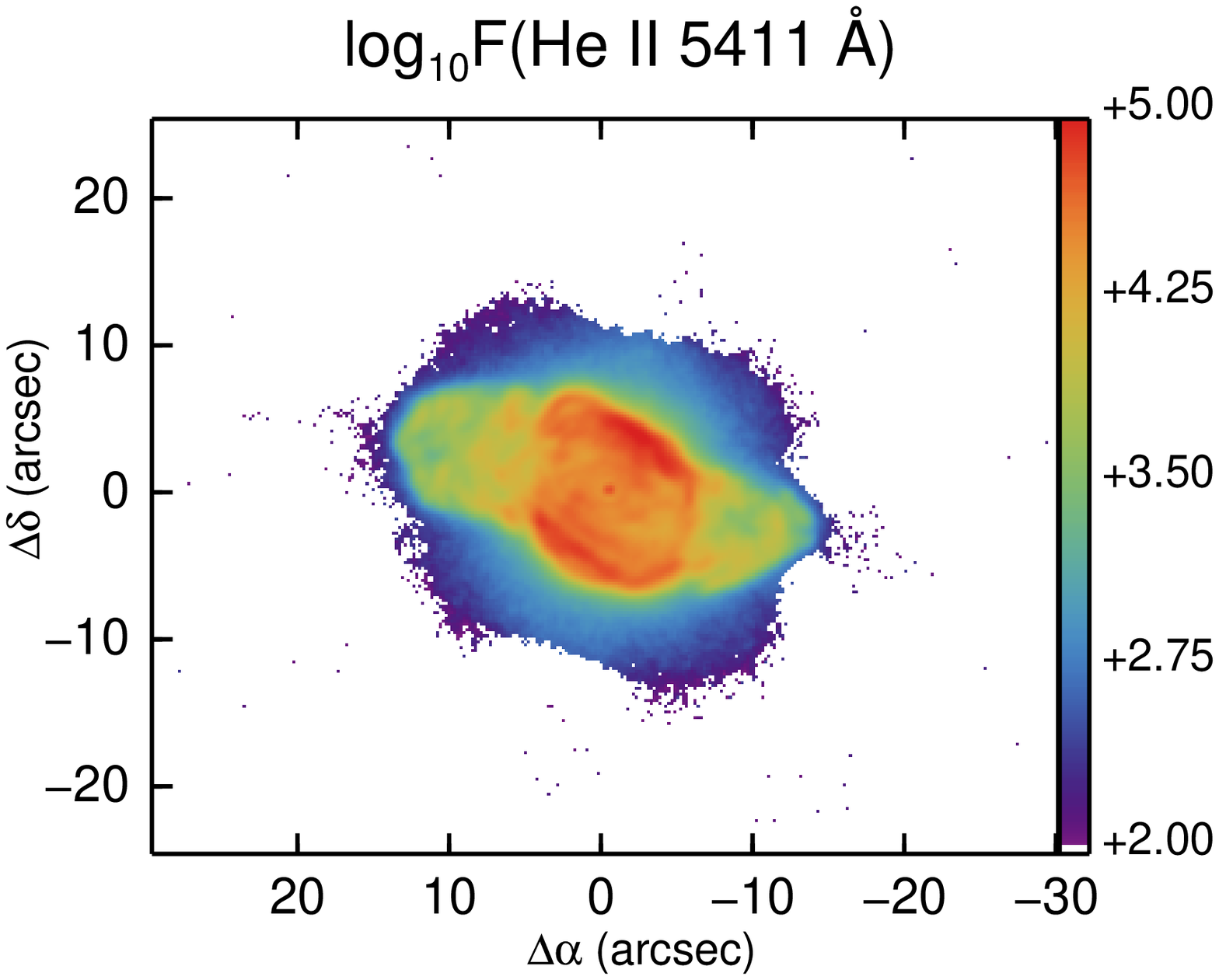}
}
\caption{Images of NGC~7009 in H and He emission line images:
H$^{+}$ as sampled by Balmer 4--2 H$\beta$ 4861.3\,\AA\ from both the 10s and 120s 
cubes; He$^{+}$ from the triplet (2P 3d~--~3D 2s) \ion{He}{I} 5875.6\,\AA\ line (120s cube); 
and He$^{++}$ from \ion{He}{II} 5411.5\,\AA\ 7--4 line (120s cube).
}
\label{fig:hhemaps}
\end{figure*}

Since the velocity extent of the emission line components in NGC~7009 
is $\sim$60 kms$^{-1}$ \citep[Fig. A.1 of][]{Schoenberner2014}, then, at the 
MUSE spectral resolution of $\sim$100 kms$^{-1}$, there is little 
velocity information on the kinematic structure: the maps of the emission 
line flux therefore encapsulate most of the information content. This
study is primarily devoted to the emission line imaging; a following work will
present analysis of integrated spectra of various regions. From the emission line 
fitting of each line (Sect. \ref{Linemis}), images were constructed 
for each fitted line (a total of 60 such maps were produced). Figure 
\ref{fig:omaps} shows the three observed lines of O: O$^{0}$ (from the 
[\ion{O}{I}]6300.3\,\AA\ line, 120s cube), O$^{+}$ (from the 
[\ion{O}{II}]7330.2\,\AA\ line, 120s cube) and O$^{++}$ (from the 
[\ion{O}{III}]4958.9\,\AA\ line, 10s cube, since the 
line is saturated in the 120s cube); O$^{+++}$ lines are not observed in the 
MUSE wavelength range. (All quoted wavelengths are as measured in air.) This 
set of images clearly shows the morphology by ionization, from the highly ionized
more central region in [\ion{O}{III}] through to the striking knots of ionization bounded 
emission in [\ion{O}{I}]. Figure \ref{fig:hhemaps} then shows some H and He line images:
H$^{+}$ as sampled by Balmer 4--2 H$\beta$ 4861.3\,\AA\ from both the 10s and 120s 
cubes; He$^{+}$ from the triplet (2P 3d~--~3D 2s) \ion{He}{I} 5875.6\,\AA\ line 
(120s cube); and He$^{++}$ from the \ion{He}{II} 5411.5\,\AA\ 7-4 line (120s cube). 
The H$\beta$ image shows the full
range of ionized gas structures from the brightest inner shell with its strong
rim, the converging extensions of the ansae, the minor axis knots (the 
northern one stronger) and the halo with its multiple rims~--~six can be counted
\citep{Corradi2004}; see \citet{Balick1994} and \cite{Sabbadin2004} for the 
nomenclature of the nebula features. The fainter outer extensions found on the 
deepest ground-based images \citep{Moreno-Corral1998} are not detectable on these 
images and this region was treated as part of the sky background. 

\begin{figure*}
\centering
\resizebox{\hsize}{!}{
\includegraphics[width=0.97\textwidth,angle=0,clip, bb=30 95 530 485]{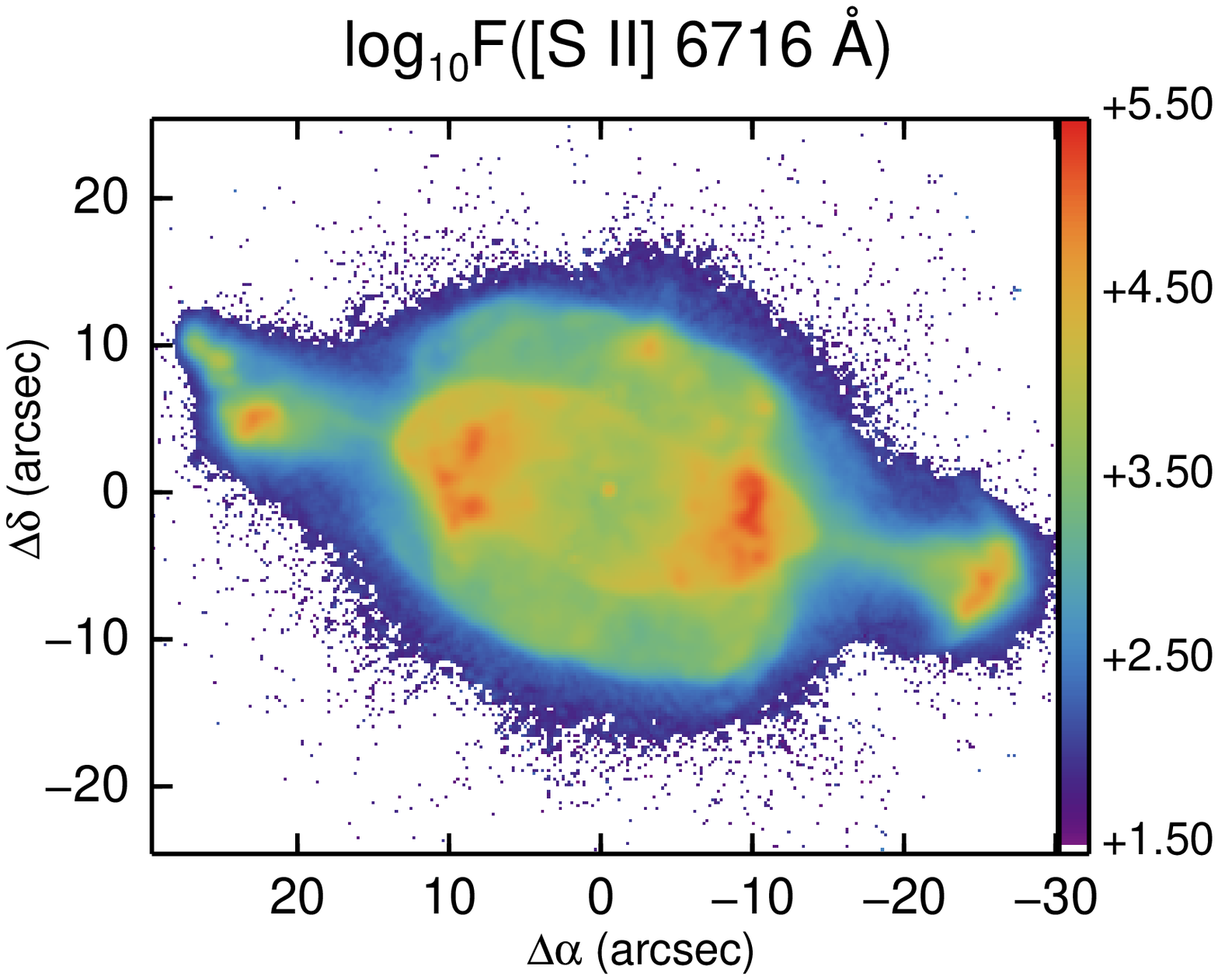}
\hspace{0.2truecm}
\includegraphics[width=0.97\textwidth,angle=0,clip, bb=30 95 530 485]{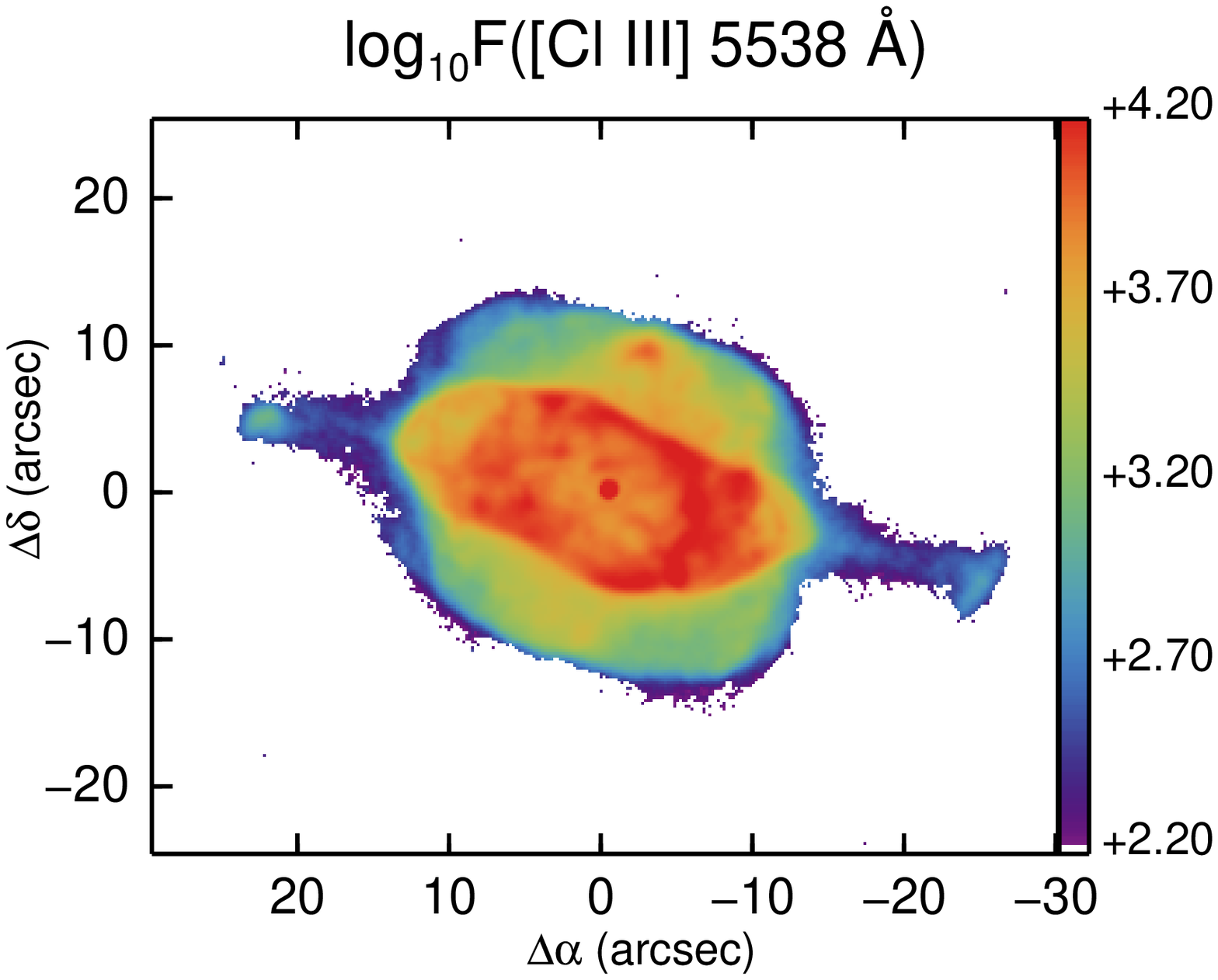}
}
\resizebox{\hsize}{!}{
\includegraphics[width=0.97\textwidth,angle=0,clip, bb=30 95 530 485]{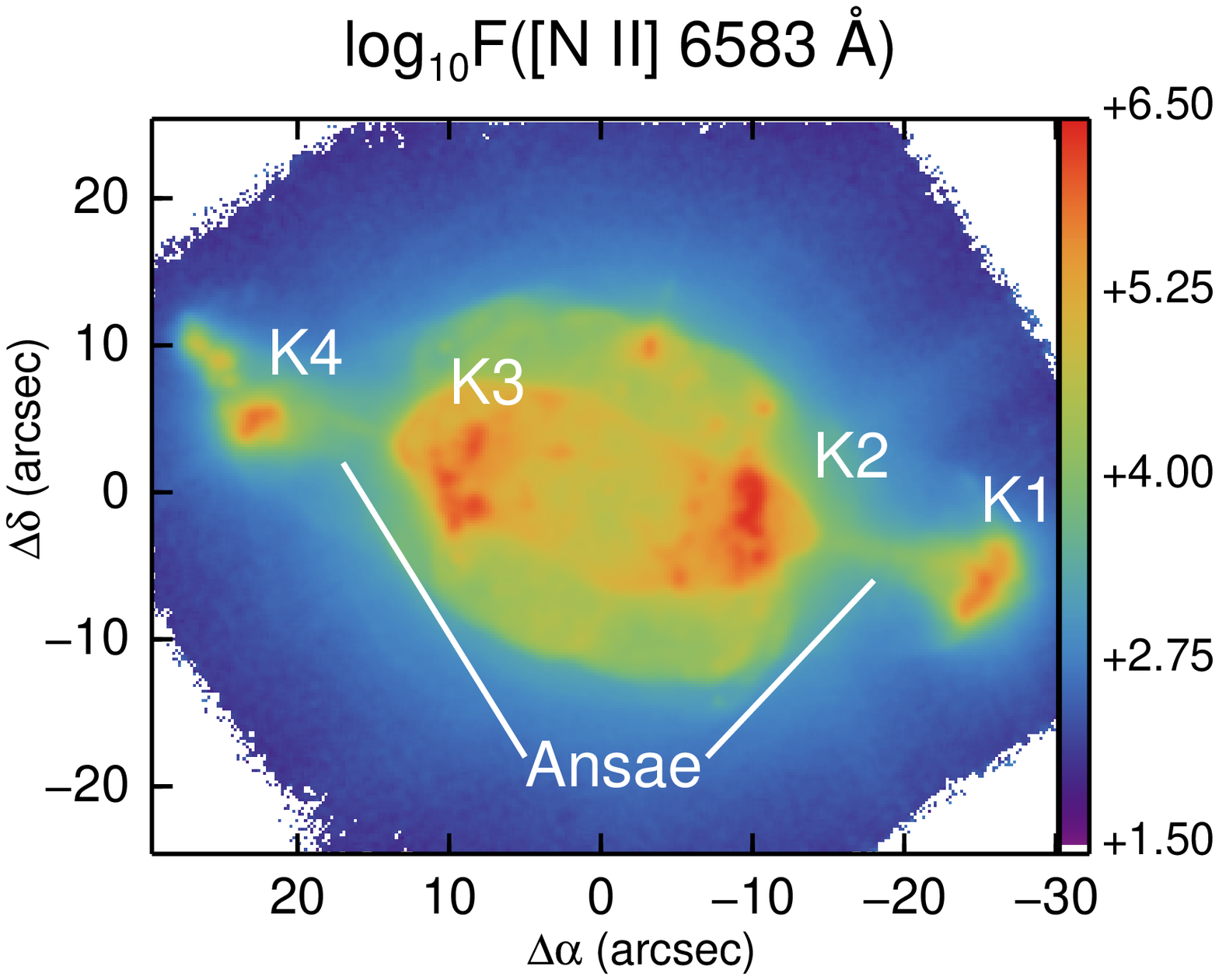}
\hspace{0.2truecm}
\includegraphics[width=0.97\textwidth,angle=0,clip, bb=30 95 530 485]{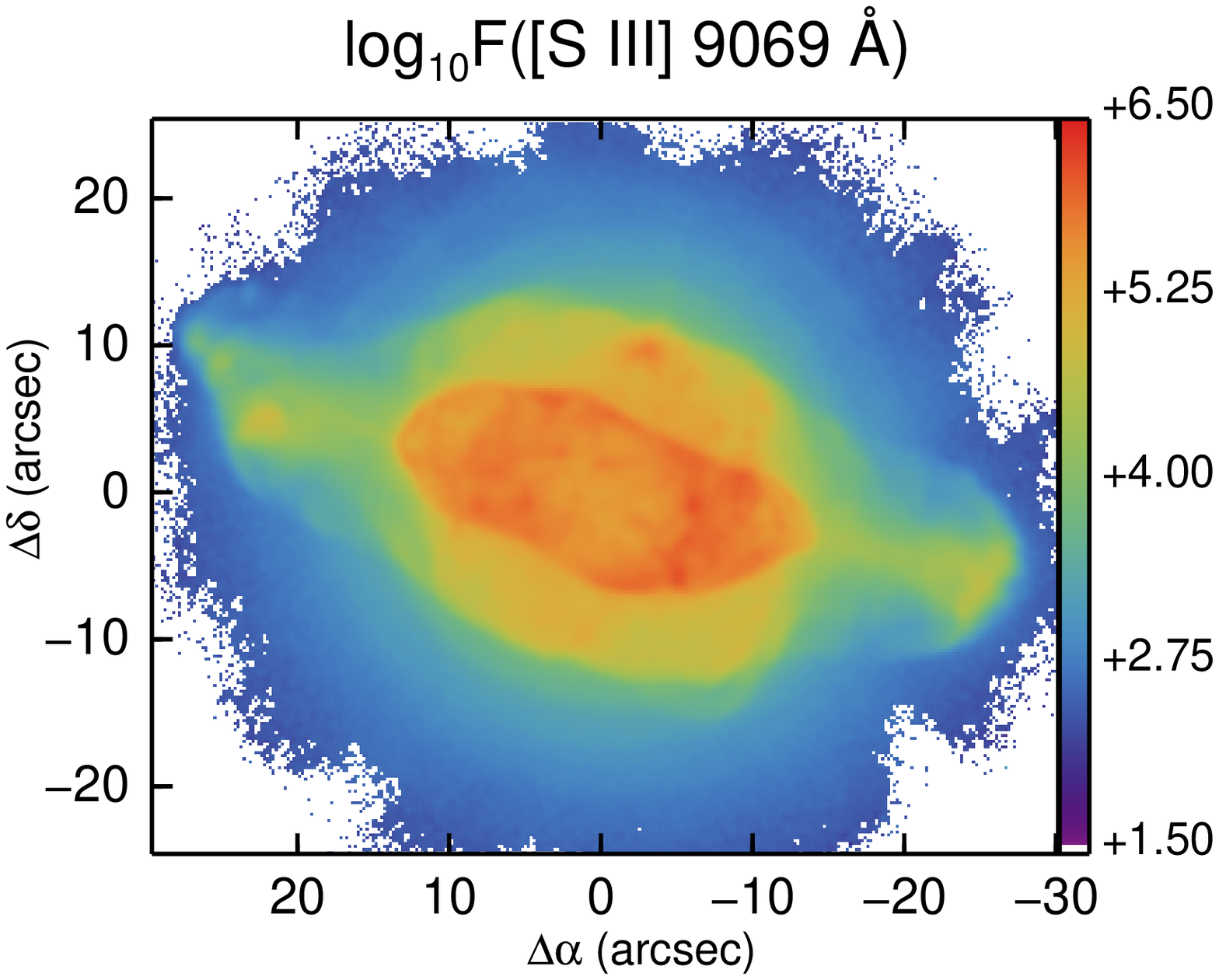}
}
\caption{Images if NGC~7009 in $N_{\rm e}$ and $T_{\rm e}$ sensitive 
emission lines: [\ion{S}{II}]6716.4\,\AA\ line for $N_{\rm e}$; 
[\ion{Cl}{III}]5537.9\,\AA\ for $N_{\rm e}$; [\ion{N}{II}]6583.5\,\AA\ for 
$T_{\rm e}$; and [\ion{S}{III}]9068.6\,\AA\ for $T_{\rm e}$. On the 
[\ion{N}{II}] image the ansae are indicated and the designation of the 
various knots from \cite{Goncalves2003}.
}
\label{fig:nsclmaps}
\end{figure*}

While the morphology of the \ion{He}{I} image is very similar to H$\beta$ (Fig. 
\ref{fig:hhemaps}), the high ionization \ion{He}{II} image differs strongly. The 
\ion{He}{II} emission is mostly confined within the bright rim and is strongest 
along the minor axis. However between the rim and the ansae along the major axis 
there is patchy extended emission indicating a preferential escape of high 
ionization photons along the major axis as compared to the minor axis. The 
ansae (designated Knots 1 and 4 by \citet{Goncalves2003} and indicated 
on Fig. \ref{fig:nsclmaps}), while seen in He$^{+}$ and H$^{+}$, are not 
detected in He$^{++}$ emission. The inner shell is however not entirely 
optically thick to high ionization photons as there is He$^{++}$ emission in 
the outer shell, which has a boxy appearance in Fig. \ref{fig:hhemaps}.
   
%
%

Figure \ref{fig:nsclmaps} shows a variety of other line emission images used in
deriving diagnostics: the [\ion{S}{II}]6716.4\,\AA\ line for $N_{\rm e}$ from the 
6716.4/6730.8\,\AA\ ratio; [\ion{Cl}{III}]5537.9\,\AA\ for $N_{\rm e}$ from the 
5517.7/5537.9\,\AA\ ratio; [\ion{N}{II}]6583.5\,\AA\ for $T_{\rm e}$ from the 
5754.6/6583.5\,\AA\ ratio; and [\ion{S}{III}]9068.6\,\AA\ for $T_{\rm e}$ from 
the 6312.1/9068.6\,\AA\ ratio. 

\begin{figure*}
\centering
\resizebox{\hsize}{!}{
\includegraphics[width=0.90\textwidth,angle=0,clip, bb=30 95 550 485]{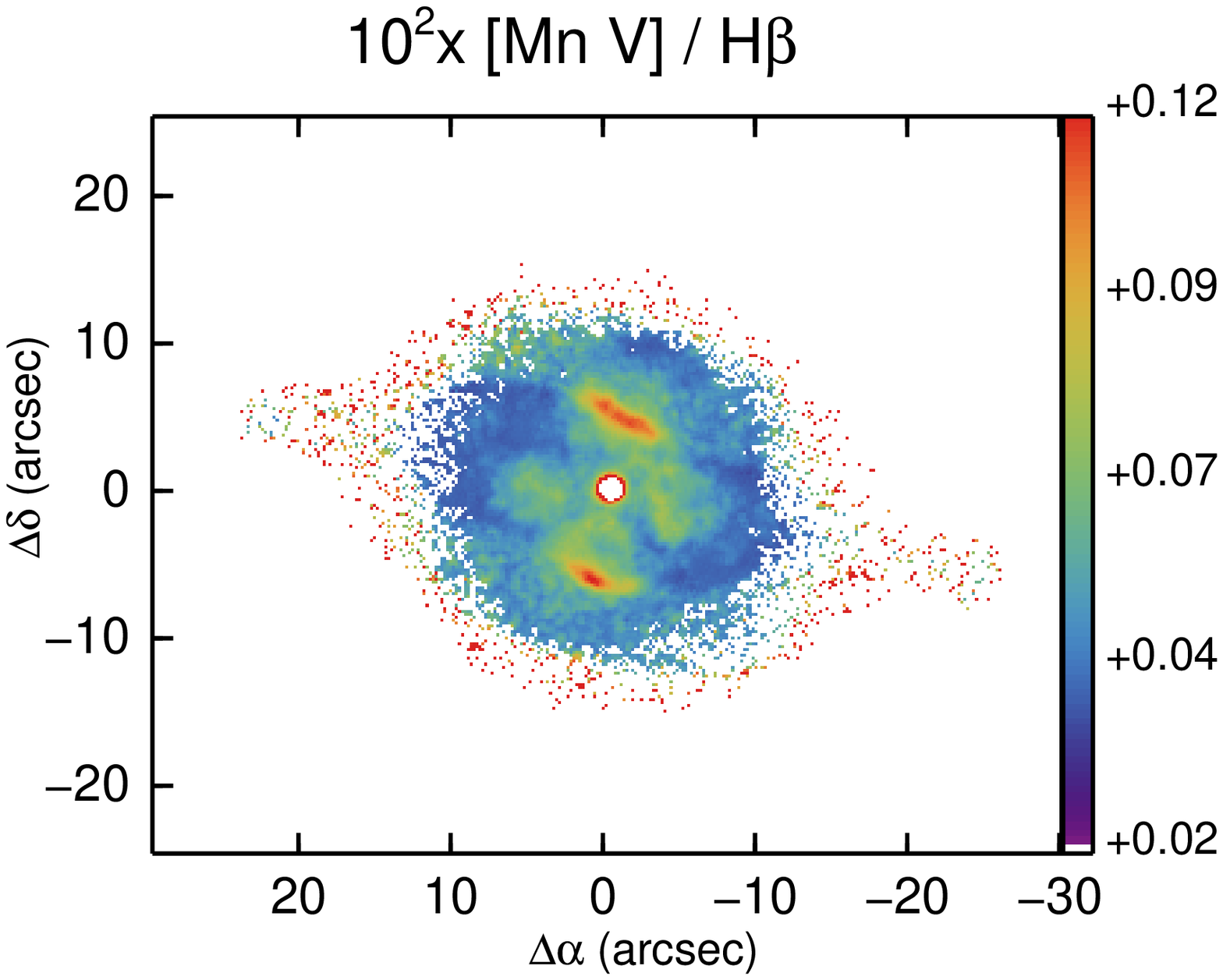}
\hspace{0.2truecm}
\includegraphics[width=0.90\textwidth,angle=0,clip, bb=30 95 550 485]{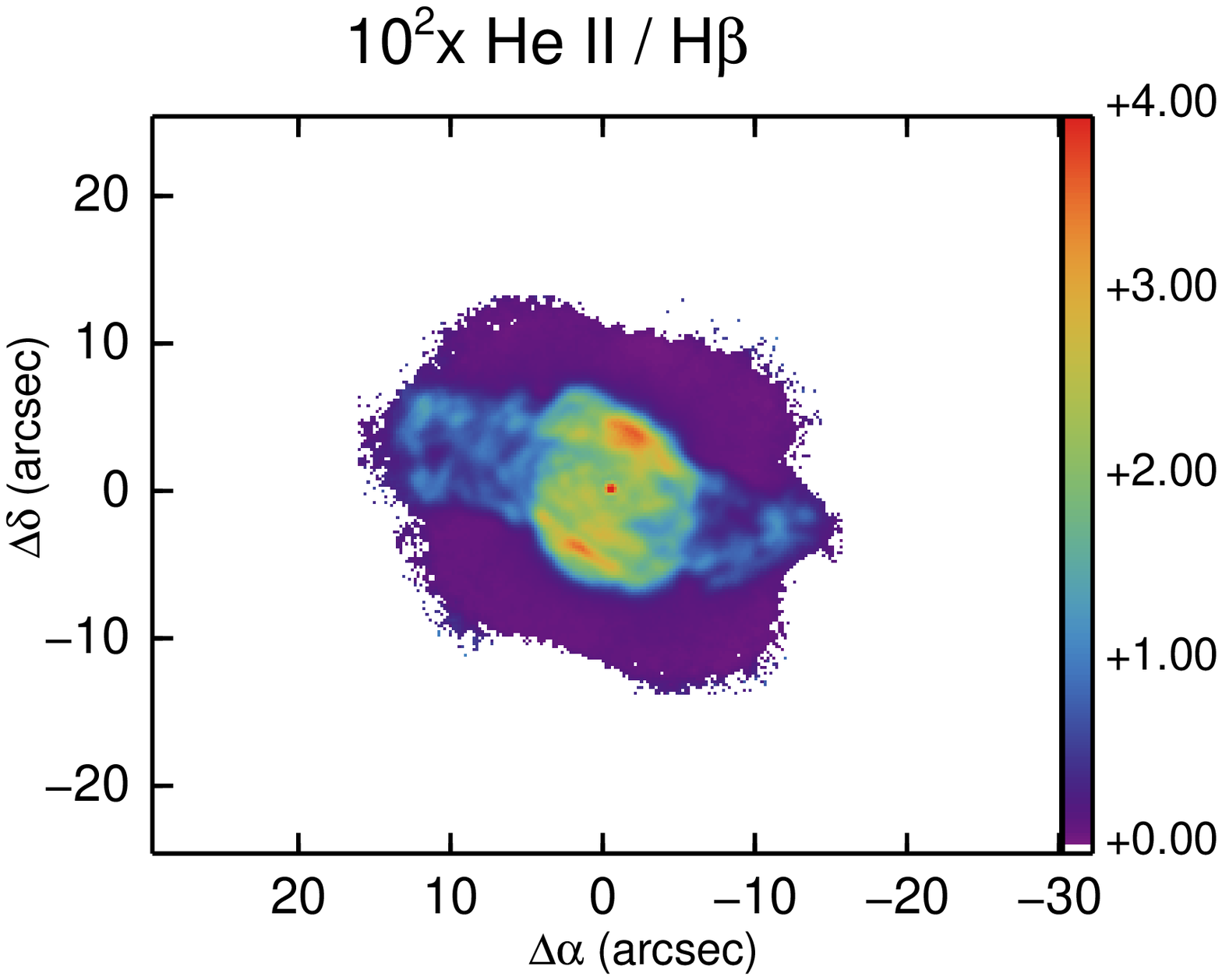}
}
\resizebox{\hsize}{!}{
\includegraphics[width=0.90\textwidth,angle=0,clip, bb=30 95 550 485]{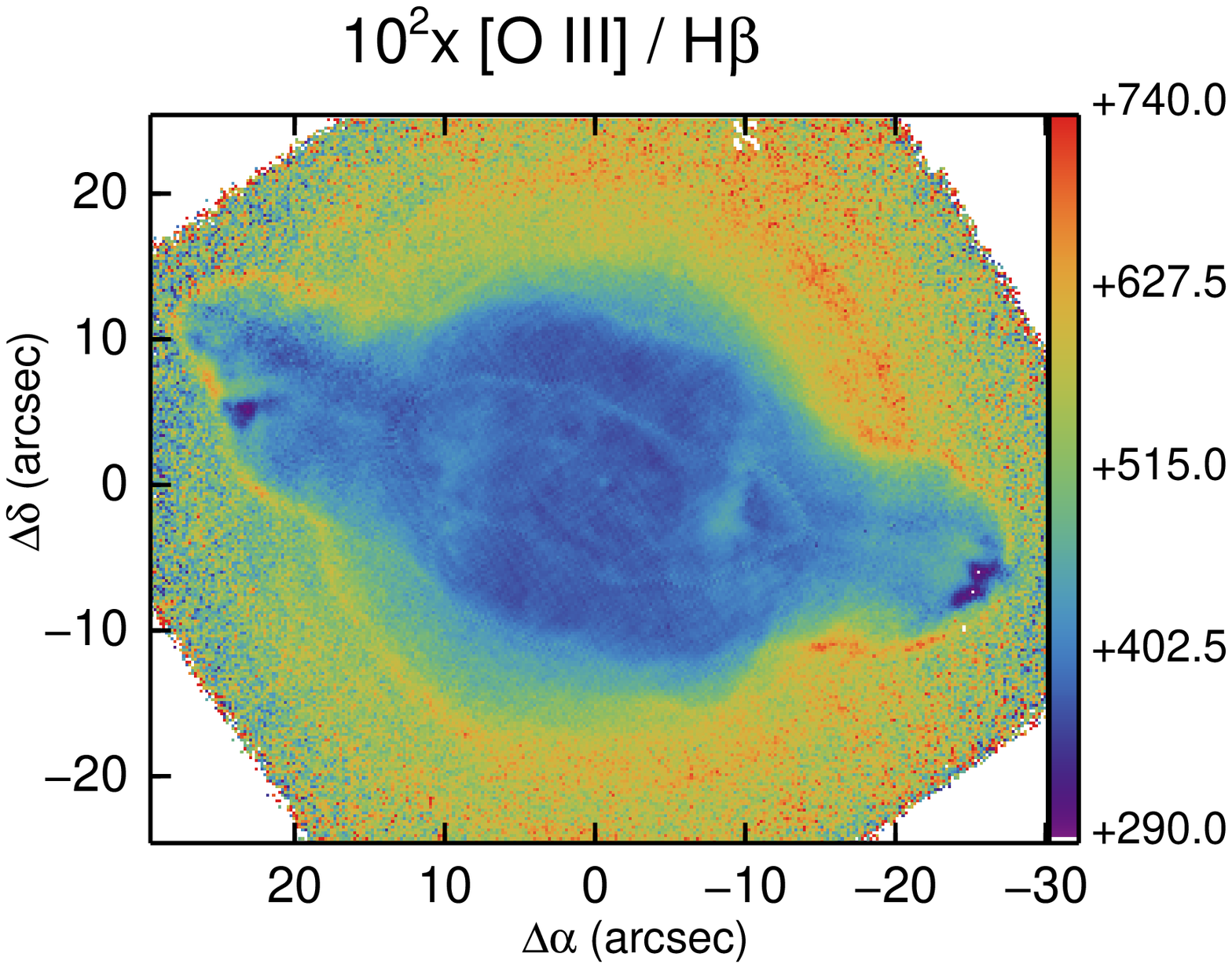}
\hspace{0.2truecm}
\includegraphics[width=0.90\textwidth,angle=0,clip, bb=30 95 550 485]{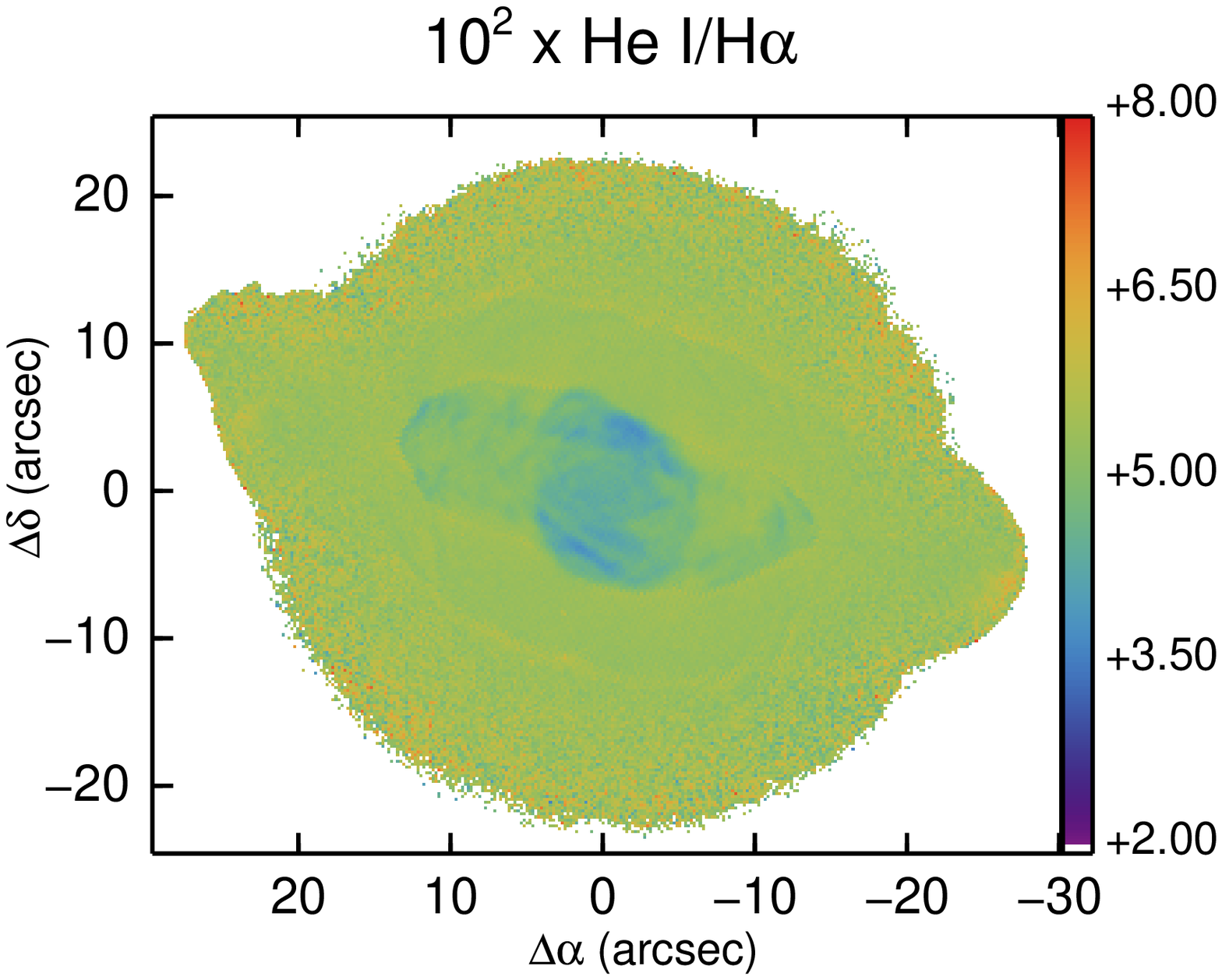}
}
\caption{Simple observed emission line ratio images (linear scale, $\times$ 100) 
with respect to an H line for NGC~7009, ordered by decreasing ionization 
potential of the nominator image: 
[\ion{Mn}{V}]6393.5\,\AA / H$\beta$; \ion{He}{II} 5411.5\,\AA / H$\beta$; 
[\ion{O}{III}]4958.9\,\AA / H$\beta$; He~I 5875.6\,\AA / H$\alpha$;
[\ion{O}{II}]7330.2\,\AA / H$\alpha$; [\ion{N}{II}]6583.5\,\AA / H$\alpha$;
[\ion{S}{II}]6730.8\,\AA / H$\alpha$ and [\ion{O}{I}]6300.3\,\AA / H$\alpha$.
The 3$\times$ 3-$\sigma$ clipped means on the signal-to-error 
above 3$\times$ the error on the ratio are: 8, 35, 47, 60, 26, 41, 34 and 
29 respectively.
}
\label{fig:ratmaps}
\end{figure*}

Figure \ref{fig:ratmaps} displays some ratio maps formed by dividing the 
observed emission line images over a range of ionization to demonstrate the 
pattern of excitation (of thermal and/or shock origin) and ionization 
from centre to edge, as well as the distinct lower ionization features. 
The low extinction (Paper I) has little effect on the morphology displayed 
in these images and the ratio maps are made with respect to a nearby H line. 
Figure \ref{fig:ratmaps} groups the ratio maps in order of decreasing 
ionization: from the highest ionization potential for which a good quality map 
could be derived (i.e., without a large fraction of unfitted spaxels falling 
below the 2.5 $\sigma$ cut) [\ion{Mn}{V}]6393.5\,\AA / H$\beta$; 
\ion{He}{II} 5411.5\,\AA / H$\beta$; [\ion{O}{III}]4958.9\,\AA / H$\beta$; 
He~I 5875.6\,\AA / H$\alpha$; [\ion{O}{II}]7330.2\,\AA / H$\alpha$; 
[\ion{N}{II}]6583.5\,\AA / H$\alpha$; [\ion{S}{II}]6730.8\,\AA / H$\alpha$ and 
[\ion{O}{I}]6300.3\,\AA / H$\alpha$. The mean signal-to-error ratios over 
these maps are reported in the caption to Fig. \ref{fig:ratmaps}; the peak values
can reach $>$100 depending on the images composing the line ratio, and the
peak value for the [\ion{O}{III}]4958.9\,\AA / H$\beta$ image is 207.

The \ion{He}{II}/H$\beta$ image differs most markedly from the other images 
and shows the loops on the minor axis prominently. The 
\ion{He}{I}/H$\alpha$ image forms the complement of the \ion{He}{II}/H$\beta$ 
one, except that the outer shell is more prominent. [\ion{O}{III}]/H$\beta$ 
displays a higher ratio over the halo than the central shell regions; the 
positions of the ansae tips show enhanced [\ion{O}{III}]/H$\beta$ but, 
coincident with the strongest [\ion{N}{II}] emission, show a decrease in 
[\ion{O}{III}]/H$\beta$. The [\ion{O}{III}]/H$\beta$ and \ion{He}{I}/H$\alpha$
images show general correspondence of features over the inner shell but more 
structure in the former over the outer shell. The [\ion{O}{II}]/H$\alpha$ 
image is rather similar to the [\ion{N}{II}]/H$\alpha$ one except that the 
inner shell has more [\ion{O}{II}] features\footnote{The [\ion{O}{II}]7330.7\,\AA\ 
line (from the $^{2}P_{3/2}$~--~$^{2}D_{3/2}$ transition) has a much higher collisional 
de-excitation density (5.1 $\times$ 10$^{6}$ cm$^{-3}$ at 10\,000 K) compared 
to the [\ion{N}{II}]6583.5\,\AA\ ($^{1}D_{2}$~--~$^{3}P_{2}$) line (9.0 $\times$ 
10$^{4}$ cm$^{-3}$ ), so the [\ion{O}{II}] image may be influenced 
by the presence of higher density low ionization gas over the inner shell.}.

All the low ionization line images -- [\ion{O}{II}], [\ion{N}{II}],
[\ion{S}{II}] and [\ion{O}{I}] show a remarkably similar morphology
and a number of well-known compact structures in the 
vicinity of Knots 2 and 3 \citep{Goncalves2003}, indicated on
Fig. \ref{fig:nsclmaps} lower left \citep[called 'caps' by][]{Balick1994}. 
In the [\ion{N}{II}]/H$\alpha$ image, besides the ansae 
(K1 and K4 on Fig. \ref{fig:nsclmaps}), knots K2 and K3 are very 
prominent; the northern minor axis polar knot is also strong and its 
elongation is aligned along the vector to the central star (on HST images 
this is resolved into two sub-knots oriented radially). Similar to
K1 and K4, knots K2 and K3 show enhanced [\ion{O}{III}]/H$\beta$, but
the peaks of [\ion{N}{II}]/H$\beta$ emission are slightly displaced 
(away from the central star) from the [\ion{O}{III}]/H$\beta$ peaks . 
The [\ion{O}{I}]/H$\alpha$ image is similar to [\ion{N}{II}]/H$\alpha$ 
and [\ion{S}{II}]/H$\alpha$, except that the low ionization knots have higher 
contrast (diffuse [\ion{O}{I}] is very weak). The [\ion{N}{I}]5197.9, 
5200.3\,\AA\ doublet is about six times fainter than [\ion{O}{I}]6300.3\,\AA\ 
and, while it was detected in the low ionization knots with similar morphology 
to [\ion{O}{I}], the S/N was not sufficient to derive a spaxel map without
spatial binning.

%
%

\begin{figure*}
\centering
\resizebox{\hsize}{!}{
\includegraphics[width=0.90\textwidth,angle=0,clip, bb=30 95 550 485]{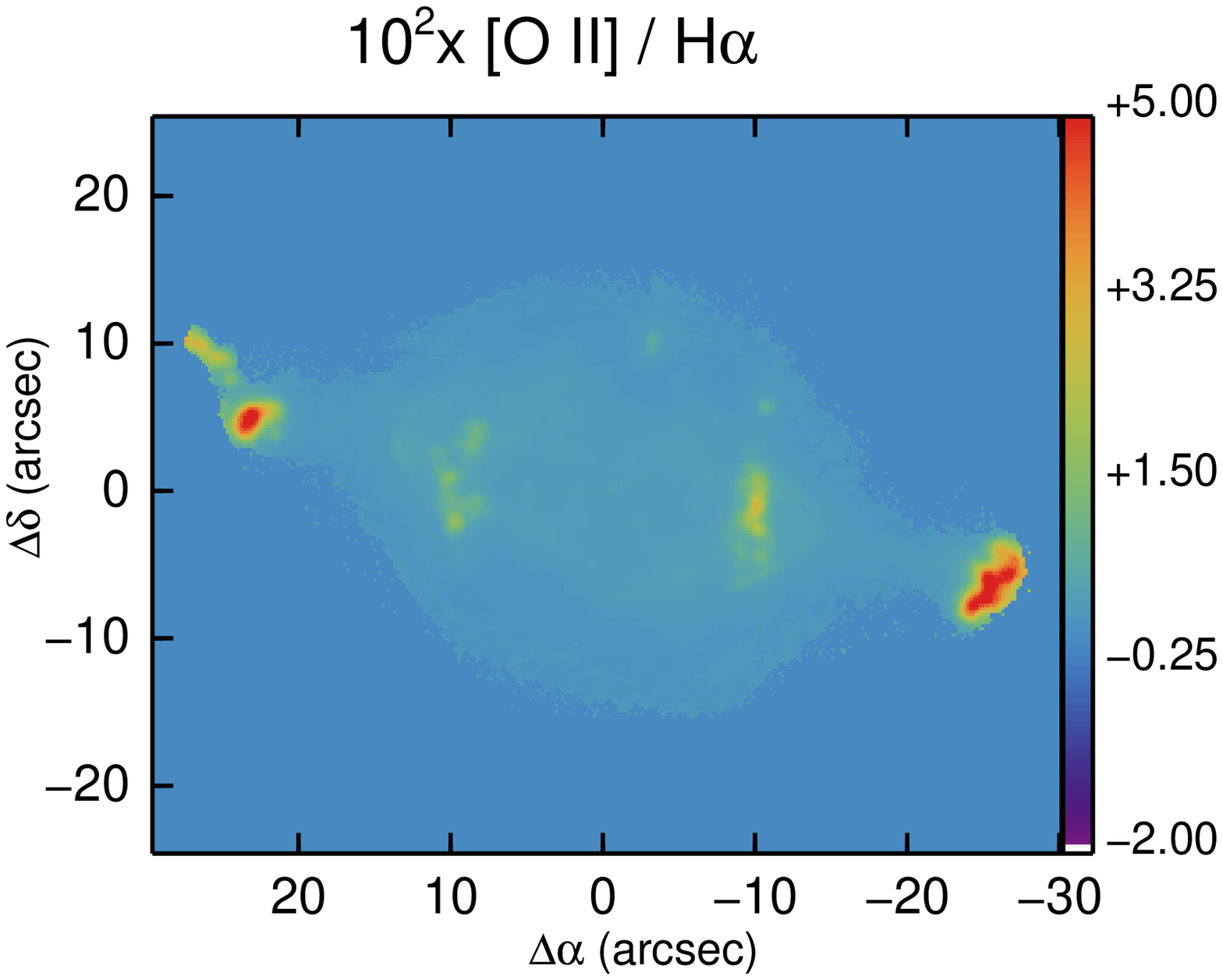}
\hspace{0.2truecm}
\includegraphics[width=0.90\textwidth,angle=0,clip, bb=30 95 550 485]{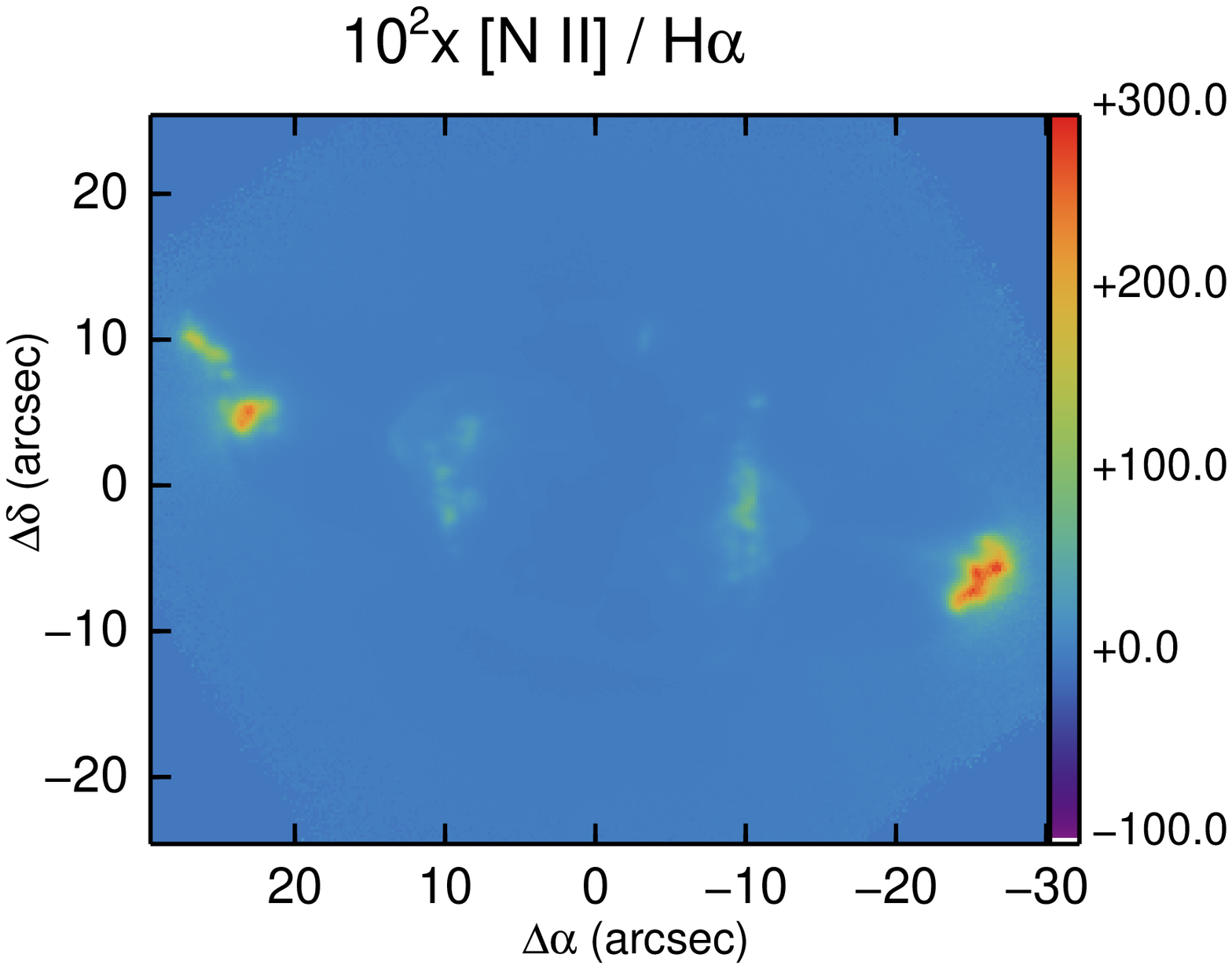}
}
\resizebox{\hsize}{!}{
\includegraphics[width=0.90\textwidth,angle=0,clip, bb=30 95 550 485]{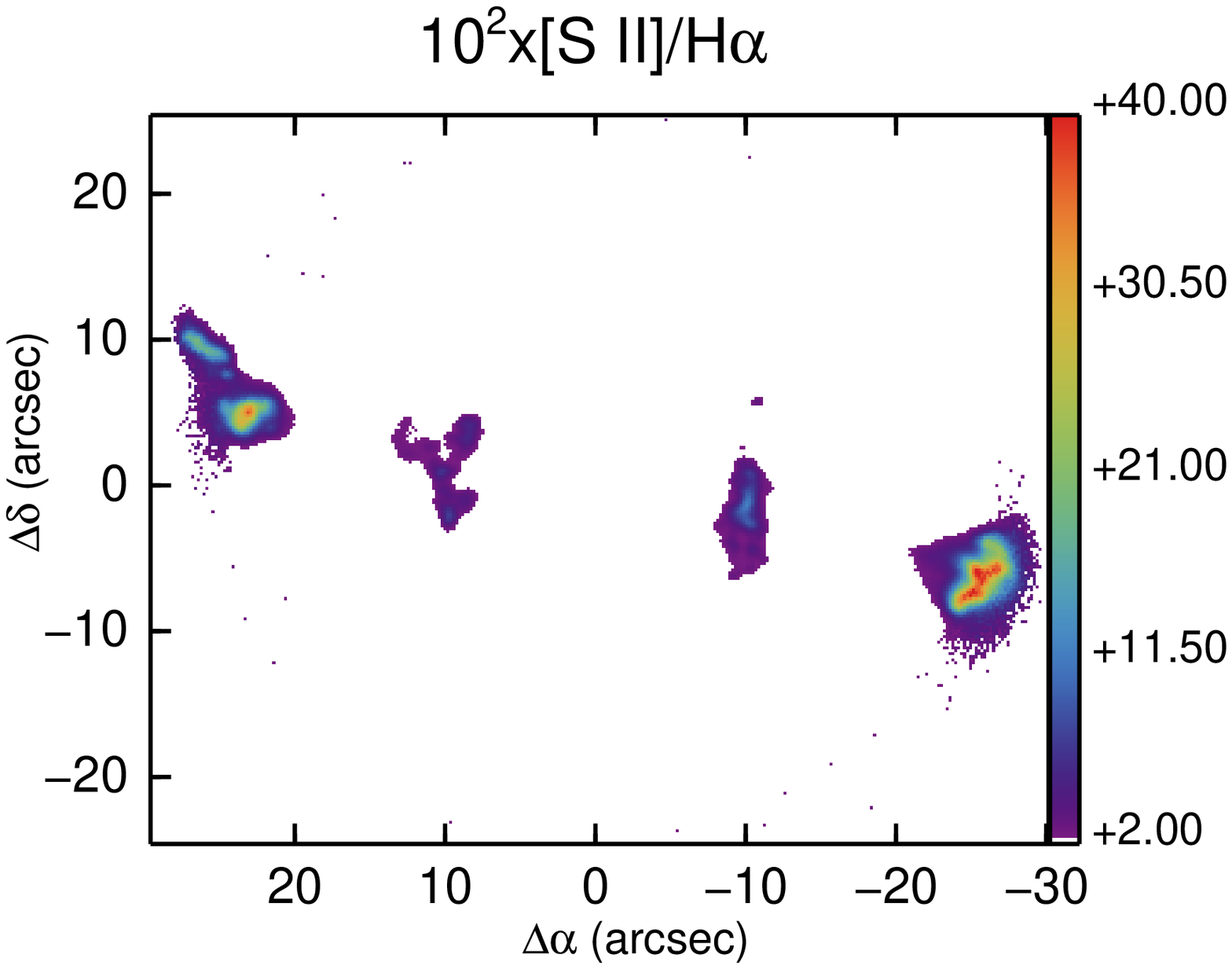}
\hspace{0.2truecm}
\includegraphics[width=0.90\textwidth,angle=0,clip, bb=30 95 550 485]{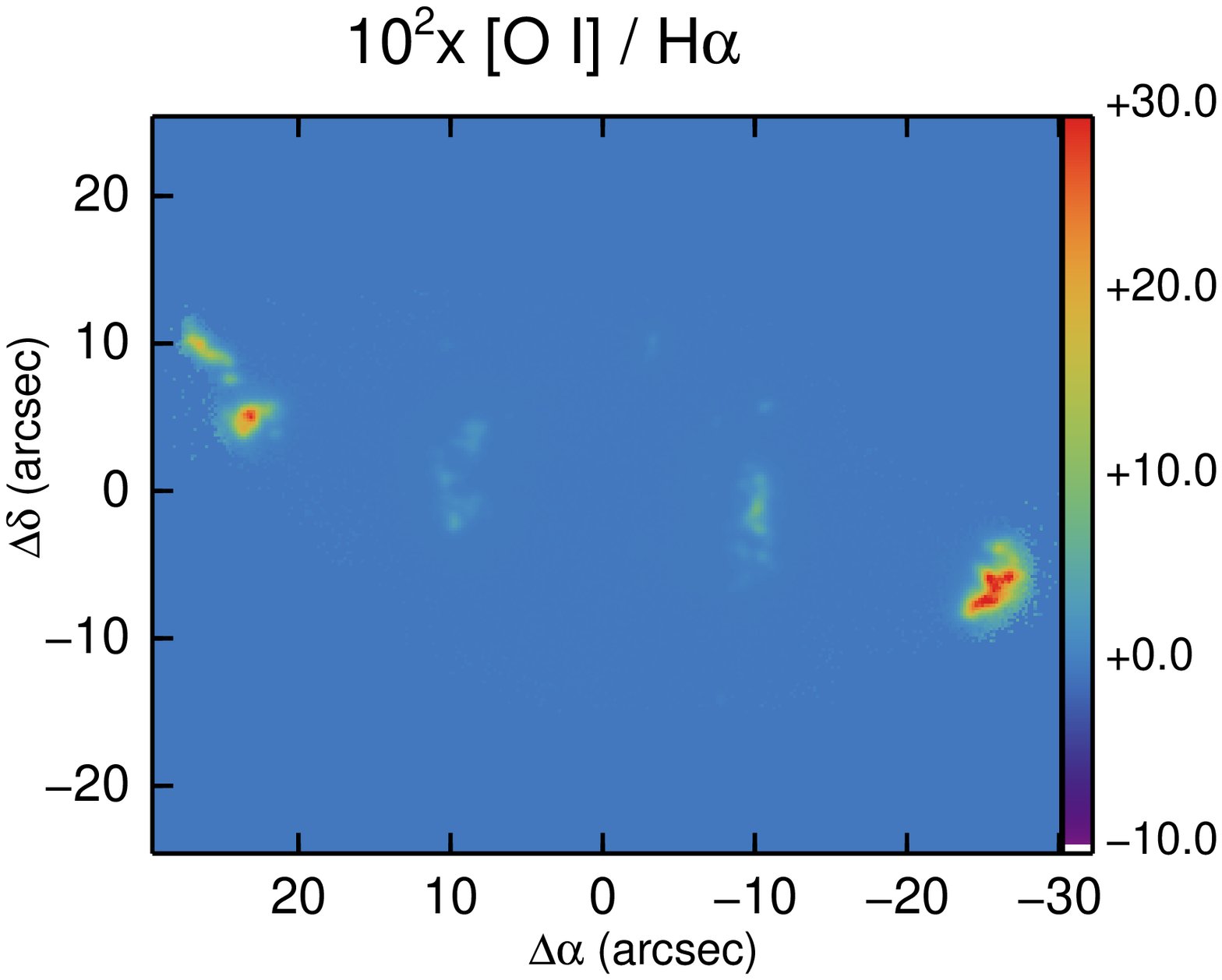}
}
\text{Fig. \ref{fig:ratmaps}. Continued.}
\end{figure*}

Although the spatial resolution of the MUSE images in Fig. \ref{fig:ratmaps} 
is about four times lower than the Hubble Space Telescope (HST) Wide
Field Planetary Camera 2 (WF/PC2) narrow band images, comparison of both
sets is interesting \citep[see][]{Balick1998, Rubin2002}. The HST images 
from Programmes 6117 (PI B. Balick) and 8114 (PI R. Rubin) allow 
emission line ratio images in [\ion{O}{III}]/H$\beta$ (although from 
5006.9\,\AA\ and not 4958.9\,\AA\ as for MUSE data), [\ion{O}{I}]/H$\alpha$, 
[\ion{N}{II}]/H$\alpha$ and [\ion{S}{II}]/H$\alpha$ to be formed for 
direct comparison with some of the images in Fig. \ref{fig:ratmaps}. The same 
knots and features are seen in both sets, but with higher resolution 
of the individual features in HST images. The outward 
increase in [\ion{O}{III}]/H$\beta$ into the outer shell is seen in both 
sets but the outermost shells are more clearly revealed in the MUSE images. 
Similarly the [\ion{N}{II}]/H$\alpha$ images are strikingly similar with a 
hint that knot K4 has developed a protuberance pointing SW in the more 
recent MUSE data (HST images from 1996). The level of [\ion{N}{II}]/H$\alpha$
over the high ionization centre of the inner shell is $\sim$3 times higher 
in the HST image, presumably on account of the continuum contribution or 
uncorrected leakage of H$\alpha$ into the [\ion{N}{II}] passband 
(WF/PC2 F658N filter). Comparison of [\ion{O}{I}]/H$\alpha$ images 
show excellent correspondence but again the HST image shows significant 
flux over the central core which is undetected in the pure emission line 
MUSE map.

\begin{figure*}
\centering
\resizebox{\hsize}{!}{
\includegraphics[width=0.97\textwidth,angle=0,clip]{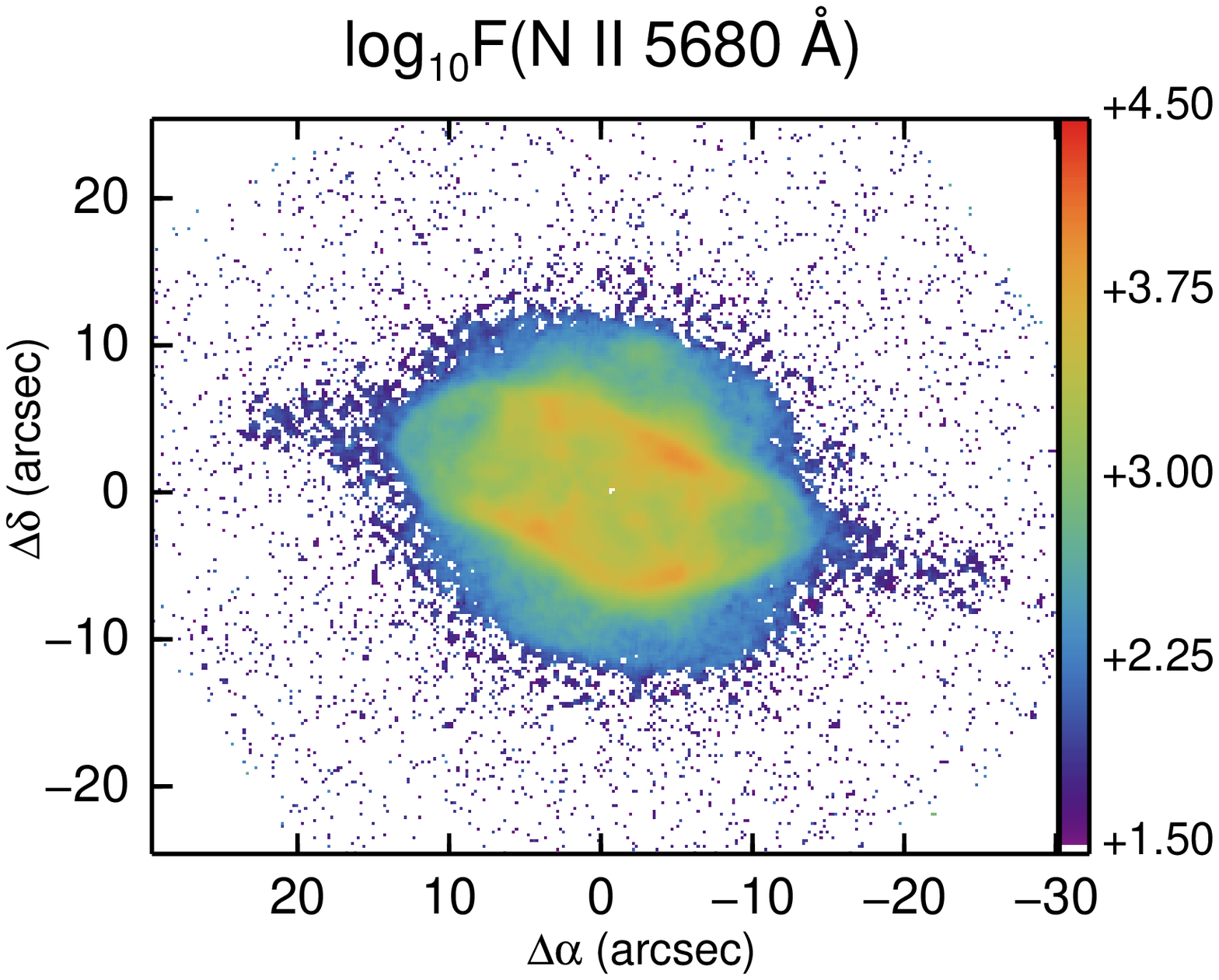}
\hspace{0.2truecm}
\includegraphics[width=0.97\textwidth,angle=0,clip]{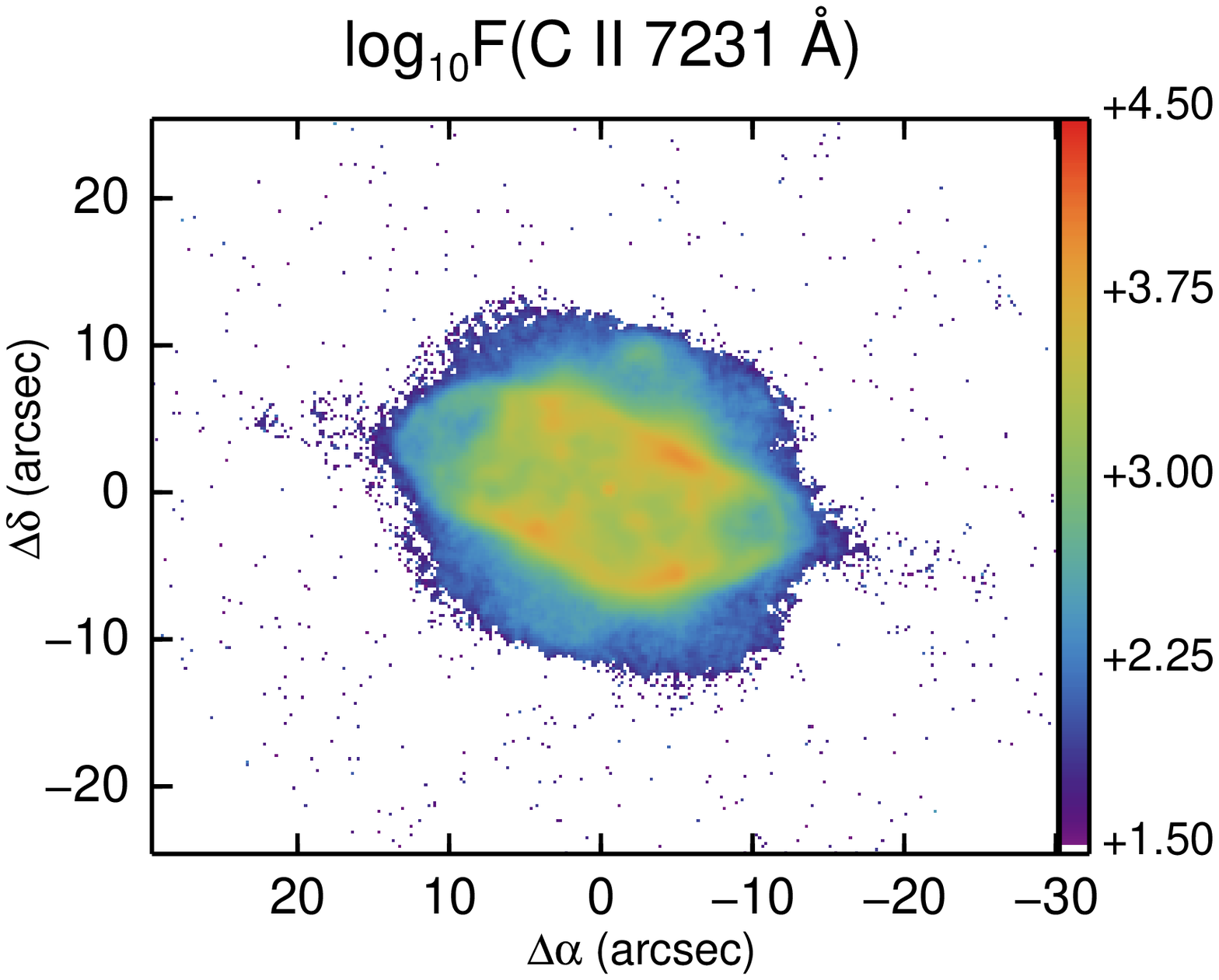}
}
\caption{Maps of two detected optical recombination lines are
shown. Left: \ion{N}{II} V3 multiplet 3p\,$^{3}$D~--~3s\,$^{3}$P line at 
5679.6\,\AA; Right: \ion{C}{II} 3d\,$^{2}$D~--~3p\,$^{2}$P line at 7231.3\,\AA.
}
\label{fig:orlmaps}
\end{figure*}

Despite the absence of the brightest \ion{O}{II} ORL's (around 4650\,\AA)
in the MUSE standard range, several of the strongest \ion{N}{II} and 
\ion{C}{II} recombination lines are captured. For \ion{N}{II}, the
multiplet 2s$^{2}$2p3p\,$^{3}$D~--~2s$^{2}$2p3s\,$^{3}$P$^{o}$ 
(V3) lines at 5666~--~5730\,\AA\ are bright enough to be
detected on a spaxel-by-spaxel basis in the 120s cube. The
(2-1) 5666.6\,\AA, (1-0) 5676.0\,\AA, (3-2) 5679.6\,\AA\ and
(2-2) 5710.8\,\AA\ could be detected on many spaxels and were 
fitted. Figure \ref{fig:orlmaps} shows the map of the strongest line
(5679.6\,\AA).

The strongest \ion{C}{II} ORL in the spectral range is 3p\,2P~--~3s\,2S 
at 6578.1\,\AA. This line was well detected in many spaxels but over the 
knots K2 and K3, where the [\ion{N}{II}]6583.5\,\AA\ is very strong, 
the fitting was not reliable as the \ion{C}{II} line lies in the wing of the
[\ion{N}{II}]6583.5\,\AA\ line which is $>$100 times stronger. In addition
the other \ion{C}{II} line from the same multiplet at 6582.9\,\AA\ makes
for challenging line fitting at the measured spectral resolution 
of 2.8\,\AA. The \ion{C}{II} 3d\,$^{2}$D~--~3p\,$^{2}$P line at 7231.3\,\AA\
is however free from nearby bright lines and an adequate map could
be produced, also shown in Fig. \ref{fig:orlmaps}.

\section{Mapping electron temperature and density}
\label{NeTemaps}

\subsection{$T_{\rm e}$ and $N_{\rm e}$ from collisionally excited species}
\label{CELTeNe}

\begin{figure*}
\centering
\resizebox{\hsize}{!}{
\includegraphics[width=0.97\textwidth,angle=0,clip]{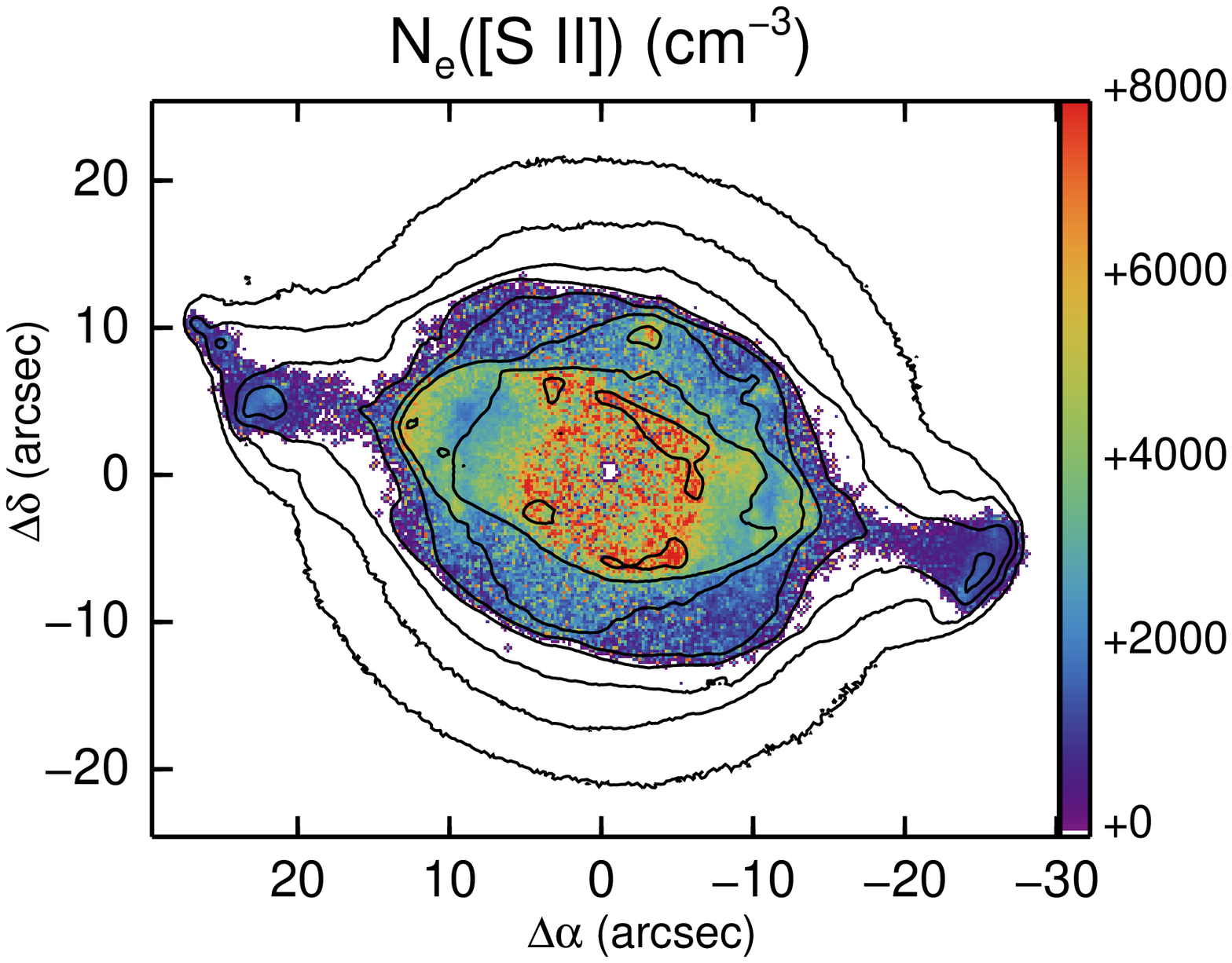}
\hspace{0.2truecm}
\includegraphics[width=0.97\textwidth,angle=0,clip]{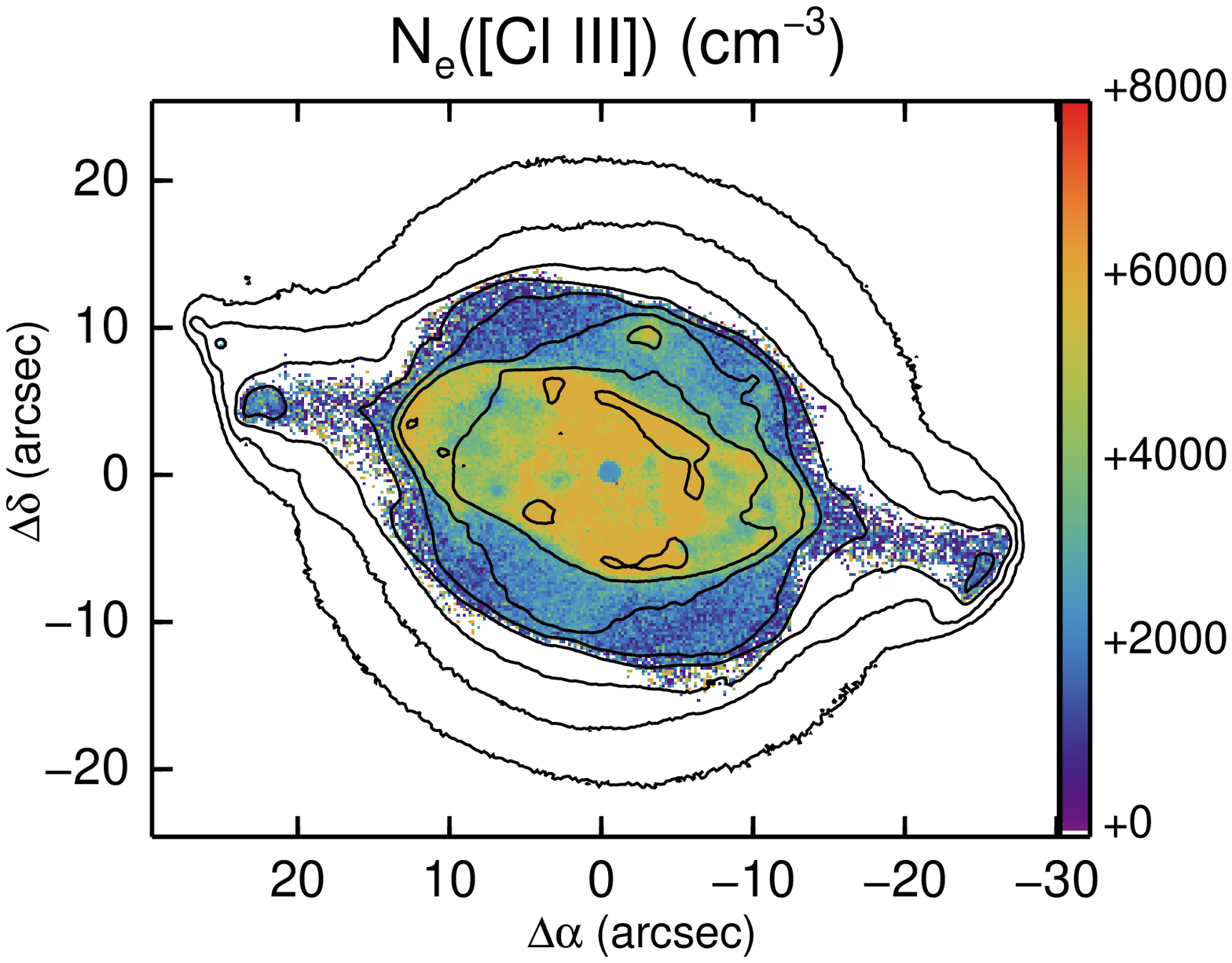}
}
\caption{Maps of $N_{\rm e}$ determined from [\ion{S}{II}] (left) and 
[\ion{Cl}{III}] (right). The contours correspond to the observed 
log F(H$\beta$) image shown in Fig. \ref{fig:hhemaps} (left), with 
contours set at log$_{10}$ F(H$\beta$) surface brightness 
(ergs cm$^{-2}$ s$^{-1}$ arcsec$^{-2}$) from -15.0 to -11.8 in 
steps of $+$0.4. The electron densities for collisional de-excitation 
of the [\ion{S}{II}] and [\ion{Cl}{III}] $^{2}D_{3/2}$ levels for these 
diagnostics are 3.1 $\times$ 10$^{3}$ and 2.4 $\times$ 10$^{4}$ 
cm$^{-3}$ at 10$^{4}$K respectively. The simple means on the 
signal-to-error value over the two $N_{\rm e}$ maps are 6 and 4,
respectively.
}
\label{fig:nemaps}
\end{figure*}

\begin{figure*}
\centering
\resizebox{\hsize}{!}{
\includegraphics[width=0.97\textwidth,angle=0,clip]{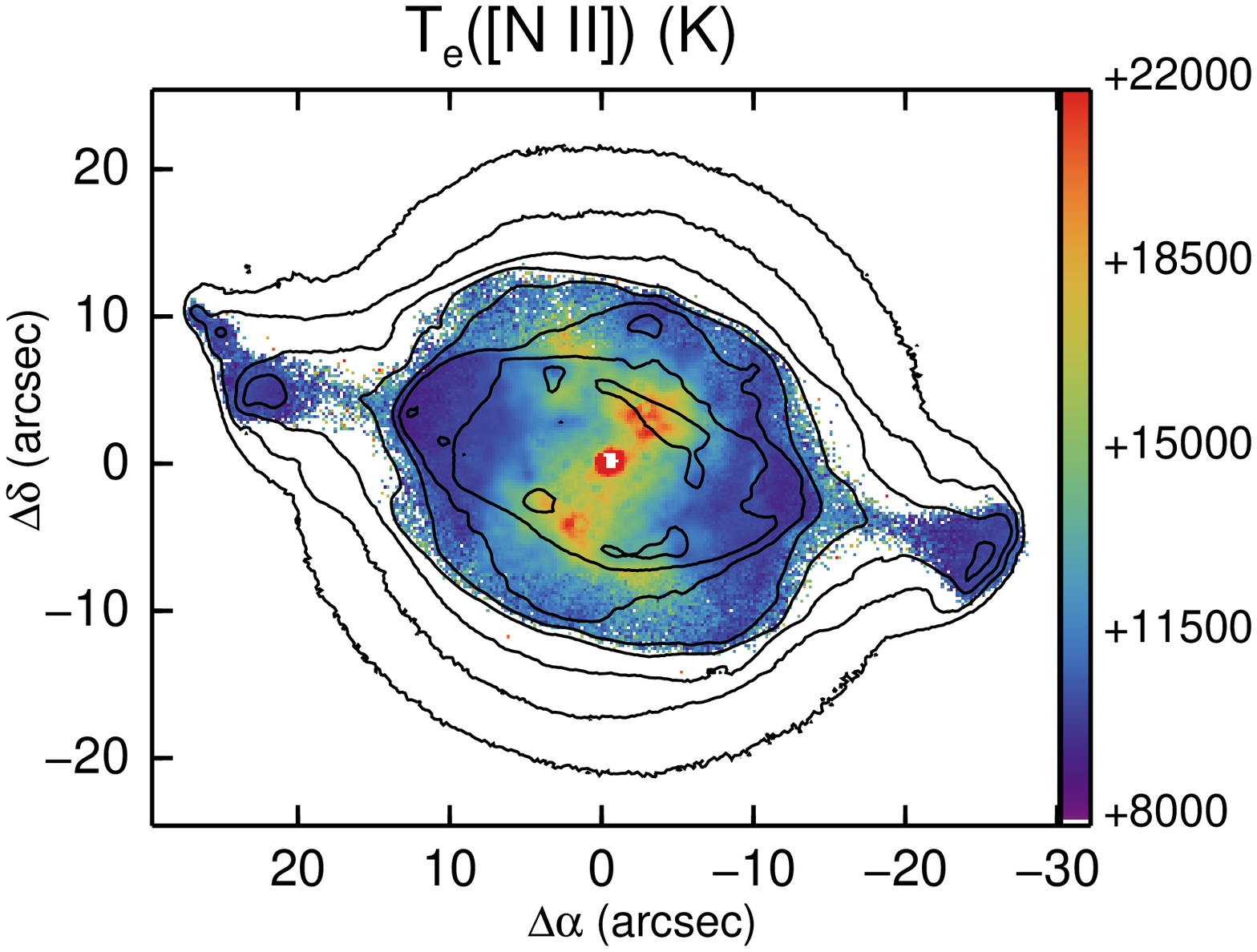}
\hspace{0.2truecm}
\includegraphics[width=0.97\textwidth,angle=0,clip]{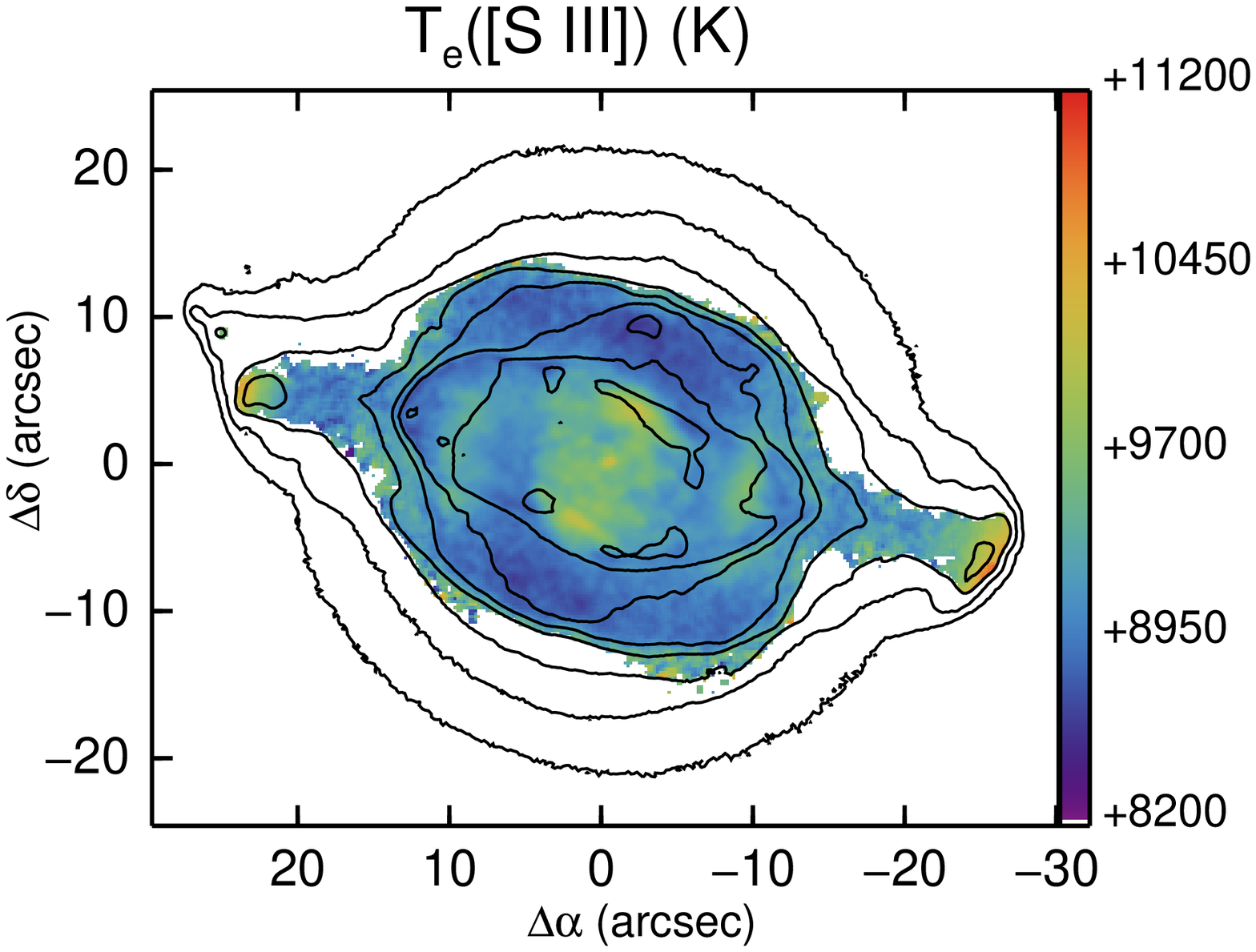}
}
\caption{Maps of $T_{\rm e}$ determined from [\ion{N}{II}] (left) and 
[\ion{S}{III}] (right). The elevated values in the central region for
the [\ion{N}{II}] $T_{\rm e}$ are assumed to be caused by the contribution of
N$^{++}$ recombination to the [\ion{N}{II}]5754.6\,\AA\ line; see text for 
a consideration. The log F(H$\beta$) surface brightness contours are 
as in Fig. \ref{fig:nemaps}. The simple means on the 
signal-to-error value over the two $T_{\rm e}$ maps are 52 and 21,
respectively.
}
\label{fig:temaps}
\end{figure*}

Within the MUSE wavelength coverage, there are a number of sets of 
collisionally excited lines (CELs) that can be employed for electron 
temperature and/or density determination, depending on the sensitivity of 
the particular lines to $T_{\rm e}$ and $N_{\rm e}$.
Among these are: [\ion{O}{I}], [\ion{N}{II}], [\ion{S}{III}],
[\ion{Ar}{III}] for $T_{\rm e}$ and [\ion{N}{I}], 
[\ion{S}{II}], [\ion{Cl}{III}] for $N_{\rm e}$.
Among this set, the emission line ratios with sufficient signal-to-noise
over the bright shells of the nebula, without binning the spaxels, to map 
$T_{\rm e}$ and $N_{\rm e}$, are presented: $T_{\rm e}$ from [\ion{N}{II}] and
[\ion{S}{III}] (Fig. \ref{fig:temaps}); $N_{\rm e}$ from [\ion{S}{II}] 
and [\ion{Cl}{III}] (Fig. \ref{fig:nemaps}).

Diagnostic CEL maps were produced using the $Python$ $PyNeb$ package 
\citep{Luridiana2015} using the default set of atomic data (transition 
probabilities and collision cross sections, in $PyNeb$ version 1.1.3).
$PyNeb$ performs a five, or more, level atom solution of CEL level populations
and thus explicitly treats the effect of collisional excitation and
de-excitation dependent on electron density.
The emission line flux maps were extinction corrected using the 
extinction map determined from the H Balmer and Paschen lines, as 
described in Paper I and using the Galactic extinction law
\citep{Seaton1979, Howarth1983} and the ratio of total 
to selective extinction, $R$, of 3.1. Line ratio maps were then formed 
from these extinction corrected images. The dereddened line ratio maps 
were masked in order that diagnostics were only calculated for spaxels 
with S/N of $>$3.0 on both lines and the error on the ratio per spaxel 
was calculated by Gaussian propagation of the errors on the emission line 
fits. The $diags.getCrossTemDen$ task was employed to calculate simultaneously
$T_{\rm e}$ and $N_{\rm e}$ from [\ion{N}{II}]6583.5/5754.6\,\AA\ and 
[\ion{S}{II}]6716.4/6730.8\,\AA\ for the lower ionization plasma, and
[\ion{S}{III}]6312.1/9068.6\,\AA\ and [\ion{Cl}{III}]5517.7/5537.9\,\AA\ for the 
higher ionization. The maximum number of iterations in 
$diags.getCrossTemDen$ for the calculation of $T_{\rm e}$ and $N_{\rm e}$
was set to 5, with the maximum error set to 0.1\%. Errors on
$T_{\rm e}$ and $N_{\rm e}$ were also computed using a Monte Carlo
approach with 50 trials of $diags.getCrossTemDen$ assuming the 
errors on both sets of line ratios were Gaussian. Figures 
\ref{fig:nemaps} and \ref{fig:temaps} show the individual
maps of $N_{\rm e}$ and $T_{\rm e}$ respectively, and the 
means on the signal-to-error value over the maps are reported 
in the captions.

The [\ion{N}{II}]5754.6\,\AA\ auroral line can be affected by recombination 
(direct and dielectronic recombination) to the $1S_{0}$ level of 
N$^{+}$, leading to apparent enhancement of $T_{\rm e}$ from
estimation by the [\ion{N}{II}]6583.5/5754.6\,\AA\ ratio \citep{Rubin1986}
\footnote{The [\ion{S}{III}]6312.1\,\AA\ line can also suffer from
recombination contribution from S$^{+++}$ in the same way as [\ion{N}{II}].
However there are currently no calculations of the recombination 
contribution to the [\ion{S}{III}] $1S_{0}$ level available. Given that the 
[\ion{S}{III}] $T_{\rm e}$ appears to be lower than [\ion{O}{III}]
$T_{\rm e}$ \citep{FangLiu2011}, in contrast to [\ion{N}{II}] $T_{\rm e}$ 
over the inner shell, suggests that any non-collisional component to 
[\ion{S}{III}]6312.1\,\AA\ emission cannot be very large, but obviously 
this cannot be confirmed}. The [\ion{N}{II}]6583.5\,\AA\ 
line is also affected by collisional 
de-excitation at $N_{\rm e}$ $^{>}_{\sim}$ 10$^{4}$ cm$^{-3}$, but such 
high densities are not apparent from the [\ion{S}{III}]6312.1/9068.6\,\AA\ 
and [\ion{Cl}{III}]5517.7/5537.9\,\AA\ ratio maps (Fig. \ref{fig:nemaps}).
The contribution by recombination can be estimated using the empirical 
formulae of \citet{Liu2000}. In order to employ this correction,
N$^{++}$ needs to be estimated (on a spaxel-by-spaxel basis) but 
no lines of [\ion{N}{III}] are available in the optical range. A possible
alternative is to use an estimate of N$^{++}$ from a recombination
line (\ion{N}{II}); however the resulting ionic abundance may not 
match the CEL N$^{++}$ abundance on account of the well-known abundance 
discrepancy factor between optical recombination lines (ORLs) and
CELs; see \citet{Liu2006} for a detailed discussion of this topic. 

One approach tried was to estimate the N$^{++}$ ionic abundance
from N$^{+}$ and an empirical ionization correction factor for
N$^{++}$ based on the O$^{++}$/O$^{+}$ ratio\footnote{This assumption
was found to be valid within $\sim$10\% by running CLOUDY 
models matching the central star parameters and nebula abundances 
(see Sect. \ref{Cloudy1D}).}. Since the derivation of O$^{+}$ from 
[\ion{O}{II}]7320.0,7330.2\,\AA\ is also affected by recombination of 
O$^{++}$, a correction was made; see Sect. \ref{Ionmaps}. Even applying 
this approach the resulting value of $T_{\rm e}$ over the central
region (i.e. a Z-shaped zone inside the main shell) was very large, 
mean 14\,600 K, somewhat lower than without any correction of 
$\sim$15\,200 K. Outside this Z-shaped region of elevated [\ion{N}{II}]
$T_{\rm e}$, the electron temperature is well-behaved and matches 
$T_{\rm e}$ from [\ion{S}{III}] very well. Applying a multiplicative
correction to the N$^{++}$ abundance (on the assumption that
the empirical correction by O$^{++}$/O$^{+}$ to N$^{+}$ is 
underestimated), can indeed reduce $T_{\rm e}$ over the central
region, but at the expense of depressing the values outside this
region. Applying a factor to N$^{++}$ abundance large enough
to decrease $T_{\rm e}$ in the central region to values
similar to [\ion{S}{III}] $T_{\rm e}$, however depresses $T_{\rm e}$ 
in the outer regions to values $<$5000 K, which cannot be 
justified (at least in comparison with the [\ion{S}{III}] $T_{\rm e}$). 

\subsection{$T_{\rm e}$ and $N_{\rm e}$ from recombination species}
\label{ORLTeNe}

The abundance discrepancy factor (ADF) for NGC~7009 is 3--5, 
the value depending on the ion \citep{Liuetal1995, FangLiu2013}.
Since it is known that the ORL temperature and density indicators 
differ from the CEL ones for integrated long slit spectra, their 
spatial differences may prove helpful for better understanding of 
the ADF problem. Several diagnostic ratios of line and continuum 
from the recombination lines of H$^{+}$, He$^{+}$ and He$^{++}$ in 
the MUSE wavelength range can be employed for electron density and 
electron temperature determination. These include the ratio of the high 
Balmer or Paschen lines as an electron density estimator 
\citep[c.f.,][]{Zhang2004}, the magnitude of the \ion{H}{I} Balmer or 
Paschen continuum jump at 3646 and 8204\,\AA\ respectively, the 
magnitude of the \ion{He}{II} continuum jump at 5875\,\AA\ and the 
ratios of He$^{+}$ singlets, which are primarily sensitive to $T_{\rm e}$. 
For the MUSE range the higher Paschen lines and the Paschen jump 
are accessible for \ion{H}{I} and the \ion{He}{II} and 
\ion{He}{I} diagnostics are useful too. The \ion{He}{II} Pfund jump
at 5694\,\AA\ is however too weak to measure in the MUSE data except 
in heavily co-added spectra, such as the long slit spectrum studied by
\citet{FangLiu2011}. The signal-to-noise in the maps is also not sufficient 
to determine $T_{\rm e}$ and $N_{\rm e}$ diagnostics from \ion{N}{II} and 
\ion{C}{III} recombination lines on a spaxel-by-spaxel basis and the 
lines of \ion{O}{II} are not available in the standard MUSE range.    

\subsubsection{$T_{\rm e}$ and $N_{\rm e}$ from \ion{He}{I}}
\label{TeHe1}
\cite{Zhang2005} showed how the ratio of the \ion{He}{I} 7281.4/6678.2\,\AA\ 
lines provides a very suitable diagnostic of the He$^{+}$ electron 
temperature and thus an interesting probe of the ORL temperature. 
They are both singlet lines, hence little affected by optical depth 
effects of the 2s $^{3}$S level, and are relatively close in wavelength 
(hence the influence of reddening on the ratio is small). The
determination of $T_{\rm e}$ from 7281.4/6678.2\,\AA\ was based 
on the analytic fits of \citet{Benjamin1999}, applicable at 
$ 5000 < T_{\rm e} < 20\,000$K, together with a fit to $T_{\rm e} <$ 
5000 K \citep{Zhang2005}. However modern determinations of \ion{He}{I} 
emissivities are based on the work of \citet{Porter2012, Porter2013}, 
which are only tabulated for $ 5000 < T_{\rm e} < 25\,000$K, thus the 3d 
$^{1}$D~--~2p $^{1}$P$^{0}$ (6678.2\,\AA) and 3s $^{1}$S~--~2p $^{1}$P$^{0}$ 
(7281.4\,\AA) emissivities were linearly extrapolated in log $T_{\rm e}$
to temperatures below 5000 K from the Porter et al. values to 3000 K 
for the tabulated log $N_{\rm e}$ of 1.0 to 7.0.

\begin{figure*}
\centering
\resizebox{\hsize}{!}{
\includegraphics[width=0.97\textwidth,angle=0,clip]{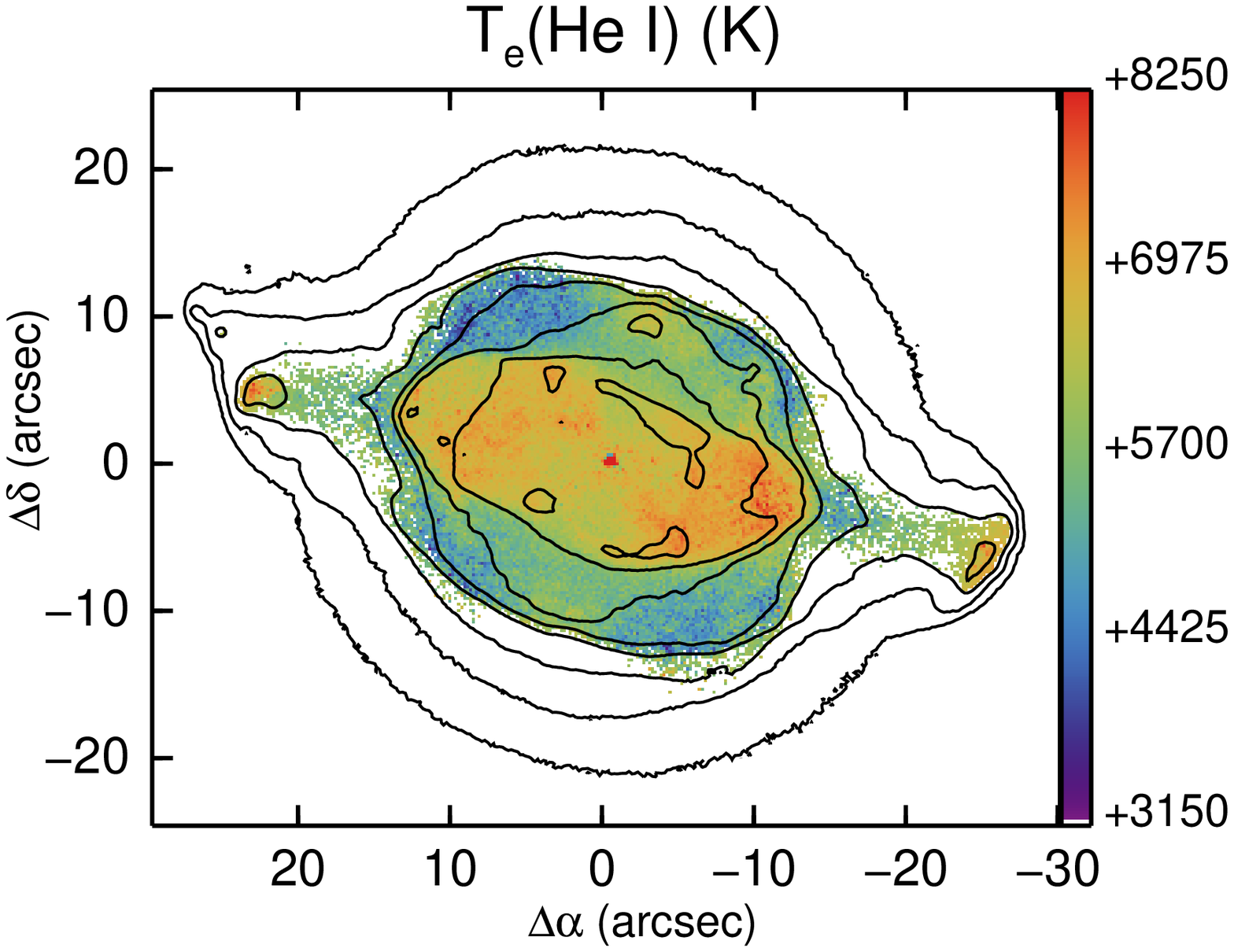}
\hspace{0.2truecm}
\includegraphics[width=0.97\textwidth,angle=0,clip]{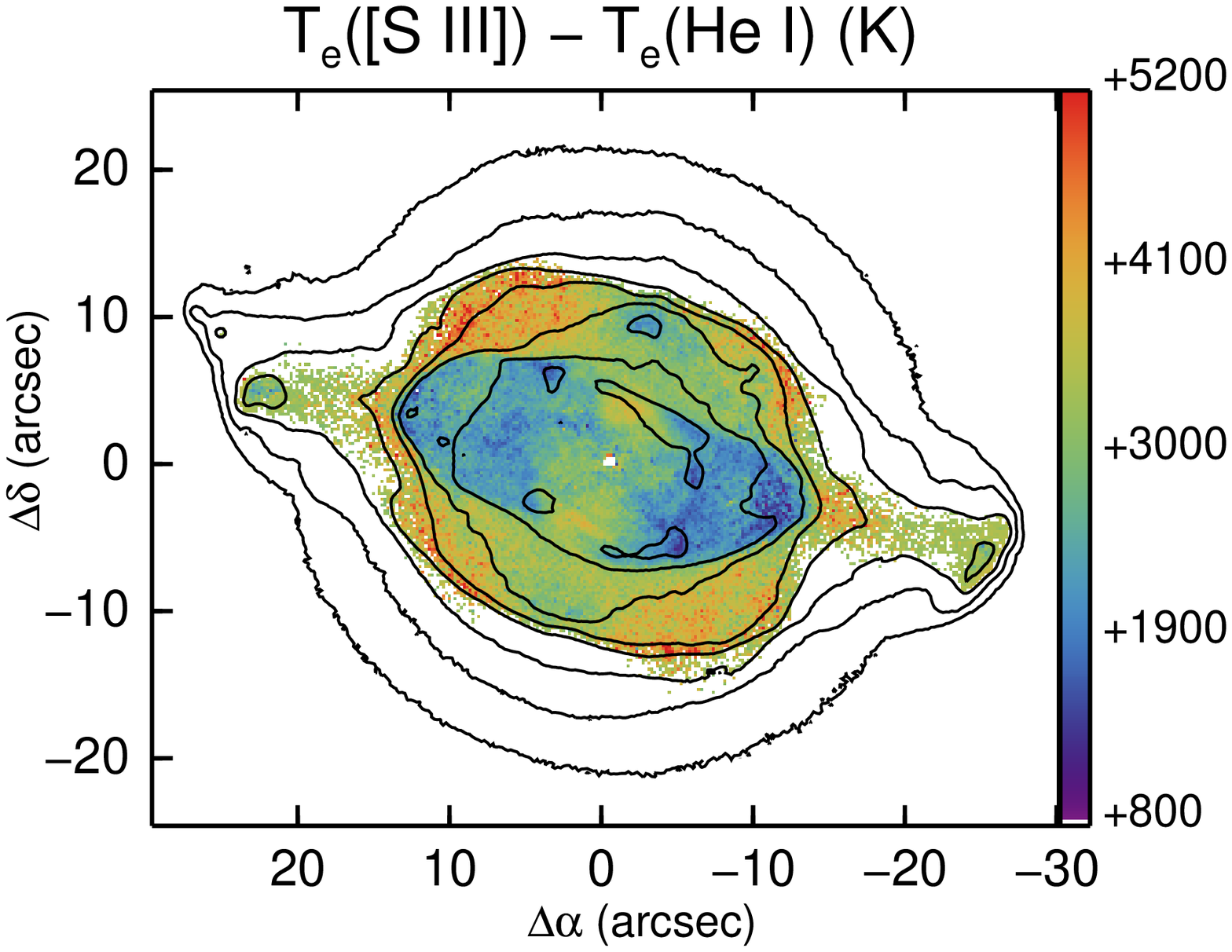}
}
\caption{The map of He$^{+}$ $T_{\rm e}$ (left) derived from the ratio 
of the \ion{He}{I} singlet lines 7281.4/6678.2\,\AA. The mean signal-to-error 
value is 11.
\newline
At right is shown the difference map between the [\ion{S}{III}]
and \ion{He}{I} $T_{\rm e}$ maps. The log F(H$\beta$) surface 
brightness contours are as in Fig. \ref{fig:nemaps}. 
}
\label{fig:He+Temaps}
\end{figure*}

The observed dereddened \ion{He}{I} 7281.4/6678.2\,\AA\ ratio was converted to 
$T_{\rm e}$ and $N_{\rm e}$ minimizing the residual with the theoretical 
ratio. A first guess for the value of $T_{\rm e}$ was provided by the 
mean dereddened 7281.4/6678.2\,\AA\ ratio for the whole image of 0.16,
corresponding to $T_{\rm e} \sim $ 6500 K for an assumed $N_{\rm e}$ 
of 3000 cm$^{-3}$. 
The ratio is much more sensitive to $T_{\rm e}$ than $N_{\rm e}$ and
the initial estimate for the He$^{+}$ electron temperature of 0.65 $\times$ 
the [\ion{S}{III}] $T_{\rm e}$ was adopted; $N_{\rm e}$ was held fixed within
narrow constraints during the minimization. Figure \ref{fig:He+Temaps} 
shows the resulting He$^{+}$ $T_{\rm e}$ map, which displays a distinct 
plateau inside the inner shell with mean $T_{\rm e}$ of 6200 K and, in 
the outer shell, lower values with mean 5400 K, extending to $<$4000 K.

\subsubsection{$T_{\rm e}$ from \ion{H}{I} Paschen Jump}
\label{TePasJum}

The magnitude of the series continuum jump for bound-free ($b-f$) 
transitions of \ion{H}{I} is sensitive to electron temperature. 
Following \citet{Peimbert1971}, \citet{LiuDanziger1993} developed the 
method using the flux difference across the Balmer jump normalized to 
the flux of a high Balmer line (H$\beta$ in this case) and computed  
$T_{\rm e}$ for several PNe. \citet{Zhang2004} also 
presented the same methodology applied to 48 Galactic PNe and extended
the method to the flux across the Paschen jump using the Paschen 20 
(8392.4\,\AA) line to normalize the flux across the jump for four PNe 
(including NGC~7009). It should be noted that the ratio of continuum 
fluxes of the blue to the red side of the jump is primarily dependent 
on $T_{\rm e}$ (a slight dependence on $N_{\rm e}$ arising from 
the two-photon emission, see Appendix \ref{Appendix}), whilst the 
normalization by an emission line, with its emissivity dependent on 
$T_{\rm e}$ and $N_{\rm e}$, introduces a more significant dependence 
of electron density into the fitting of the jump.

The usual method to derive the flux difference across the jump is to
fit the continuum spectrum, with the lines masked, in a dereddened spectrum
by the theoretical continuum computed using \ion{H}{I} recombination theory 
\citep[c.f.,][]{Zhang2004}. \citet{FangLiu2011} applied the method to a deep
integrated spectrum of NGC~7009 and provide a fitting formula for 
$T_{\rm e}$ as a function of the Paschen jump (defined as $F$(8194\,\AA) - 
$F$(8269\,\AA)) normalized by the Paschen 11 flux (8862.8\,\AA), all 
dereddened. In order 
to handle the tens of thousands of spaxels in the MUSE data, a method 
tuned to the dispersion of the MUSE spectra was developed. The 
mean continuum in several line-free windows on both sides of the Paschen 
Jump was measured and a custom conversion of this difference with respect 
to the flux of the P11 line was applied as a function of $T_{\rm e}$ and 
$N_{\rm e}$ and the fractions of He$^{+}$ and He$^{++}$ with respect to 
H$^{+}$ (see Sect. \ref{He+} and \ref{He++} respectively). A higher series 
Paschen line
could have been chosen minimizing the uncertainty from reddening correction, 
but P11 has the advantage of being a strong line well separated from 
neighbouring lines and also occurring in a relatively clear region of the 
telluric absorption spectrum \citep{Noll2012}. Appendix \ref{Appendix} gives 
details of the selection of the continuum windows and the computation of 
the conversion from measured jump to $T_{\rm e}$, using recent computation 
of b--f emissivities from \citet{ErcolanoStorey2006} and new computation 
of the thermally averaged free-free Gaunt factor from \citet{vanHoof2014}.

\begin{figure}
\resizebox{\hsize}{!}{
\includegraphics[width=0.97\textwidth,angle=0,clip]{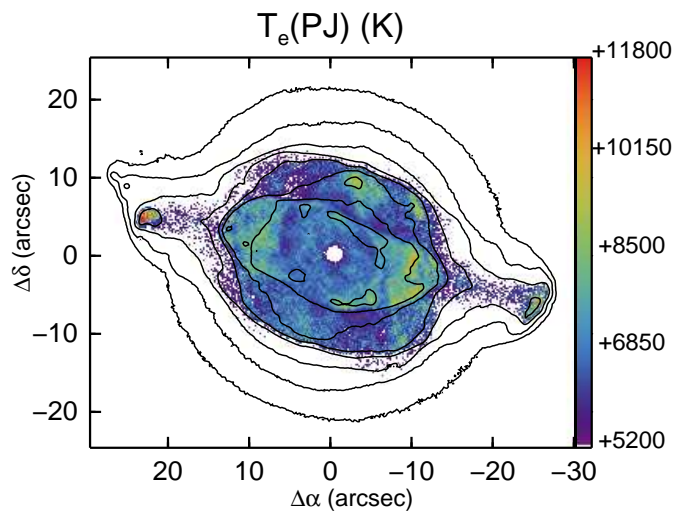}
}
\caption{Map of $T_{\rm e}$ derived from the 
magnitude of the Paschen continuum jump at 8250\,\AA\ ratioed by the 
dereddened \ion{H}{I} Paschen 11 (8862.8\,\AA) emission line strength. 
Initial estimates of $T_{\rm e}$ and $N_{\rm e}$ from the [\ion{S}{III}] 
and [\ion{Cl}{III}]
ratio maps (Sect. \ref{CELTeNe}) were applied, followed by two iterations
with the derived $T_{\rm e}$ map from the Paschen Jump. See text for
details. The low $T_{\rm e}$ values over the central star are not 
representative and arise from the strong stellar continuum. The log
F(H$\beta$) surface brightness contours are as in Fig. \ref{fig:nemaps}.
The mean signal-to-error is 13.
}
\label{fig:PJTemap}
\end{figure}

No correction for the presence of stellar continuum on the Paschen
Jump was applied \citep[c.f.,][]{Zhang2004}. However since the Paschen
Jump $T_{\rm e}$ is determined per spaxel, then spaxels over the 
seeing disk of the central star will be affected; the increase of the 
jump with increasing blue stellar continuum (and also when ratioed by P11)
results in very low values of $T_{\rm e}$ and these can be clearly discerned 
in the map. No other localized large deviations (positive or negative)
of $T_{\rm e}$, not correlated with \ion{H}{I} emission line morphology, were found, 
indicating no other significant sources of non-nebular continuum are present. 
Figure \ref{fig:PJTemap} shows the resulting map of PJ $T_{\rm e}$. The
mean value is 6610 K with a root mean square (RMS) of 1010 K (3 $\times$ 3 $\sigma$ 
clipped mean); the mean signal-to-error over the map, based on 100 Monte Carlo 
trials using the propagated errors on the measured PJ, is 13.

\subsubsection{$N_{\rm e}$ from \ion{H}{I} high Paschen lines}

\citet{Zhang2004} describe a method for determining $T_{\rm e}$
and $N_{\rm e}$ from the decrement of the higher Balmer lines,
and a similar method was applied to the high Paschen lines. Fitting
reddening corrected images of P15~--~P26 (8545.4, 8502.5, 8467.3,
8438.0, 8413.3, 8392.4, 8374.8, 8359.0, 
8345.5, 8333.8, 8323.4 and 8314.3\,\AA) and minimizing the residuals 
against the Case B values for initial values of $T_{\rm e}$ and
$N_{\rm e}$ from [\ion{S}{III}] and [\ion{Cl}{III}] (Sect. \ref{CELTeNe})
enables estimates of the \ion{H}{I} recombination density to be estimated. 
Higher Paschen series lines than P26 begin to be significantly
blended at the MUSE resolution of $\sim$3\,\AA\ and were not used. 
The set of higher Paschen lines is more sensitive to $N_{\rm e}$ than 
$T_{\rm e}$ \citep[c.f. Fig.1 in][]{Zhang2004}, so a given $T_{\rm e}$
was adopted per spaxel from [\ion{S}{III}] and [\ion{Cl}{III}] 
(Sect. \ref{CELTeNe}), although $T_{\rm e}$ from the Paschen 
decrement (Sect. \ref{TePasJum}) could have been used.

\begin{figure}
\resizebox{\hsize}{!}{
\includegraphics[width=0.97\textwidth,angle=0,clip]{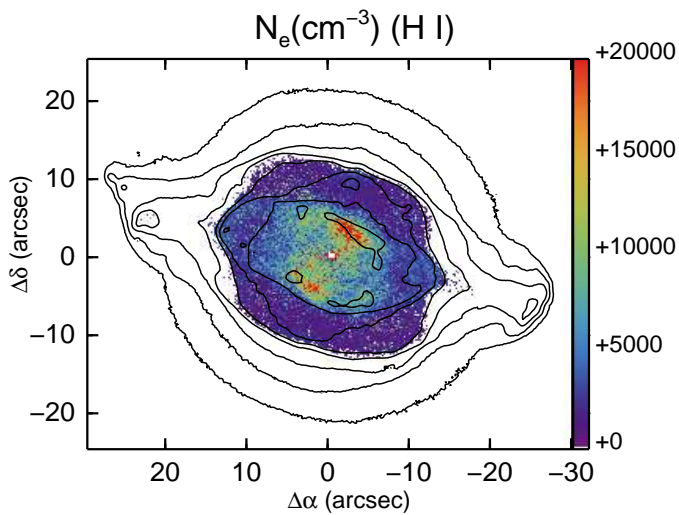}
}
\caption{Map of $N_{\rm e}$ derived by comparing the strengths
of the higher Paschen lines (P15~--~P26) with the Case B predictions,
adopting $T_{\rm e}$ values provided by the [\ion{S}{III}] ratio.
The log F(H$\beta$) surface brightness contours are as in Fig. 
\ref{fig:nemaps}. The mean signal-to-error is 2.
}
\label{fig:hinPasNemap}
\end{figure}

It was noted that P22 (8359.0\,\AA) has a much higher strength than predicted 
by Case B at reasonable values of $T_{\rm e}$ and $N_{\rm e}$; the same 
behaviour was found by \citet{FangLiu2011} (see their figure 7) from their
long slit spectrum of NGC~7009. In their data P22 does not appear to be 
much above Case B within the errors, but for an integrated spectrum of the
bright rim to the NW of the central star, the MUSE data clearly show 
P22 to be unusually strong (factor 1.8). No obvious \ion{He}{I} 
or \ion{He}{II} recombination
lines coincide with the wavelength of \ion{H}{I} P22, so another species 
may be contaminating this line (or the apparent strength arises from
sky subtraction errors). P22 was therefore removed from the comparison of
observed $v.$ predicted Paschen line strengths. The Paschen
line strengths were compared to the Case B values at the value
of $T_{\rm e}$ for each spaxel, weighting by the signal-to-noise of
each line to determine $N_{\rm e}$; the initial value of $N_{\rm e}$
from the [\ion{Cl}{III}] ratio was used. The signal-to-noise ratio of the
fitted emission lines is only high enough in the highest
\ion{H}{I} surface brightness regions, so the resulting map has sparse 
coverage of the outer shell. In a number of spaxels no
solution was possible within the errors and the initial value of
$N_{\rm e}$ was adopted. Figure \ref{fig:hinPasNemap} shows the resulting
\ion{H}{I} $N_{\rm e}$ map; high values are seen at the positions of the 
minor axis knots, corresponding to strong \ion{He}{II} emission. 
The mean value of the map is 4250 
$\pm$ 3030 cm$^{-3}$, compared to the mean value for the [\ion{Cl}{III}] 
$N_{\rm e}$ of 3490 $\pm$ 1930 cm$^{-3}$ (3 $\sigma$, 3 clipped 
iterate means).

\subsection{Comparison of $T_{\rm e}$ and $N_{\rm e}$ maps}
\label{TeNecomps}
Maps have been presented of $T_{\rm e}$ from two CEL line ratios 
([\ion{S}{III}] and [\ion{N}{II}], Fig. \ref{fig:temaps}) and two ORL 
combinations -- \ion{He}{I} 7281.4/6678.2\,\AA\ singlet line ratio (Fig. 
\ref{fig:He+Temaps}) and \ion{H}{I} Paschen Jump (Fig. \ref{fig:PJTemap}) 
and maps of $N_{\rm e}$ from the two CEL ratios ([\ion{Cl}{III}] and 
[\ion{S}{II}], Fig. \ref{fig:nemaps}) and one ORL diagnostic from the 
high series \ion{H}{I} Paschen lines (Fig. \ref{fig:hinPasNemap}).
These data provide a unique opportunity to spatially compare the
various diagnostics of CEL and ORL origin. No information
about the line-of-sight (radial velocity ) variation of CEL and ORL 
diagnostics is however available from these maps given the low spectral 
resolution of MUSE. 

$T_{\rm e}$ from the [\ion{N}{II}] line ratio is strongly 
affected by recombination to the upper level (auroral) 
5754.6\,\AA\ line and the resulting $T_{\rm e}$'s are 
too high, particularly in the central region; thus the CEL comparisons 
can only reliably be made with the [\ion{S}{III}] $T_{\rm e}$ map. 
\citet{FangLiu2011} measured a $T_{\rm e}$ value from the [\ion{S}{III}] 
ratio of 11500 K, in comparison with 10940 K from the [\ion{O}{III}] ratio 
(their table 4). Lacking measurement of [\ion{O}{III}]4363.2\,\AA\ and a 
map of $T_{\rm e}$(O$^{++}$) with MUSE, then $T_{\rm e}$ from the 
[\ion{S}{III}] ratio is the best CEL $T_{\rm e}$ surrogate for probing 
the bulk of the ionized emission.

The CEL [\ion{S}{III}] $T_{\rm e}$ map (Fig. \ref{fig:temaps}) shows
the highest values of 10300 K in the inner shell minor axis lobes, 
which correspond to the strongest \ion{He}{II} emission (and, as shown 
in Sect. \ref{He++}, the highest He$^{++}$/H$^{+}$). Within 
a circular plateau of radius 4.5$''$ around the central star, the 
$T_{\rm e}$ values are high. Inside the inner shell
along the major axis there are regions of lower $T_{\rm e}$ by 
$\sim$500 K except for two almost symmetrical stripes 
perpendicularly across the major axis (offset radii 9.5$''$) with 
slightly elevated $T_{\rm e}$ of $\sim$ 450 K above their surroundings.
These elevations in $T_{\rm e}$ correspond to knots K2 and K3 
\citep{Goncalves2003}, indicated on Fig. \ref{fig:nsclmaps}. The 
outer shell has a lower and more uniform $T_{\rm e}$ appearance with 
a value $\sim$ 8900 K, including the ansae, but there are
two regions along the minor axis which have $T_{\rm e}$ depressed
by $\sim$100 K. The ansae, however, notably show higher 
$T_{\rm e}$ with values around 10600 K, similar for both ansae, with
values increasing to the extremities (to 11000 K). All these
comparisons show differences larger than the propagated errors 
on $T_{\rm e}$ (Sect. \ref{CELTeNe}).

Although the central region with higher ionization in the [\ion{N}{II}] 
$T_{\rm e}$ map is strongly affected by the N$^{++}$ recombination 
contribution, the outer regions should be less affected (see Fig. 
\ref{fig:temaps}). The ends of the major axis show $T_{\rm e}$ values
elevated by only about 1000 K with respect to [\ion{S}{III}] $T_{\rm e}$ 
but the outer shell displays values higher by 1000's of K suggesting that 
N$^{++}$ recombination also contributes in this region.
The extremities of the ansae show [\ion{N}{II}] $T_{\rm e}$ notably lower than
from [\ion{S}{III}], while the emission extends further along the minor axis 
than S$^{++}$.

Comparison of the \ion{He}{I} singlet line ratio and \ion{H}{I} Paschen 
Jump $T_{\rm e}$ maps (Figs. \ref{fig:He+Temaps} and \ref{fig:PJTemap}, 
respectively) show many similarities with a lower temperature
circular region around the central star and elevations at the ends of the 
major axis; the contrast between the higher $T_{\rm e}$ inner shell
and the outer shell is greater (by $\sim$500 K) for the \ion{He}{I} $T_{\rm e}$
map, and in general the \ion{He}{I} map is smoother.
The \ion{He}{I} $T_{\rm e}$ map is very different in value and appearance 
from the [\ion{S}{III}] $T_{\rm e}$ map (Fig. \ref{fig:temaps}). The 
mean temperature difference in the inner shell is 2500 K whilst
in the outer shell it is 3500 K. There is a distinct step in the 
temperature difference ($\Delta$ $T_{\rm e}$ ([\ion{S}{III}] - \ion{He}{I})) 
$\sim$ 1000 K at the outer edge of the main shell. Both $T_{\rm e}$ 
maps show a central circular depression in $T_{\rm e}$ with 
elevations towards the ends of the major axis, with some 
general correspondence to the positions of the K2 and K3 knots. 
Both $T_{\rm e}$ maps show elevations
up to $\sim$1500 K over the minor axis polar knots, well above the
errors. Very prominent is the elevation in $T_{\rm e}$ at the 
extremities of the ansae, with values increasing from the \ion{He}{I} 
$T_{\rm e}$ map ($\sim$ 7200 K) to the \ion{H}{I} PJ map ($\sim$ 9200 K), 
compared to the values of $\sim$ 10500 K in the [\ion{S}{III}] 
$T_{\rm e}$ image.

The $N_{\rm e}$ maps from the [\ion{Cl}{III}] and [\ion{S}{II}] line ratios (Fig.
\ref{fig:nemaps}) show some similarities, such as the higher density 
inner circular region around the central star, with $N_{\rm e}$ $\sim$
7000 cm$^{-3}$, but also some differences. Several notable holes in
the higher ionization $N_{\rm e}$ map with sizes $^{<}_{\sim}$ 1$''$
are apparent in the inner shell, but not in the lower ionization 
density map. The edges of the inner shell on the minor axis
display prominently elevated $N_{\rm e}$ (to 8000 cm$^{-3}$). 
However the western filament (K2) shows as a depression in
[\ion{S}{II}] $N_{\rm e}$, although not in [\ion{Cl}{III}] $N_{\rm e}$. 
Both maps show the northern polar knot over the outer shell with 
density peaking at above 6000 cm$^{-3}$. The ansae are undistinguished
in both CEL $N_{\rm e}$ maps with values averaging 2200 cm$^{-3}$.

The ORL $N_{\rm e}$ map from the ratio among the high Paschen 
lines (Fig. \ref{fig:hinPasNemap}), although of rather low significance, 
does show general similarities to the [\ion{Cl}{III}] $N_{\rm e}$ map. 
The central circular region however has $N_{\rm e}$ elevated by $\sim$3000 
cm$^{-3}$ with respect to the [\ion{Cl}{III}] density map and the features 
along the minor axis have higher density with a bi-triangular morphology. 
There are similarities in this morphology with the \ion{He}{II}/H$\beta$ 
image (Fig. \ref{fig:ratmaps}) showing that the regions of highest ionization
are of high density. An incidental similarity also occurs with the 
[\ion{N}{II}] $T_{\rm e}$ map (Fig. \ref{fig:temaps}) which probably arises 
from the large contribution to [\ion{N}{II}]5754.6\,\AA\ emission from 
recombination of N$^{++}$ (Sect. \ref{CELTeNe}) in this high ionization 
region.

It is pertinent that regions distinguished by their emission line
surface brightness, in addition to the obvious shell structure, such 
as the K2 and K3 filaments, the minor axis polar knots and the 
ansae, also display distinct physical properties, such as 
elevated, or depressed, $T_{\rm e}$ and $N_{\rm e}$. This suggests a
fundamental link between emission line morphology and physical
conditions, which might indeed have been expected, but is 
nevertheless re-assuring and lends physical justification to the many
decades of emission line morphological studies of PNe.  

\subsection{A map of the temperature fluctuation parameter $t^{2}$}
\label{t2mapping}
Comparison of the maps of $T_{\rm e}$ from the CEL line ratio 
for [\ion{S}{III}] and the ORL Paschen Jump provide a method to study the 
spatial variation of the temperature fluctuation parameter, $t^{2}$,
introduced by \citet{Peimbert1967}. Given that there may be
temperature variations within a nebular volume, \citet{Peimbert1967} 
showed how the difference between $T_{\rm e}$ values from the ratio 
of the Balmer jump to H$\beta$, compared to $T_{\rm e}$ from the 
[\ion{O}{III}]5006.9/4363.2\,\AA\ ratio, can be used to derive the mean 
temperature $T0$ and the root mean square temperature fluctuation, 
$t^{2}$, by series expansion to 2nd order. 
\citet{Zhang2004} applied the method to several nebulae including 
NGC~7009, considering also the $T_{\rm e}$ independence of the 
far-infrared [\ion{O}{III}]88.33/51.80 $\mu$m ratio.

The MUSE spectra do not include the Balmer jump at 3646\,\AA\ or the 
[\ion{O}{III}]4363.2\,\AA\ line, but the problem can be reformulated following 
\citet{Peimbert1967} using the [\ion{S}{III}]9068.6/6312.1\,\AA\ ratio and the 
Paschen Jump to Paschen 11 ratio (see Appendix \ref{Appendix}). The
equivalent dependencies are (c.f., \citet{Peimbert1967}, eqns. 18 
and 16 respectively):
\begin{equation}
T_{\rm e}([\ion{S}{III}]) = T0\left[1 + \frac{1}{2} \left(\frac{5.537\times10^{4}}{T0} 
- 3\right)t^{2}\right]
\end{equation}
and
\begin{equation}
T_{\rm e}(PJ/P11) = T0\left(1 - 2.311t^{2}\right)
\end{equation}
where the latter relation was determined by fits to the variation
of PJ and P11 emissivity with $T_{\rm e}$ at a density of 3500 cm$^{-3}$
(see Appendix \ref{Appendix} and \citet{StoreyHummer1995}).

\begin{figure*}
\centering
\resizebox{\hsize}{!}{
\includegraphics[width=0.97\textwidth,angle=0,clip]{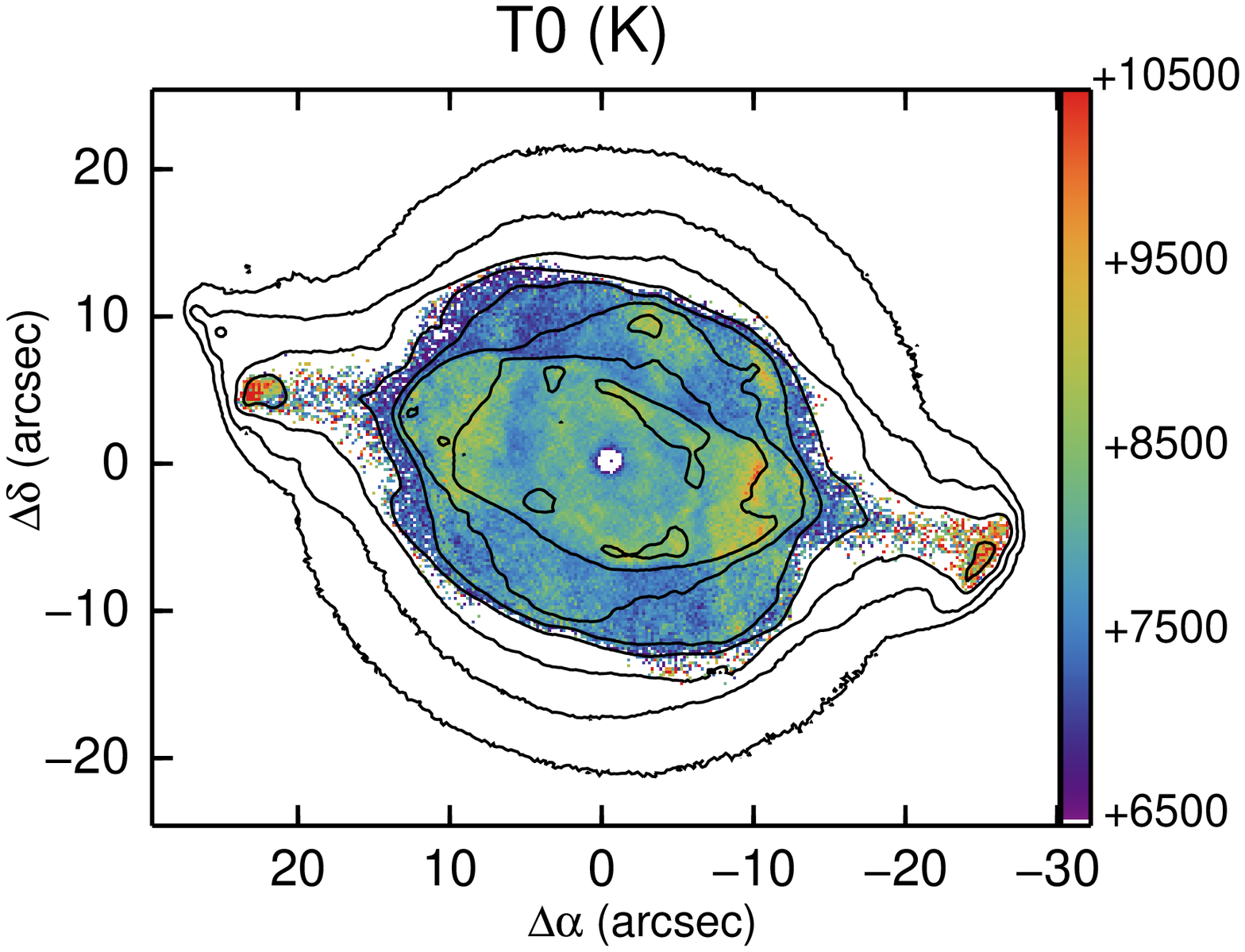}
\hspace{0.2truecm}
\includegraphics[width=0.97\textwidth,angle=0,clip]{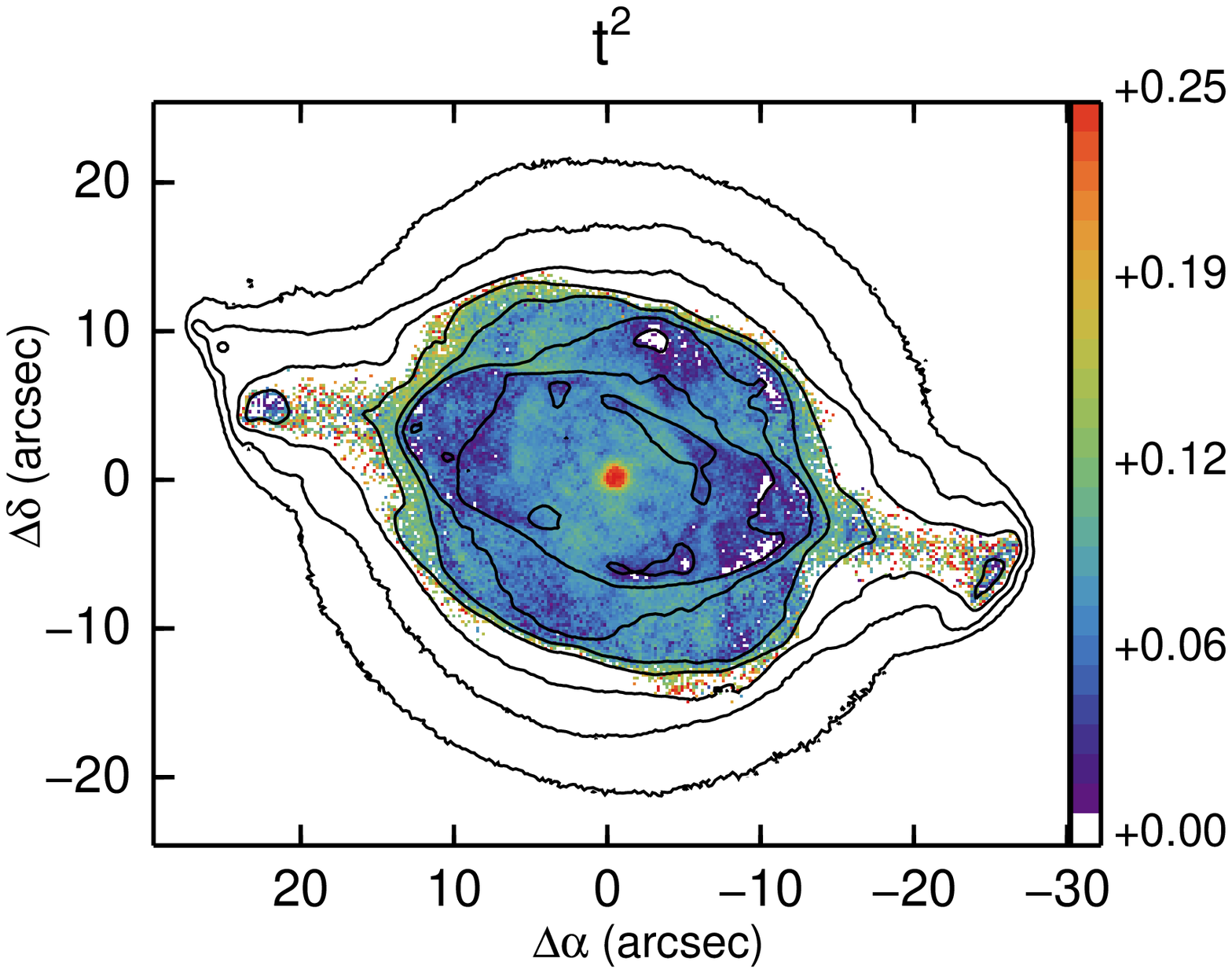}
}
\caption{Maps of the average temperature ($T0$, left) and the 
temperature fluctuation parameter ($t^{2}$, right) determined from 
the [\ion{S}{III}] and the Paschen Jump $T_{\rm e}$ maps. The 
log F(H$\beta$) surface brightness contours are as in Fig. 
\ref{fig:nemaps}. 
The mean signal-to-error over the $T0$ map is 45 and for the 
$t^{2}$ map is 15.
}
\label{fig:T0t2maps}
\end{figure*}

From the [\ion{S}{III}] $T_{\rm e}$ image (Fig. \ref{fig:temaps}) and the 
PJ/P11 $T_{\rm e}$ image (Fig. \ref{fig:PJTemap}), the average temperature
$T0$ and the temperature fluctuation parameter $t^{2}$ were determined
by minimizing the difference of the observed $T_{\rm e}$'s against the 
difference from equations 1 and 2. Figure \ref{fig:T0t2maps} 
shows the $T0$ and $t^{2}$ maps: the (3-$\sigma$ clipped) mean values are 
7900 K and 0.075, respectively. 
Since the [\ion{S}{III}] $T_{\rm e}$ map always has higher values, it dominates 
the morphology of the $T0$ image, but is reduced on average by $\sim$
15\% (1200 K). The $t^{2}$ map on the other hand more closely resembles 
the PJ $T_{\rm e}$ map and is by no means uniform, with a 
demarcation between the inner (lower $t^{2}$) and outer (higher $t^{2}$)
shells. In particular the central \ion{He}{II} zone has
elevated $t^{2}$, while the ends of the inner shell major axis and the 
north minor axis polar knot have lower values (by $\sim$0.05); the
ansae, particularly the eastern one, also show lower values 
of $t^{2}$, with mean 0.03. The outer rim of large $t^{2}$ values is not 
necessarily real and is accountable to large errors in [\ion{S}{III}] and 
PJ $T_{\rm e}$; the large values over the central star are also not real 
on account of stellar continuum contamination of the Paschen Jump
(Sect. \ref{TePasJum}).

Given such large values of $t^{2}$,
it can be argued that the definition of \citet{Peimbert1967} in
terms of a Taylor series expansion of $T_{\rm e}$ to account for
differences in temperature along the line of sight breaks down,
and so the description of the differences in $T_{\rm e}$ between
a collisionally excited line indicator and a recombination line/
continuum indicator in terms of $t^{2}$ is no longer strictly 
valid. A higher order temperature fluctuation parameter could be
sought for a quantitative description of this regime.


\section{Extending diagnostics to faint outer regions}
\label{Voronoi}
On account of the lower surface brightness of the emission from the halo 
beyond the outer shell, at the resolution of single spaxels the S/N is
too low to fit more than a few of the strongest emission lines (e.g., strong
H Balmer lines, [\ion{O}{III}], etc). In order to explore the physical conditions
in the halo, binning of the spaxels is required to achieve an adequate
S/N for line fitting (conservatively considered to be $>$2.5 $\sigma$). The
data can be binned either into regular $n \times m$ spaxels bins or an adaptive
signal-to-noise based criterion can be used to construct the bins, which
need not be rectangular. The adaptive binning technique based on Voronoi 
tesselations, developed by \cite{Capellari2003} specifically for IFS data,
provides a natural choice. The high S/N of single spaxels in the bright inner 
shell is retained, whilst some binning is applied in the outer shell and 
large and irregular shaped bins are appropriate for the halo. 

\begin{figure*}
\centering
\resizebox{\hsize}{!}{
\includegraphics[width=0.97\textwidth,angle=0,clip]{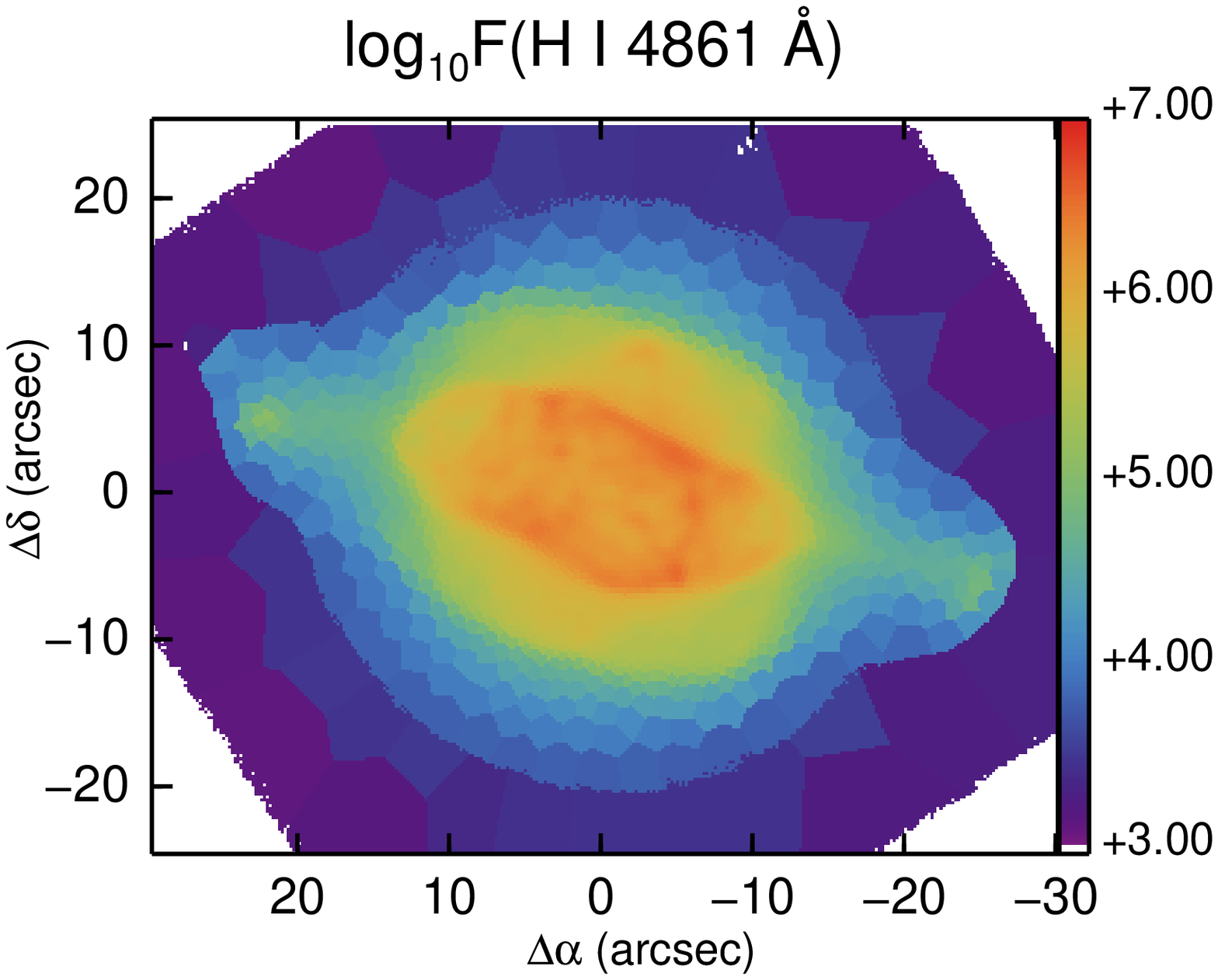}
\hspace{0.2truecm}
\includegraphics[width=0.97\textwidth,angle=0,clip]{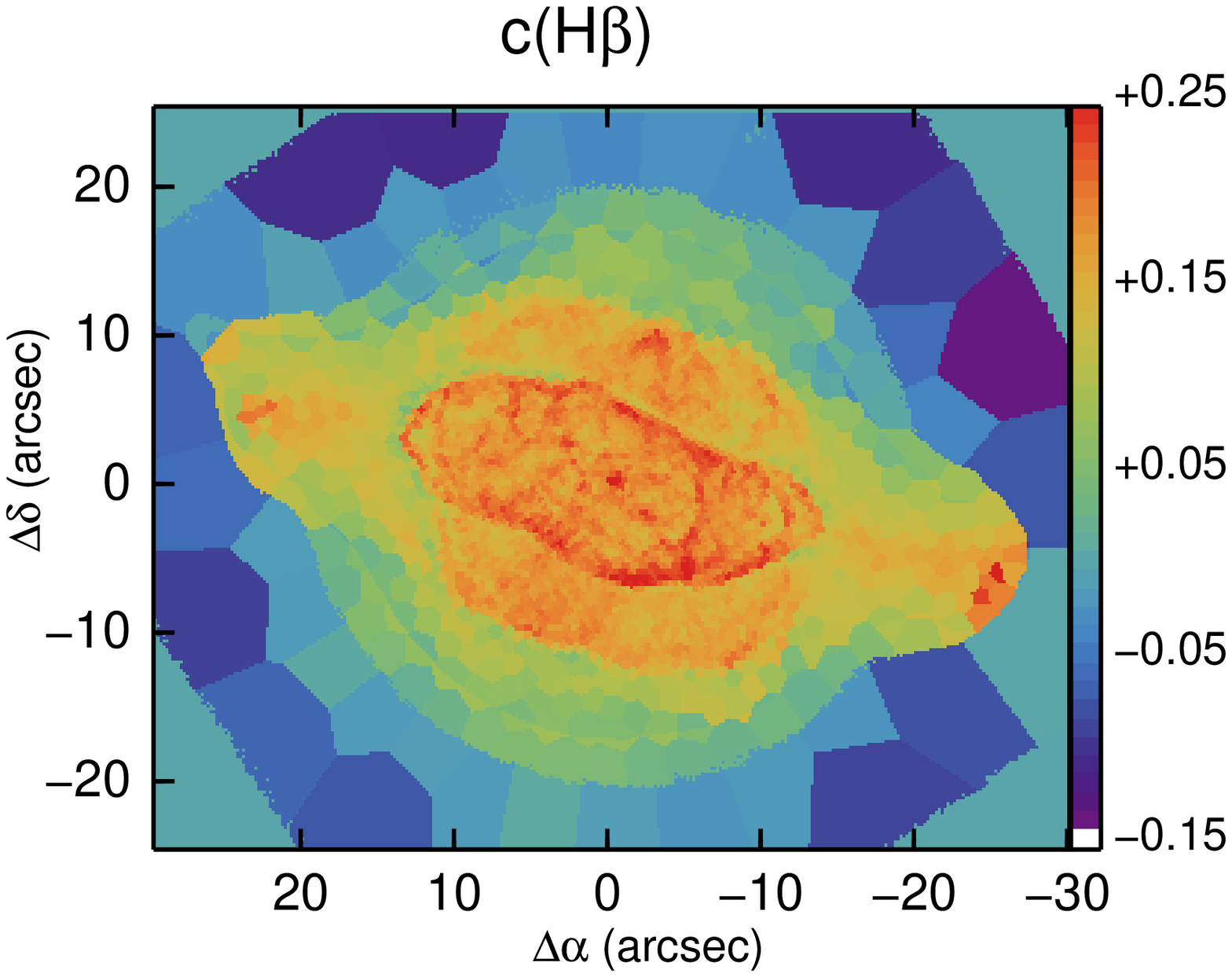}
}
\caption{Left: The Voronoi tesselated version of the 120s H$\beta$ emission line
image, in a log display showing the adaptive bins.
Right: The extinction map, $c(H\beta)$, for the Voronoi tesselated images based
on the ratio of the 120s H$\beta$ image and the 10s H$\alpha$ images.
}
\label{fig:VorHbmap}
\end{figure*}

The IDL $VORONOI\_2D\_BINNING$ code based on \citet{Capellari2003} was used 
to generate the bins based on the 120s H$\beta$
image and a target S/N of 500 was used: this choice leaves the spaxels in the
bright inner shell with low rebinning (typically 2x2 spaxels) and results in 
very large bins in the outer regions; Fig. \ref{fig:VorHbmap} shows the 
binned H$\beta$ map. The cubes were rebinned with the spectra set by
the binning but on the original pixelation. Each spaxel spectrum was then
fitted by multiple Gaussians, as described for the unbinned data (Sect.
\ref{Linemis}), and some maps of line flux and flux error were generated,
Fig. \ref{fig:VorHbmap} being an example. 

The Voronoi tesselated images were then subjected to a similar analysis to
the unbinned images: the extinction was obtained from the ratio of the
10s H$\alpha$ and 120s H$\beta$ images (since the H Paschen images have much 
lower S/N, the lines were not detectable in much of the halo so only the Balmer
line images were employed for extinction determination); the emission line
images were dereddened; line ratio maps were formed with propagated errors and
used to determine $T_{\rm e}$ and $N_{\rm e}$ from the same diagnostic
lines as the unbinned data (Sect. \ref{NeTemaps}). Since $T_{\rm e}$ and 
$N_{\rm e}$ are required to compute $c$ by comparison to the Case B H Balmer 
line ratio, an iterative procedure was adopted for the outer regions, with
initial values of $10^{4}$K and 1000 cm$^{-3}$ for $T_{\rm e}$ and 
$N_{\rm e}$ respectively. After two iterations of determining $c$, dereddening
and recomputing $T_{\rm e}$ and $N_{\rm e}$, the mean values in the outer
regions were found to be 11\,500 K and 400 cm$^{-3}$. 

The extinction map notably displays negative values in the outer 
extensions of the halo visible within the MUSE 
field of view. This finding was independently confirmed by 
fitting ellipses to the 10s H$\alpha$ and 120s H$\beta$ and examining
the ratio of surface brightnesses as a function of radius. Above a radius 
of 130 pix (26 $''$), the H$\alpha$/H$\beta$ ratio falls below the
Case B value of 2.855, appropriate for the nebula plasma conditions in
the inner shell. A negative value of $c$ cannot be derived if the 
appropriate density and temperature in the ionized gas are chosen and 
the correct optical depth condition in the Lyman continuum (Case A/B).  
The effect is most probably caused by a component of scattered Balmer 
light from the bright central nebula ($^{>}_{\sim}10^{3}$ times the 
surface brightness of the halo) and the presence of a scattering 
medium, predominantly dust, since Paper I showed the presence of dust 
within the nebula. 

Some instrumental component of scattered light from the bright inner 
shell into the outer shell must occur, although no direct evidence was seen 
by comparing the longer and shorter exposures and brighter and fainter 
emission lines between the two instrumental position angles. According to the 
MUSE instrument manual \citep{Richard2017}, both internal ghosts and 
straylight are below the 10$^{-5}$ level. \citet{Sandin2014} has carefully 
considered the role of diffuse scattered light in observations of
extended objects and derived surface brightness profiles. There must be some 
contribution to the halo light from the diffuse scattered profile of the inner 
shell of NGC~7009, but the presence of several rims in the faint halo 
\citep[e.g., ][]{Moreno-Corral1998} suggest this cannot be the dominant effect. 
The scattering problem for H$\alpha$ and H$\beta$ in a PN with a neutral dust 
shell has been treated by \citet{Gray2012}; in the case of \object{NGC~6537} 
they derived a clumpy distribution of dust in the shell. 
Linear polarization has also been measured by \citet{Leroy1986} over 
the outer halo of NGC~7009, so the prediction 
from these observations is that a higher value of polarization is expected
over the faint halo (c.f., \citet{Walsh1994} for the halo of
\object{NGC~7027}).

\section{Ionization maps}
\label{Ionmaps}
For those ORL's and CEL's with sufficient signal-to-noise to allow coverage 
of the main shell and outer shell, maps of the ionic fraction 
with respect to H$^{+}$ were formed 
using $PyNeb$. The maps of $T_{\rm e}$ from [\ion{S}{III}] and $N_{\rm e}$ from 
[\ion{Cl}{III}] (Sect. \ref{NeTemaps}) together with the dereddened line and 
H$\beta$ maps (applying the \citet{Seaton1979} extinction law with R=3.1) 
were employed to generate the ionic abundance at each spaxel.  
The following ionic maps were made: 
He$^{+}$ and He$^{++}$; N$^{+}$; O$^{0}$, O$^{+}$ and O$^{++}$; S$^{+}$ and 
S$^{++}$; Cl$^{++}$ and Cl$^{+++}$; Ar$^{++}$ and Ar$^{+++}$; and Kr$^{+++}$.

\subsection{He$^{+}$}
\label{He+}
There are a range of singlet and triplet He$^{+}$ lines in the MUSE coverage,
some of which are affected by optical depth and by collisional effects from the 
2 $^{3}$3S and 2 $^{1}$1S levels, although neither should be severe in NGC~7009. 
The He$^{+}$ abundance was calculated by $PyNeb$ using the Case B emissivities of 
\citet{Porter2012, Porter2013}. The He$^{+}$/H$^{+}$ abundance can be determined
from \ion{He}{I} 5875.6\,\AA\ (3d $^{3}$D~--~2p $^{3}$P$^{0}$, triplet), 6678.2\,\AA\ 
(3d $^{1}$D~--~2p $^{1}$P$^{0}$, singlet), 7065.7\,\AA\ (3s $^{3}$S~--~2p 
$^{3}$P$^{0}$, triplet) and 7281.4\,\AA\ (3s $^{1}$S~--~2p $^{1}$P$^{0}$, singlet)
lines. The other singlet \ion{He}{I} lines detected are 4d $^{1}$D~--~2p 
$^{1}$P 4921.9\,\AA\ and 2s $^{1}$S~--~3p $^{1}$P 5015.7\,\AA; however the 
4921.9\,\AA\ line is weaker than the red \ion{He}{I} lines and the 5015.7\,\AA\ 
line is very close to the very bright [\ion{O}{III}]5006.9\,\AA\ line, so is very 
problematic to obtain a good line flux. It was decided therefore to 
concentrate on the four red and brighter \ion{He}{I} lines to provide better 
spatial coverage of the fainter regions. Of these lines, \ion{He}{I} 7065.7\,\AA\ 
is the one most sensitive to the \ion{He}{I} $\tau$(3888.6\,\AA) optical depth 
\citep{Robbins}, followed by 5875.6\,\AA, while the singlets 6678.2\,\AA\ and 
7281.4\,\AA\ are entirely insensitive to $\tau$(3888.6\,\AA). 

The He$^{+}$ maps based on the He$^{+}$ $T_{\rm e}$ (Sect. \ref{TeHe1})
and the singlet 6678.2\,\AA\ and triplet 5875.6\,\AA\ images are very
similar in appearance and mean value, while the He$^{+}$ map from 
7065.7\,\AA\ has a much higher mean value, although similar morphology.
The difference arises since the 7065.7\,\AA\ map shows the effect of higher 
\ion{He}{I} 3888.6\,\AA\ optical depth with a morphology of a ring around 
the inner shell. 

\begin{figure}
\resizebox{\hsize}{!}{
\includegraphics[width=0.97\textwidth,angle=0,clip]{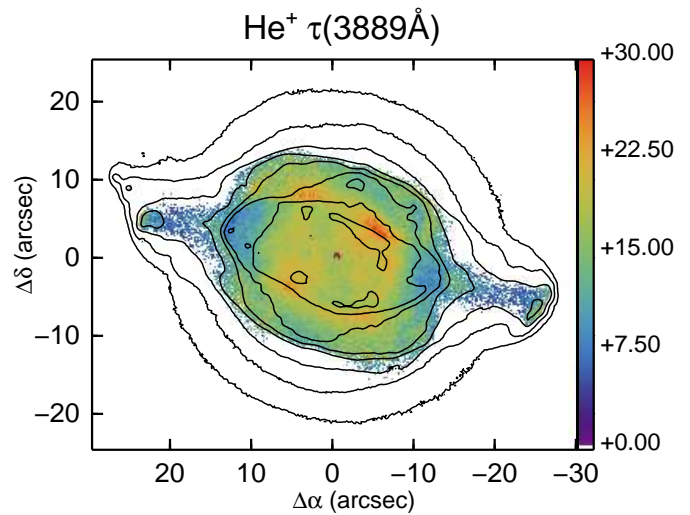}
}
\caption{$\tau$(3888.6\,\AA) from the ratio of \ion{He}{I} 7065.7\,\AA\ and 
6678.2\,\AA\ lines, calculated with the \ion{He}{I} $T_{\rm e}$ map and
a ratio of expansion velocity to thermal velocity for the nebula of 
5. The log F(H$\beta$) surface brightness contours are as in Fig. 
\ref{fig:nemaps}.
}
\label{fig:Hetaumap}
\end{figure}

A $\tau$(3888.6\,\AA) optical depth map can be calculated from the ratio 
of He$^{+}$ maps from the \ion{He}{I} 7065.7\,\AA\ and \ion{He}{I} 
6678.2\,\AA\ lines, following Eqn. 5 of \citet{MonrealIbero2013}. However in 
the case of NGC~7009, $\tau$(3888.6\,\AA) should be calculated appropriate 
for an expanding nebula (\citet{MonrealIbero2013} used a fit to the \citet{Robbins}
He$^{+}$ emissivity appropriate to a static nebula), so a new fit was made for the
case when the ratio of radial velocity to thermal velocity ($V(R)/V(TH)$ in
\citet{Robbins}, table 3) is 5, although the value in NGC~7009 is higher 
than this, $\sim$20. It should be noted however that the variation of
7065.7\,\AA\ optical depth appropriate for $T_{\rm e}$ of 10$^{4}$K was used, 
as \citet{Robbins} only tabulates the variation at two values of 
$T_{\rm e}$. The ratio of He$^{+}$/H${+}$ images from 7065.7\,\AA\
and 6678.2\,\AA\ was used to compute $\tau$(3888.6\,\AA) and the resulting map
is shown in Fig. \ref{fig:Hetaumap}. The value of $\tau$(3888.6\,\AA) is quite high,
mean 15.0. with some structure associated with the boundary of the inner rim and
outer shell. Notably there is less structure than in the same map produced from
the He$^{+}$ images with [\ion{S}{III}] $T_{\rm e}$ where there are prominent holes in the 
optical depth map on the minor axis, but the mean value of $\tau$(3888.6\,\AA) 
is lower (mean 2.0). Interestingly \citet{Robbins} deduced a (single) optical 
depth value for NGC~7009 of 15, based on $T_{\rm e}$ of 10$^{4}$K. Given the 
relatively high value of $\tau$(3888.6\,\AA) derived here, calculating 
He$^{+}$ $T_{\rm e}$ from 7281.4/5875.6\,\AA\ can be expected to deliver a 
lower value of $T_{\rm e}$ \citep[c.f.,][]{Zhang2005} since 5875.6\,\AA\ is a 
triplet line.

\begin{figure}
\resizebox{\hsize}{!}{
\includegraphics[width=0.97\textwidth,angle=0,clip]{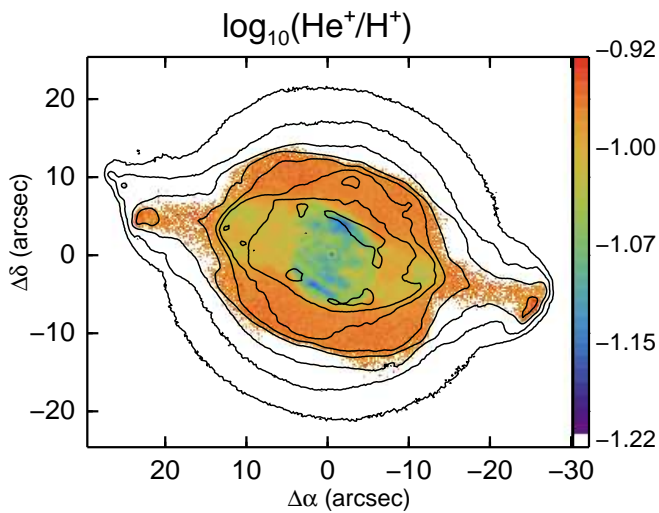}
}
\caption{Map of He$^{+}$/H$^{+}$ determined from the \ion{He}{I} 6678.2\,\AA\ 
and H$\beta$ dereddened emission line maps with the
He$^{+}$ $T_{\rm e}$ and [\ion{S}{III}] $T_{\rm e}$ maps respectively and
the single [\ion{Cl}{III}] $N_{\rm e}$ map. The log F(H$\beta$) surface brightness 
contours are as in Fig. \ref{fig:nemaps}. The mean signal-to-error ratio
on He$^{+}$/H$^{+}$ is $\sim$ 14.
}
\label{fig:He+map}
\end{figure}

The strongest \ion{He}{I} singlet line, hence showing no optical depth sensitivity, 
is 6678.2\,\AA, and Fig. \ref{fig:He+map} shows the ionic fraction He$^{+}$/H$^{+}$
using the \ion{He}{I} $T_{\rm e}$, the [\ion{S}{III}] $T_{\rm e}$ for H$^{+}$ and
the single [\ion{Cl}{III}] $N_{\rm e}$ map.

\subsection{He$^{++}$}
\label{He++}
\begin{figure}
\resizebox{\hsize}{!}{
\includegraphics[width=0.50\textwidth,angle=0,clip]{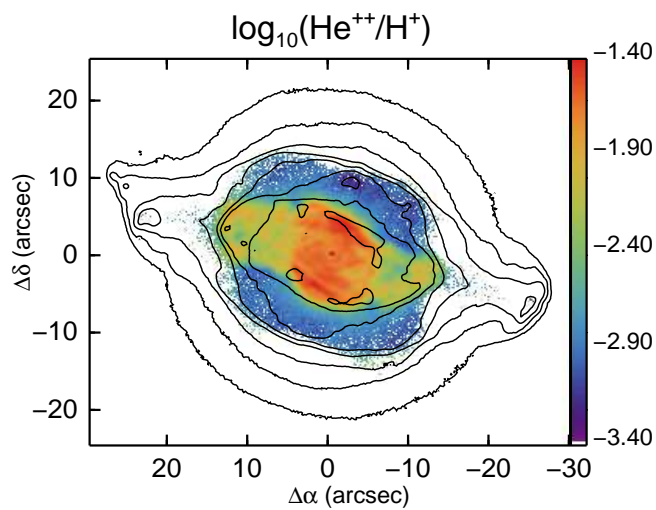}
}
\caption{Map of He$^{++}$/H$^{+}$ using the [\ion{S}{III}] $T_{\rm e}$ and
the [\ion{Cl}{III}] $N_{\rm e}$ maps. The log F(H$\beta$) surface brightness 
contours are as in Fig. \ref{fig:nemaps}. The mean value of
signal-to-error is 12.
}
\label{fig:He++map}
\end{figure}

The He$^{++}$ map produced with the dereddened {He}{II} and H$\beta$ 
images and the [\ion{S}{III}] $T_{\rm e}$ and [\ion{Cl}{III}] $N_{\rm e}$ 
maps (Sec. \ref{CELTeNe}) is shown in Fig. \ref{fig:He++map}. It 
shows the high ionization main shell with the minor axis regions prominent, 
where He$^{+}$ is notably lower. Whilst there is weak He$^{++}$ 
present in the outer shell, it is more pronounced along the major axis. 
This suggests an anisotropic escape of highly ionized photons, with much 
higher optical depth to $h\nu \ge 54.4$ eV photons along the minor axis 
than the major axis, although in other respects the major axis extensions 
are fairly unassuming with similar extinction, $T_{\rm e}$ and $N_{\rm e}$ 
to the rest of the outer shell (Sect. \ref{CELTeNe}). The extensions
of the He$^{++}$ emission beyond the inner shell contrast with the 
Chandra diffuse X-ray appearance (\citet{Kastner2012}, their Fig. 4),
which displays soft X-ray emission only inside the main shell, and 
attributed to the interface between the shocked stellar wind and
the cooler, denser expanding gas of the optical nebula. 

\subsection{O$^{0}$, O$^{+}$ and O$^{++}$}
\begin{figure*}
\centering
\resizebox{\hsize}{!}{
\includegraphics[width=0.50\textwidth,angle=0,clip]{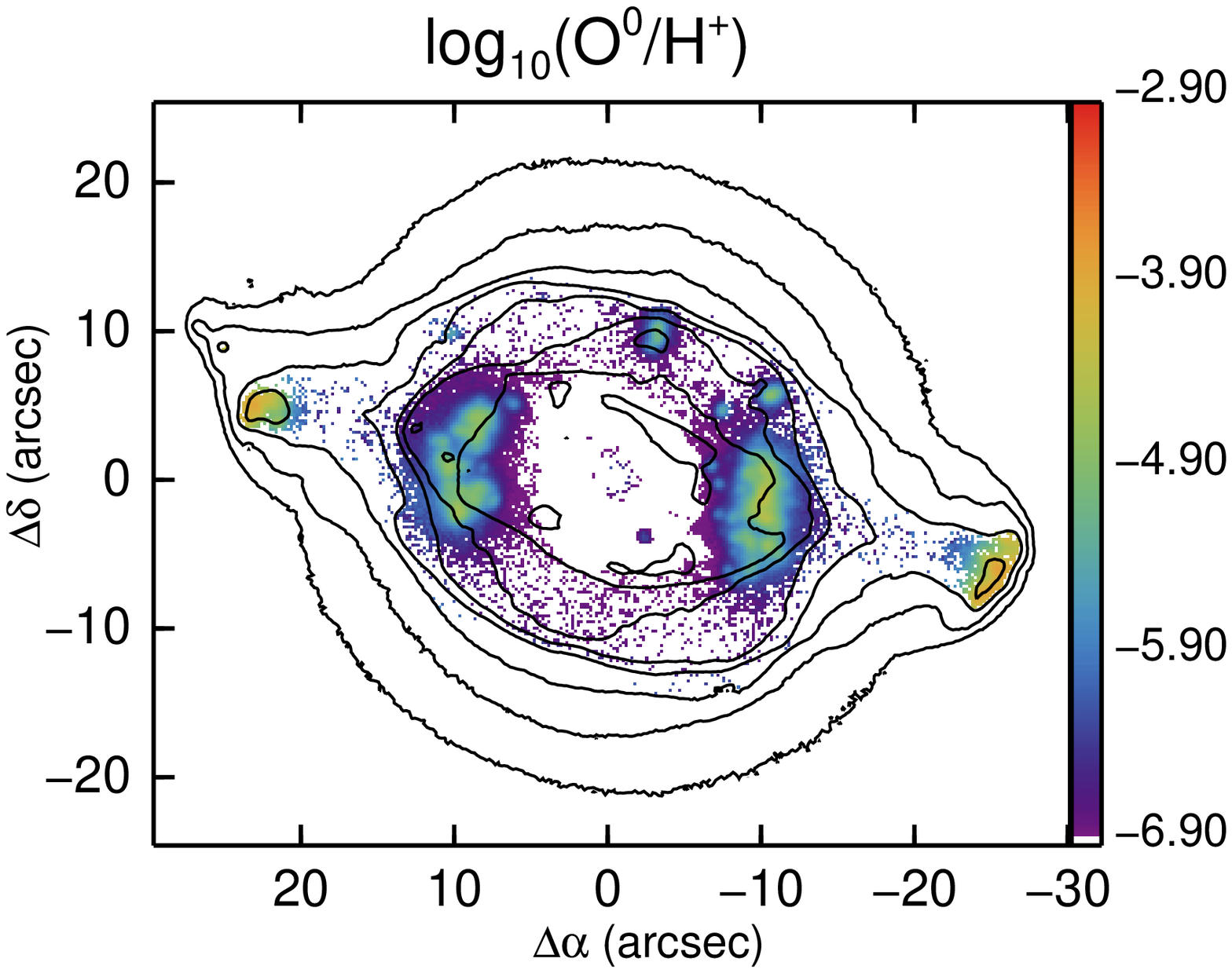}
\hspace{0.2truecm}
\includegraphics[width=0.50\textwidth,angle=0,clip]{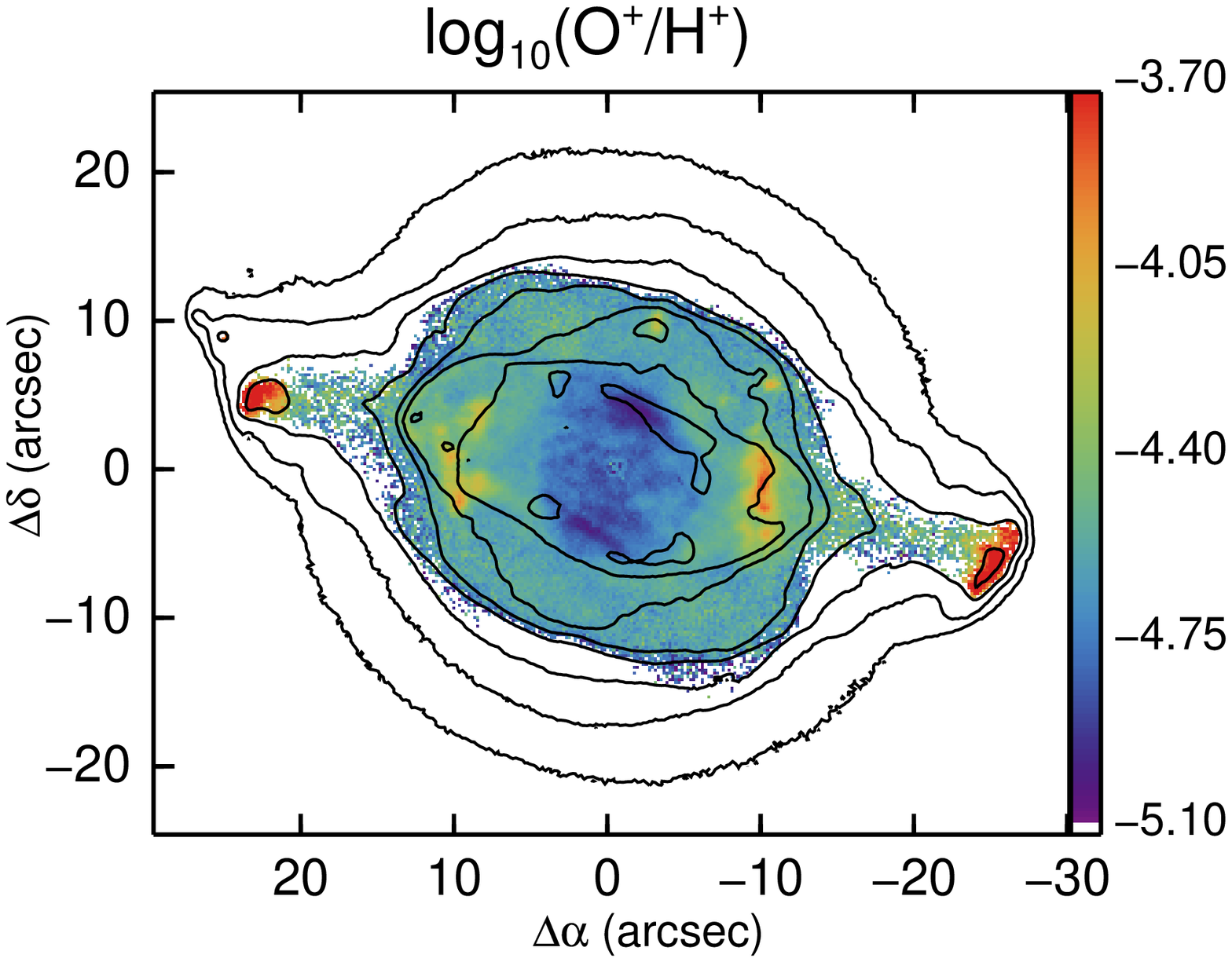}
\hspace{0.2truecm}
\includegraphics[width=0.50\textwidth,angle=0,clip]{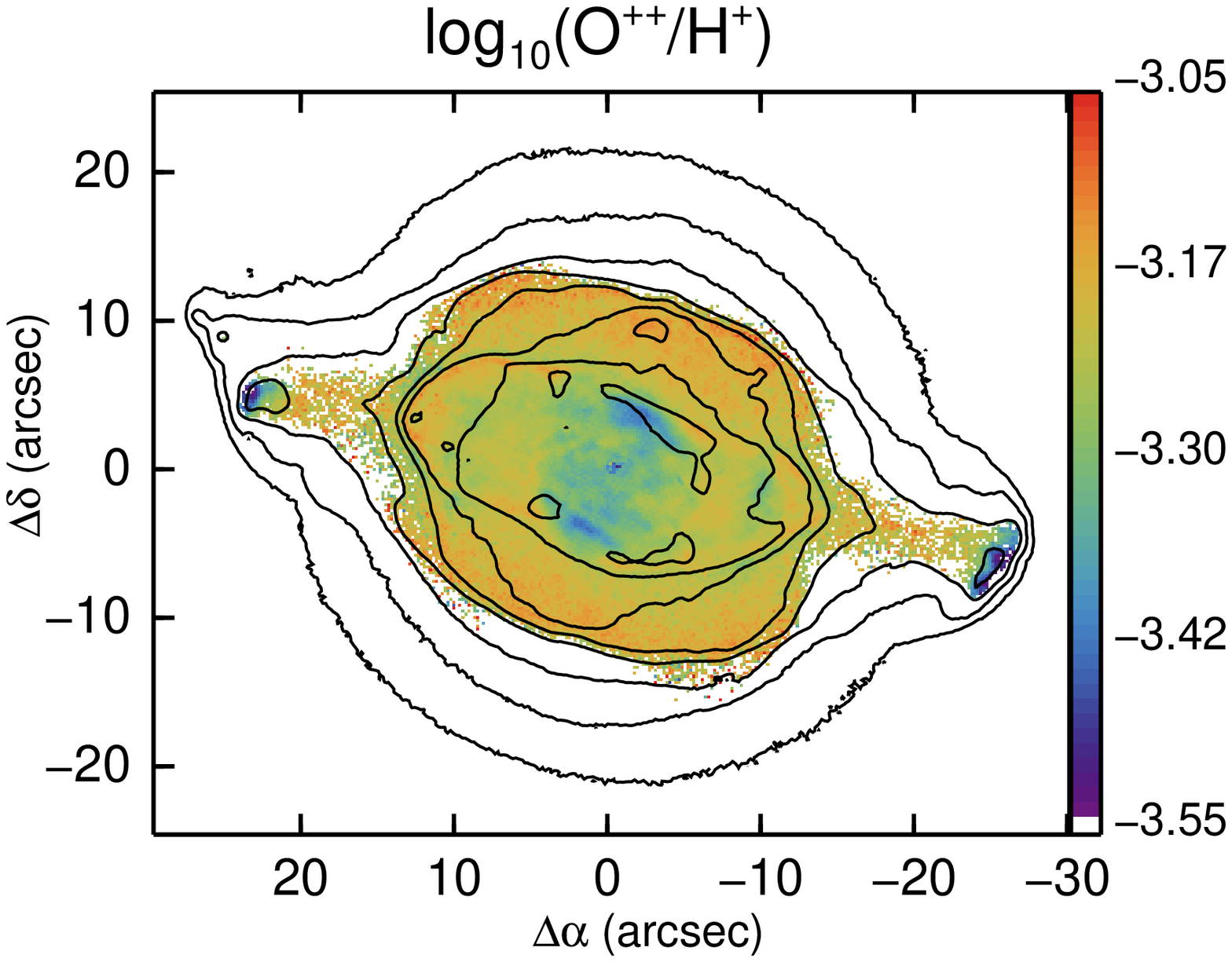}
}
\caption{Maps of O$^{0}$, O$^{+}$ and O$^{++}$/H$^{+}$ ionic abundances.
The log F(H$\beta$) surface brightness contours are as in Fig. 
\ref{fig:nemaps}. The mean signal-to-error values on O$^{0}$, O$^{+}$ 
and O$^{++}$/H$^{+}$ are 12, 21 and 13, respectively. 
}
\label{fig:Oabmaps}
\end{figure*}

Figure \ref{fig:Oabmaps} shows the O$^{0}$, O$^{+}$ and O$^{++}$ maps;
the quoted mean signal-to-error values are from Monte Carlo trials. 
The O$^{+}$ abundance was formed by correcting the dereddened 
[\ion{O}{II}]7330.2\,\AA\ line flux by the O$^{++}$ recombination contribution. 
The correction scheme proposed by \citet{Liu2000} was adopted (their eqn. 2,
detailed in their Appendix A) and the O$^{++}$/H$^{+}$ contribution was 
provided by the O$^{++}$ map based on the [\ion{S}{III}] $T_{\rm e}$ map. 
This contribution for O$^{++}$ recombination alters the mean 
value over the map of O$^{+}$/H$^{+}$ by 30\%. 

The O$^{0}$ abundance, calculated assuming pure photoionization and 
$T_{\rm e}$ and $N_{\rm e}$ from [\ion{S}{III}] and [\ion{S}{II}] respectively,
is only strong in a few knots along the major axis, in the ansae and 
knots K2 and K3. There may however be some component of fluorescence to 
[\ion{O}{I}] emission or it may be predominantly produced in photodissociation
regions \citep[c.f.,][]{Richer1991, Liuetal1995}, so the O$^{0}$ cannot 
naively be added to O$^{+}$ and O$^{++}$ to determine the total O 
abundance (although it only contributes at the level of 0.25\% of 
the total O abundance). The northern minor axis peak is strong in 
O$^{0}$ as are several compact knots in the vicinity of the western 
major axis knot K2. The map of O$^{+}$ bears 
many similarities to the O$^{0}$ but with more extended emission present 
over the inner and outer shells. The ansae are notably strong in 
O$^{+}$. The O$^{++}$ map is notably extended 
to the outer shell and towards the ansae, but depressed at the ansae
themselves. The lower O$^{++}$ regions in the inner shell correspond to 
those with highest He$^{++}$, consistent with strong O$^{+++}$ presence.
  
\subsection{N ionization maps}
\begin{figure*}
\centering
\resizebox{\hsize}{!}{
\includegraphics[width=0.97\textwidth,angle=0,clip]{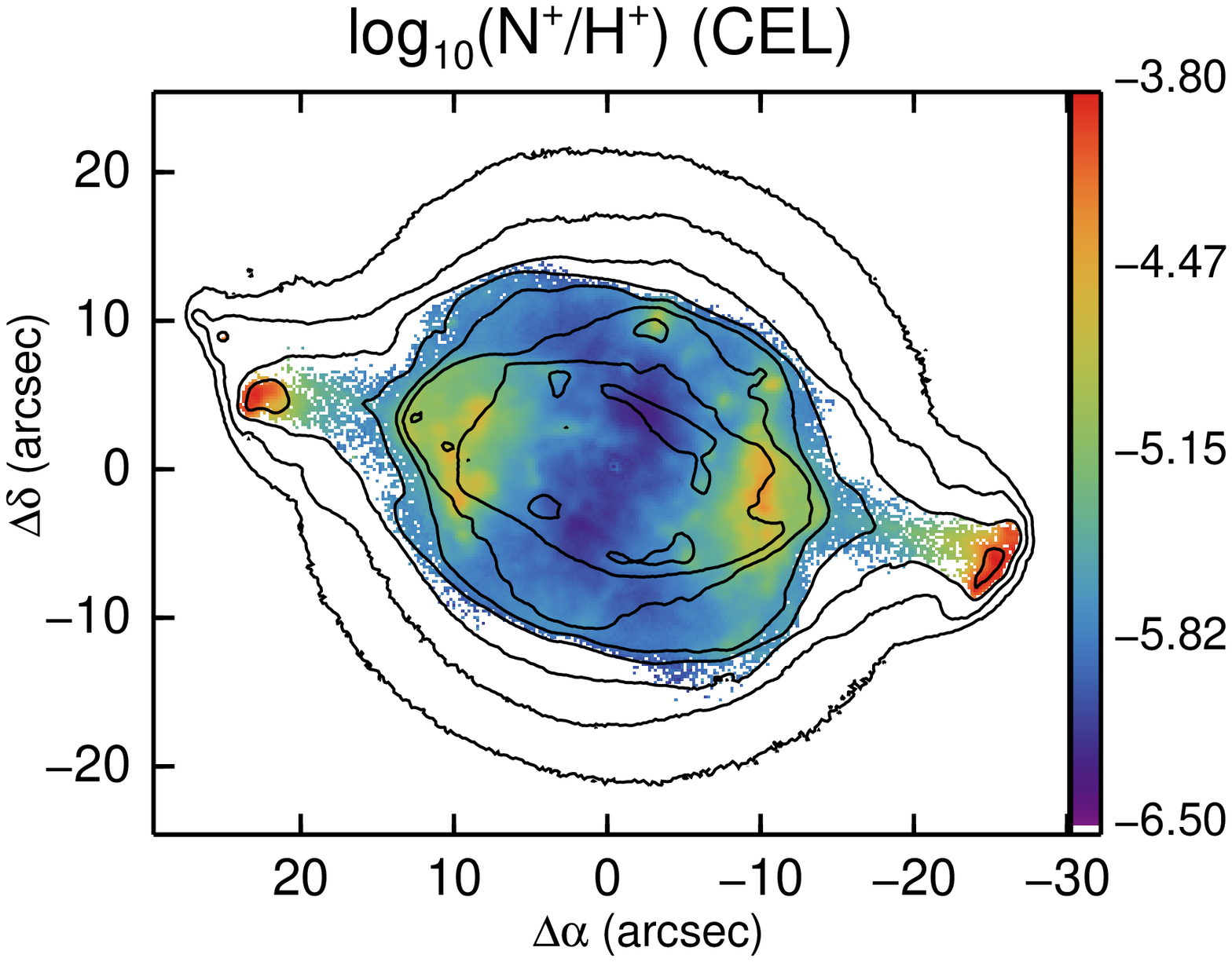}
\hspace{0.2truecm}
\includegraphics[width=0.97\textwidth,angle=0,clip]{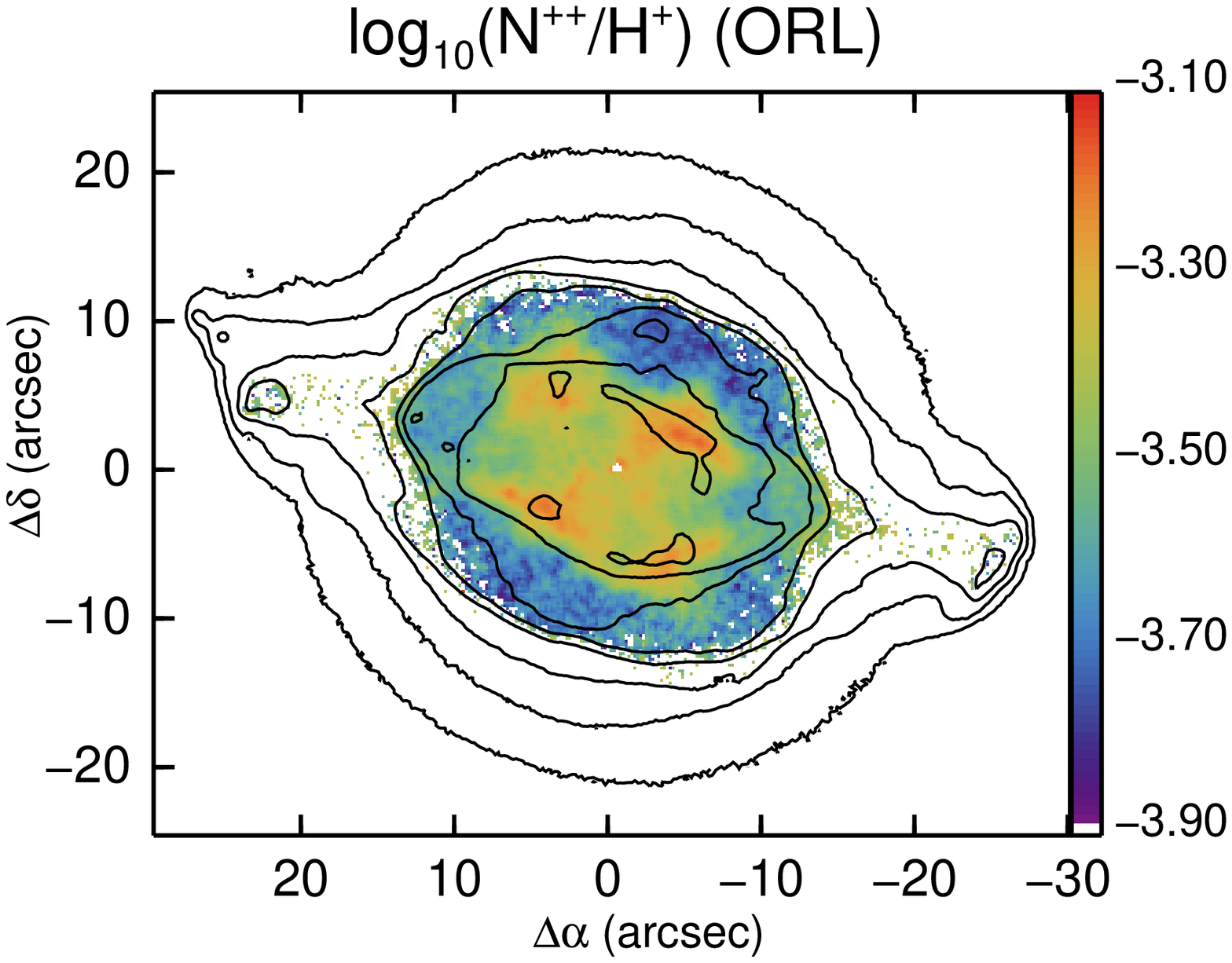}
}
\caption{Map of N$^{+}$/H$^{+}$ from the CEL [\ion{N}{II}] lines (left) and
for N$^{++}$/H$^{+}$ from the \ion{N}{II} 5679.6\,\AA\ ORL line (right). 
The log F(H$\beta$) surface brightness contours are as in Fig. 
\ref{fig:nemaps}. The mean values of signal-to-error are 15 and 5, 
respectively.
}
\label{fig:N+maps}
\end{figure*}

Only the N$^{+}$ image is available from the CEL [\ion{N}{II}] lines but, as shown
in Sect. \ref{CELTeNe}, the 5754.6\,\AA\ line is affected by recombination and so
over-estimates $T_{\rm e}$; to produce the N$^{+}$/H$^{+}$ map shown in
Fig. \ref{fig:N+maps}, $T_{\rm e}$ from [\ion{S}{III}] and $N_{\rm e}$ from 
[\ion{S}{II}]
were employed. The map resembles closely that of O$^{0}$ (Fig. \ref{fig:Oabmaps})
with the low ionization knots prominent. However the N$^{++}$/H$^{+}$ 
map from the \ion{N}{II} ORL 2s$^{2}$2p3p\,$^{3}$D$^{e}$~--~2s$^{2}$2p3s\,
$^{3}$P$^{o}$ (3-2) 5679.6\,\AA\ line (recombination of N$^{++}$ to N$^{+}$)
is markedly different in appearance (Fig. \ref{fig:N+maps}). The ORL
N$^{++}$/H$^{+}$ map was formed using the [\ion{S}{III}] $T_{\rm e}$ 
and [\ion{Cl}{III}] $N_{\rm e}$ (the latter only 
500 cm$^{-3}$ higher in the mean than the [\ion{S}{II}] $N_{\rm e}$ 
image) in the event of not having a spaxel-by-spaxel map of \ion{N}{II} ORL 
$T_{\rm e}$ (see Sect. \ref{ORLTeNe}). This ORL N$^{++}$/H$^{+}$ image 
resembles the Ar$^{+++}$ map (shown in Fig. \ref{fig:SClArmaps}), and 
the ionization potentials substantially overlap (29.6 $ <$ N$^{++} < $ 
47.4 eV; 27.6 $ <$ Ar$^{+++} < $ 40.7 eV).


\subsection{S, Cl and Ar ionization maps}
\begin{figure*}
\centering
\resizebox{\hsize}{!}{
\includegraphics[width=0.50\textwidth,angle=0,clip]{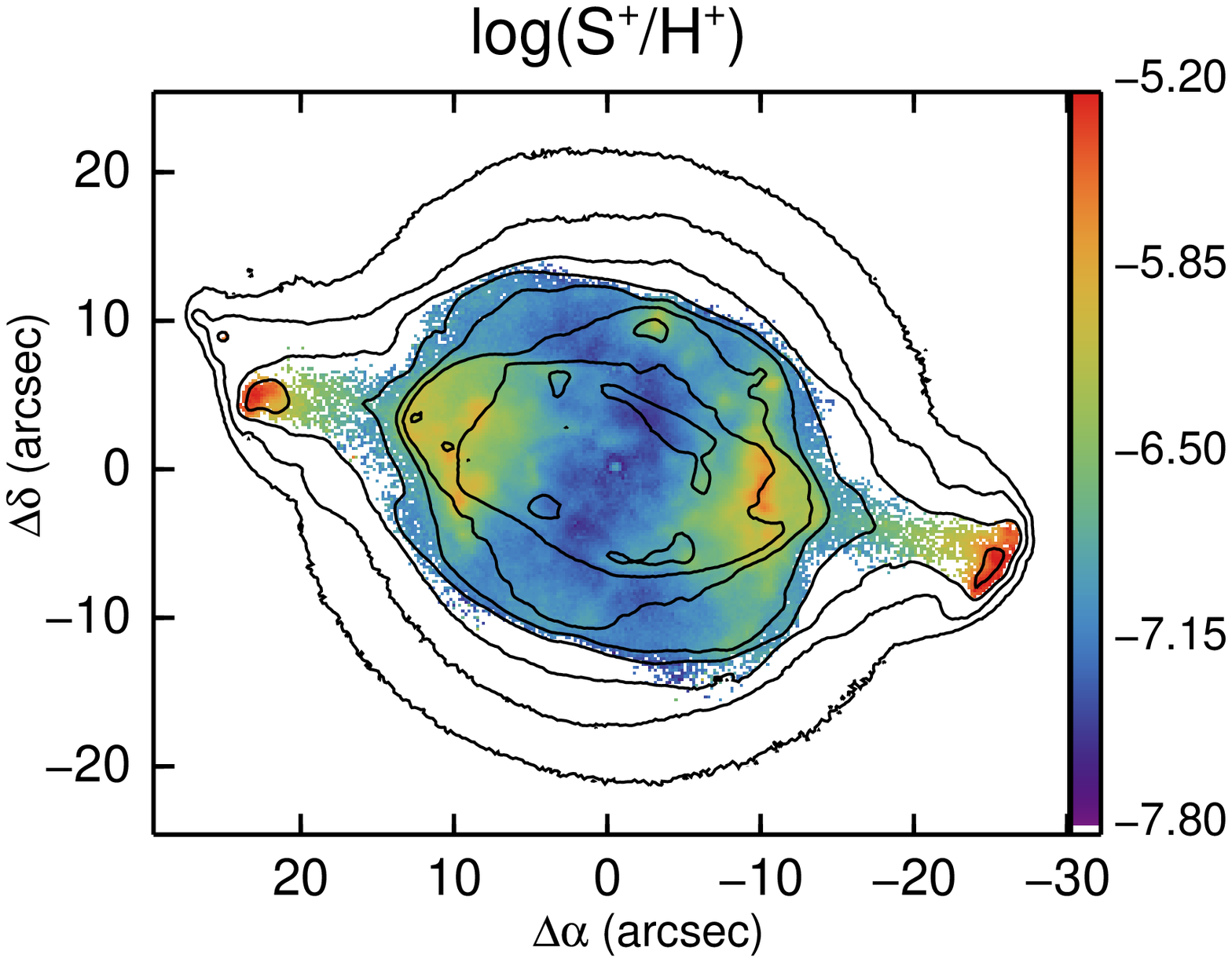}
\hspace{0.2truecm}
\includegraphics[width=0.50\textwidth,angle=0,clip]{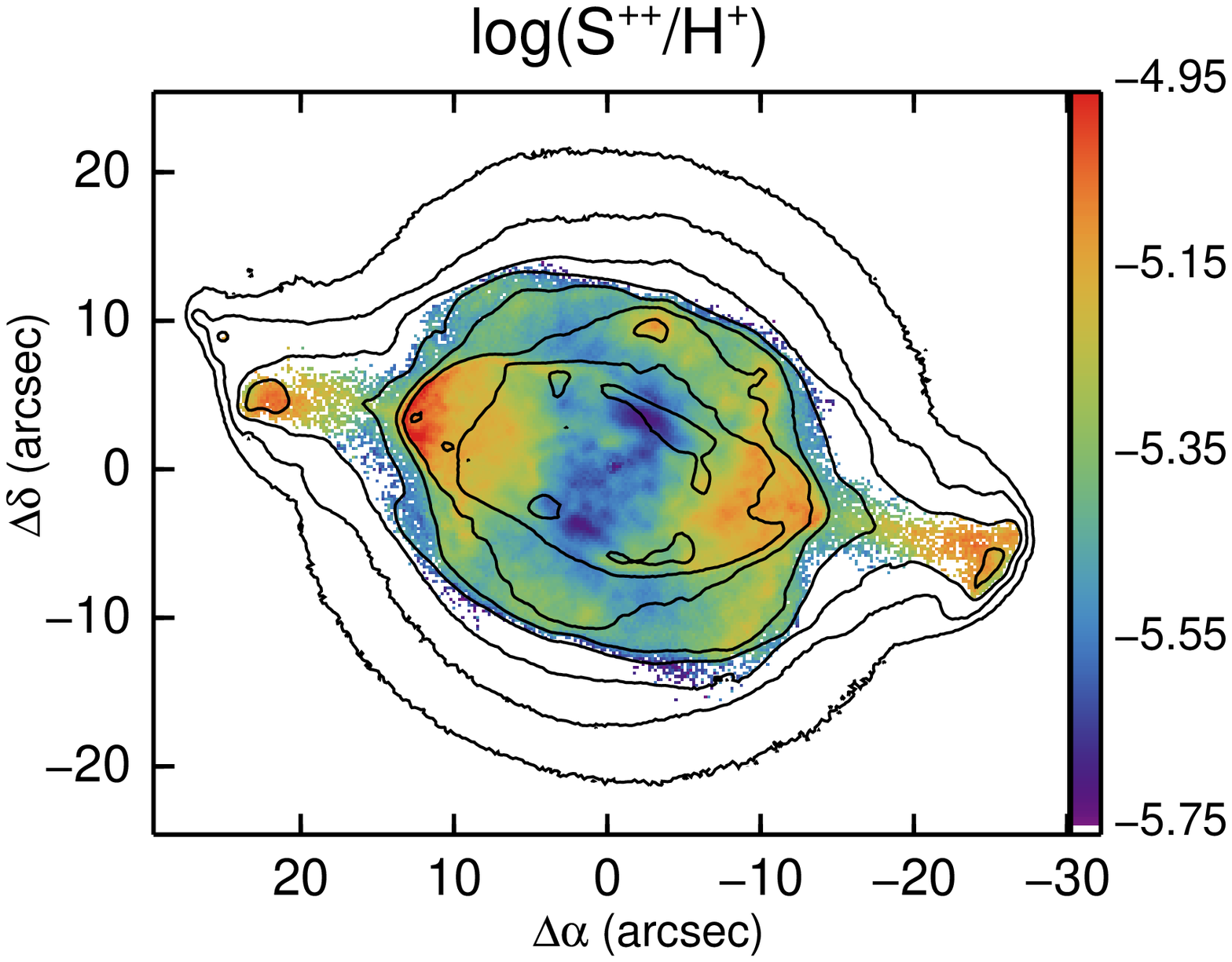}
\hspace{0.2truecm}
\includegraphics[width=0.50\textwidth,angle=0,clip]{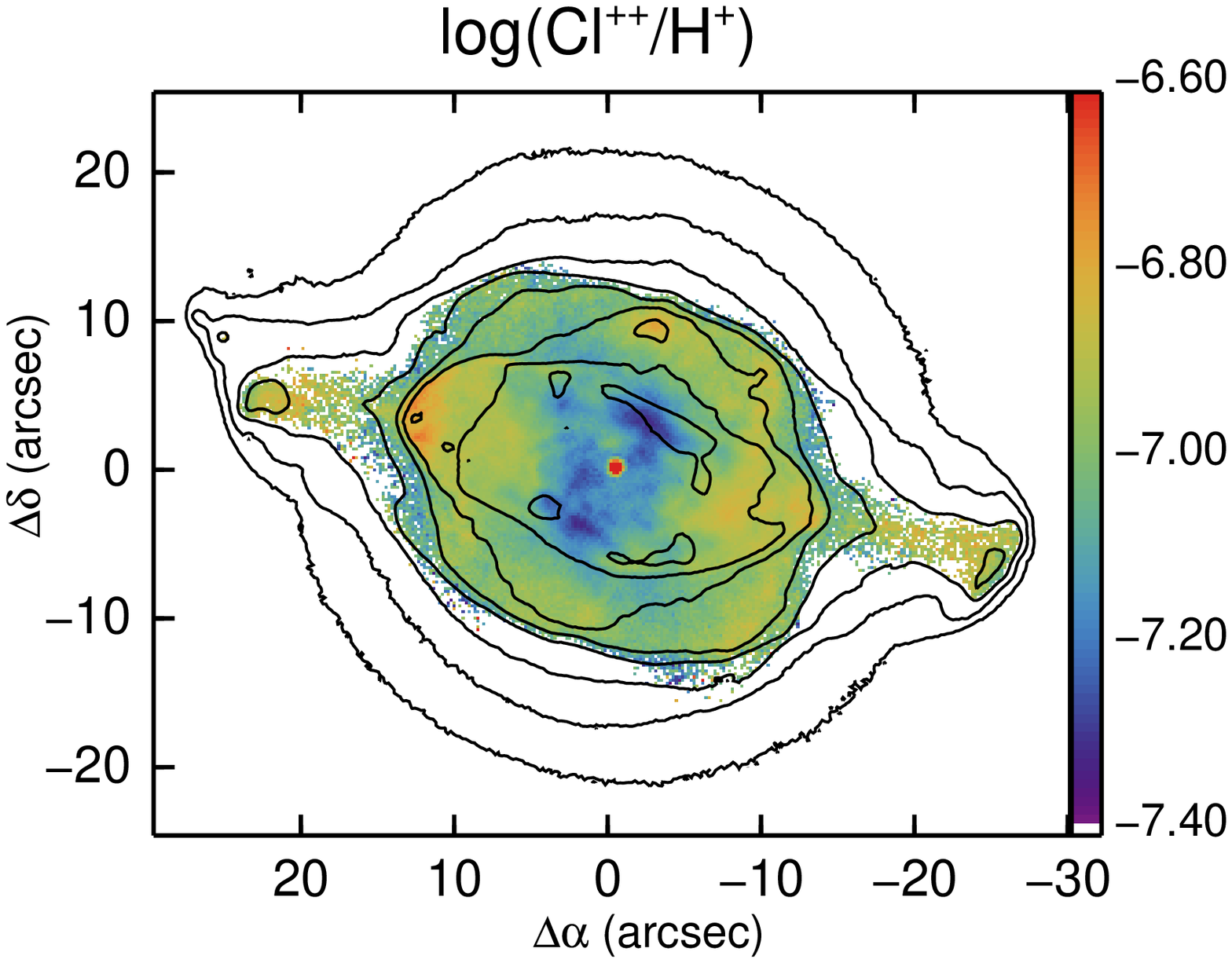}
}
\centering
\resizebox{\hsize}{!}{
\includegraphics[width=0.50\textwidth,angle=0,clip]{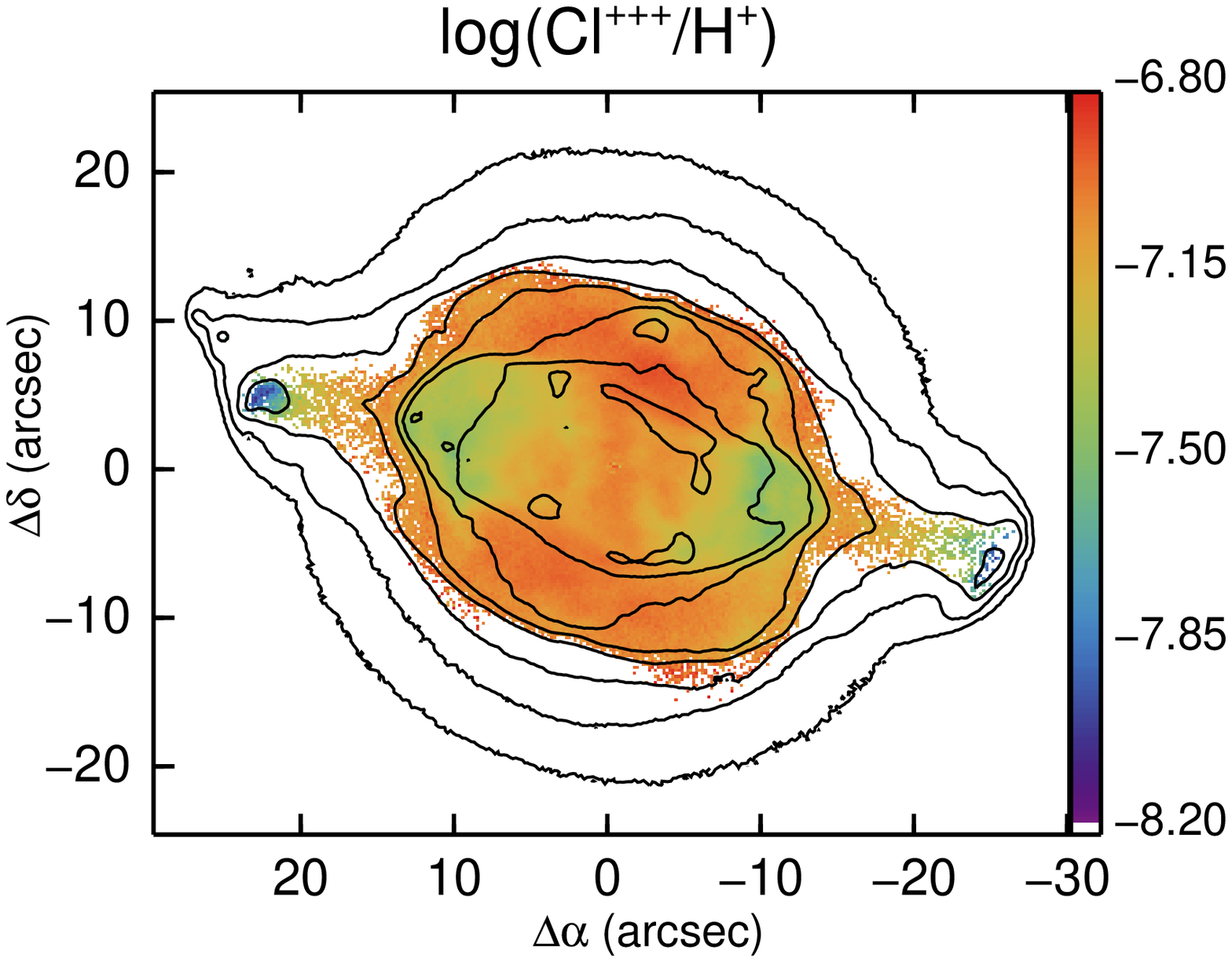}
\hspace{0.2truecm}
\includegraphics[width=0.50\textwidth,angle=0,clip]{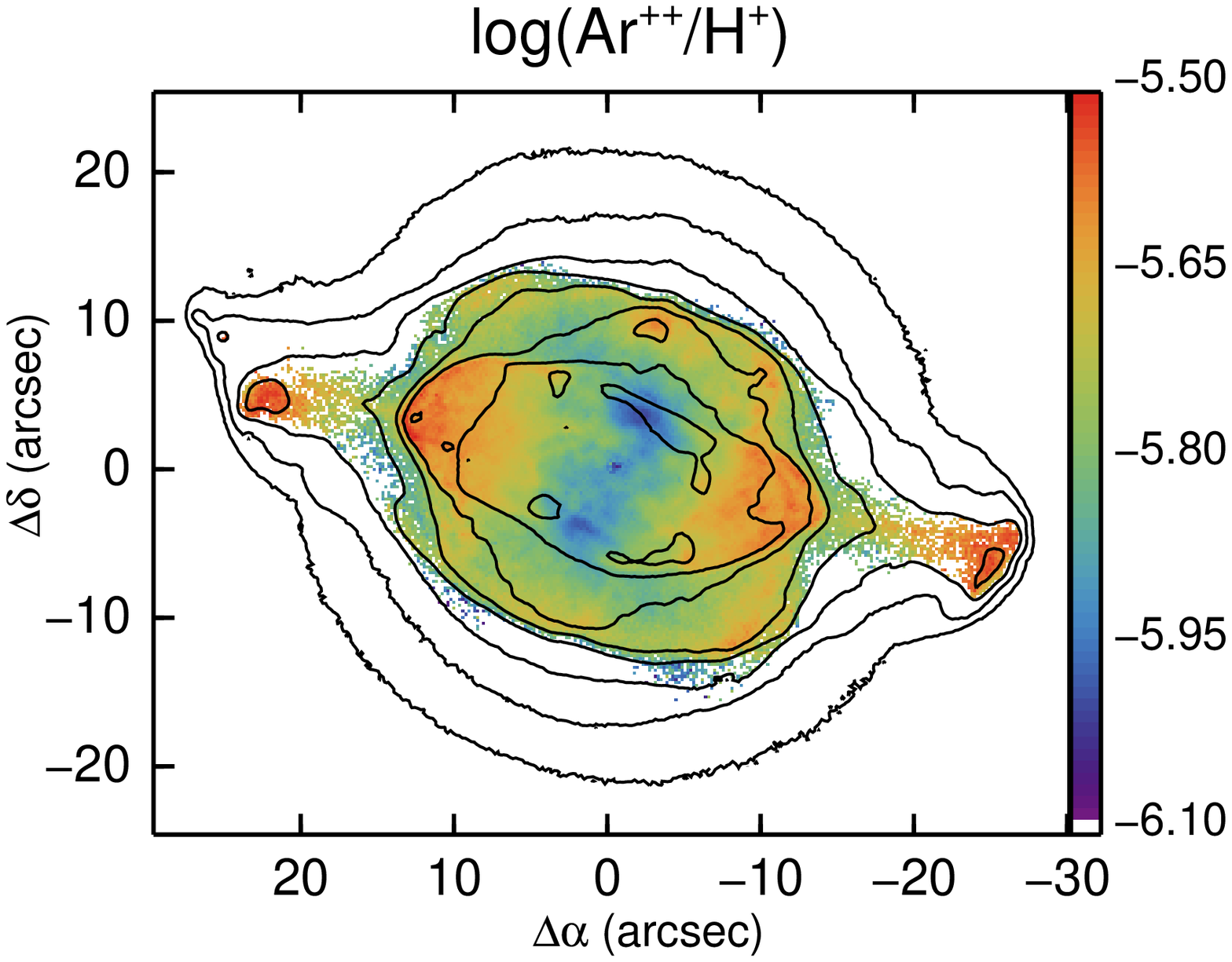}
\hspace{0.2truecm}
\includegraphics[width=0.50\textwidth,angle=0,clip]{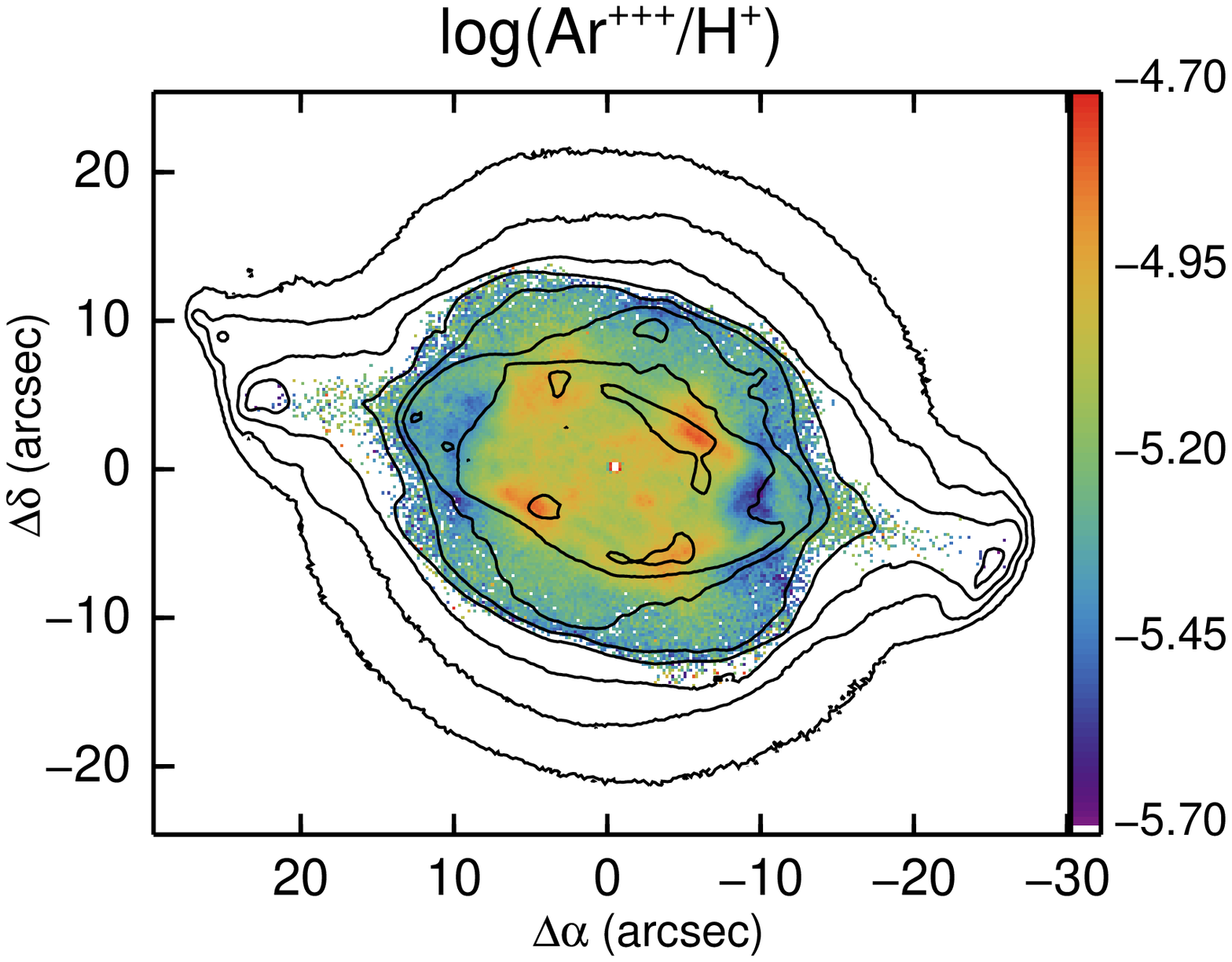}
}
\caption{Maps of S$^{+}$ and S$^{++}$, Cl$^{++}$ and Cl$^{+++}$
and Ar$^{++}$ and Ar$^{+++}$/H$^{+}$ ionic abundances. The log F(H$\beta$) 
surface brightness contours are as in Fig. \ref{fig:nemaps}.
}
\label{fig:SClArmaps}
\end{figure*}

The other species for which maps of several ionization states
from CEL lines can be formed are S (S$^{+}$, S$^{++}$ from [\ion{S}{II}]
6730.8\,\AA\ and [\ion{S}{III}]9068.6\,\AA\ respectively), Cl (from Cl$^{++}$ 
and Cl$^{+++}$ from [\ion{Cl}{III}]5537.9 and [\ion{Cl}{IV}]8045.6\,\AA) and Ar
(Ar$^{++}$ and Ar$^{+++}$ from [\ion{Ar}{III}]7134.8 and 
[\ion{Ar}{IV}]7237.4\,\AA). Figure \ref{fig:SClArmaps} shows the 
S, Cl and Ar ionization
maps. The morphology of the lower ionization species
(S$^{+}$, Cl$^{++}$ and S$^{++}$ and Ar$^{++}$, ionization potentials
10.4 to 27.6eV are quite similar, and to O$^{+}$, but the appearance
of the Cl$^{+++}$ and Ar$^{+++}$ (IP's to 39.6 and 40.7 eV respectively)
differ strongly, and also from the O$^{++}$ (Fig. \ref{fig:Oabmaps}).
Neither Cl$^{+++}$ and Ar$^{+++}$ show the central ionic emission with
a depression, as shown by O$^{++}$; the Cl$^{+++}$ has a bipolar appearance 
with the outer shell minor axes enhanced and Ar$^{+++}$ has a curious 
four-cornered (biretta) shape extending into the outer shell.
    
\section{Abundance maps}
\label{Abundmaps}

\subsection{He abundance}
The only species for which a total abundance map can be formed without
invoking use of an explicit ionization correction factor (ICF) map is He. 
Given that there is no reliable diagnostic for the fraction of 
He$^{0}$ in an ionized nebula, the default assumption is made that it 
represents a negligible proportion: the sum of He$^{+}$ and He$^{++}$ 
should therefore map the total He abundance. This assumption is
reinforced by the 1D photoionization models presented in Sect. 
\ref{Cloudy1D}, based on an integrated spectrum, where both show 
neutral He fractions $<$0.15\%. The 
simplest assumption, following Occam's razor, is that the He abundance in
the nebular shells is uniform. Helium is a general product of Big Bang 
nucleosynthesis and is supplemented by stellar H-burning phases, which 
begin on the main sequence and are first brought to the surface by the 
first dredge-up, and later, principally, during third dredge-up on the 
asymptotic giant branch. 
By comparison of $L, T_{eff}$ from \citet{Sabbadin2004} with the 
\citet{MillerBerto2016} solar abundance evolutionary tracks, an initial mass 
of 1.4M$_{\odot}$ and a current mass of 0.58M$_{\odot}$ are derived. Thus 
second dredge-up is not expected to have occurred in NGC~7009 
\citep[c.f.,][]{KarakasLatt2014}.

\begin{figure}
\resizebox{\hsize}{!}{
\includegraphics[width=0.50\textwidth,angle=0,clip]{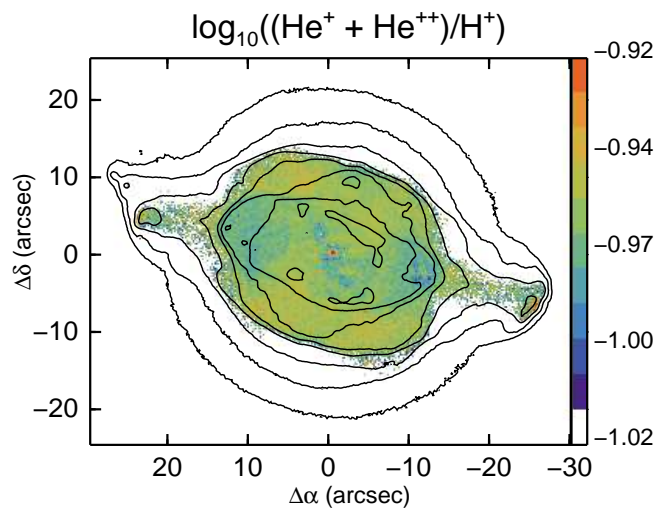}
}
\caption{Sum of He$^{+}$/H$^{+}$ and He$^{++}$/H$^{+}$ using 
the [\ion{S}{III}] $T_{\rm e}$ and [\ion{Cl}{III}] $N_{\rm e}$ maps
for He$^{+}$/H$^{+}$ and He$^{++}$/H$^{+}$ computation.
The log F(H$\beta$) surface brightness contours are as in 
Fig. \ref{fig:nemaps}.
}
\label{fig:Hetot}
\end{figure}

Figure \ref{fig:Hetot} shows the He/H$^{+}$ map formed from the 
He$^{+}$/H$^{+}$ map (Fig. \ref{fig:He+map}) and the He$^{++}$/H$^{+}$ 
map (Fig. \ref{fig:He++map}), where all (He and H) emissivities were 
calculated with $T_{\rm e}$ from [\ion{S}{III}] and $N_{\rm e}$ from 
[\ion{Cl}{III}] (Sect. \ref{CELTeNe}). There does not appear to 
be evidence that the He/H$^{+}$ values are depressed in the knots 
K1 -- K4 (Fig. \ref{fig:nsclmaps}), suggesting no enhanced contribution 
from He$^{0}$ in these low ionization regions.

The flattest map of total He abundance was produced by using 
the values of $T_{\rm e}$ from [\ion{S}{III}] (Sect. \ref{CELTeNe})
for both He$^{++}$ and He$^{+}$ with respect to H$^{+}$. Adopting 
$T_{\rm e}$ from \ion{He}{I} 7281.4/6678.2\,\AA\ (Sect. \ref{TeHe1}) for 
calculating the emissivity of both He$^{+}$ and H$^{+}$, and the 
[\ion{S}{III}] $T_{\rm e}$ map for both He$^{++}$ and H$^{+}$, leads to 
a map displaying stronger residuals from flatness, corresponding to 
the inner and outer shells and the region of He$^{++}$ emission 
(see Fig. \ref{fig:He++map}); the mean and 3-$\sigma$ 
RMS on this image was 0.1057 $\pm$ 0.0061. The map produced solely
from [\ion{S}{III}] $T_{\rm e}$ resulted in a mean and 3-$\sigma$ RMS of 
0.1079 $\pm$ 0.0025. If the Paschen Jump $T_{\rm e}$ map is used to 
calculate the He$^{+}$ and H$^{+}$ emissivities, the map is
far from flat, the He$^{++}$ regions are enhanced and the outer shell 
has values of He/H$^{+}$ of around 0.085, worryingly close to the primordial 
He/H value of 0.0745 (He$^{4}$ mass fraction 0.24709$\pm$0.00017, 
\citep{Pitrou2018} given the heavy element abundances of NGC~7009 
\citep{FangLiu2011}). 

\subsection{O abundance}
Typical ICF schemes employ the He$^{+}$ and He$^{++}$ ratios to
correct for the presence of O$^{+++}$, which is not covered by 
optical ground-based spectroscopy since [\ion{O}{IV}] lines occur in the 
ultra-violet and infrared (see \citet{Phillips2010} for the 
latter). Several ionization correction schemes have been 
developed, starting with \citet{TorresPeimbert1977}, followed by 
\citet{KingBarlow1994} and \citet{Delgado-Inglada2014},
primarily based on proximity of ionization potentials in various
ions and, in the later work, through development with photoionization 
models. Recently \citet{Morisset2017} suggested that in the
era of IFUs and 3D ionization codes, ICFs can be allowed to
vary spatially depending on the local ionization conditions,
dictated by the model. For the simplest expectation that the 
abundance of O is uniform across the nebula, then large scale
discontinuities coincident with ionization boundaries and physical 
features, such as ansae, polar knots etc., should not be expected. 
However this simple assumption should not preclude small-scale 
abundance differences, such as associated with H-poor knots in 
born again PNe, for example Abell 78 \citep{JacobyFord1983}.

\begin{figure*}
\centering
\resizebox{\hsize}{!}{
\includegraphics[width=0.50\textwidth,angle=0,clip]{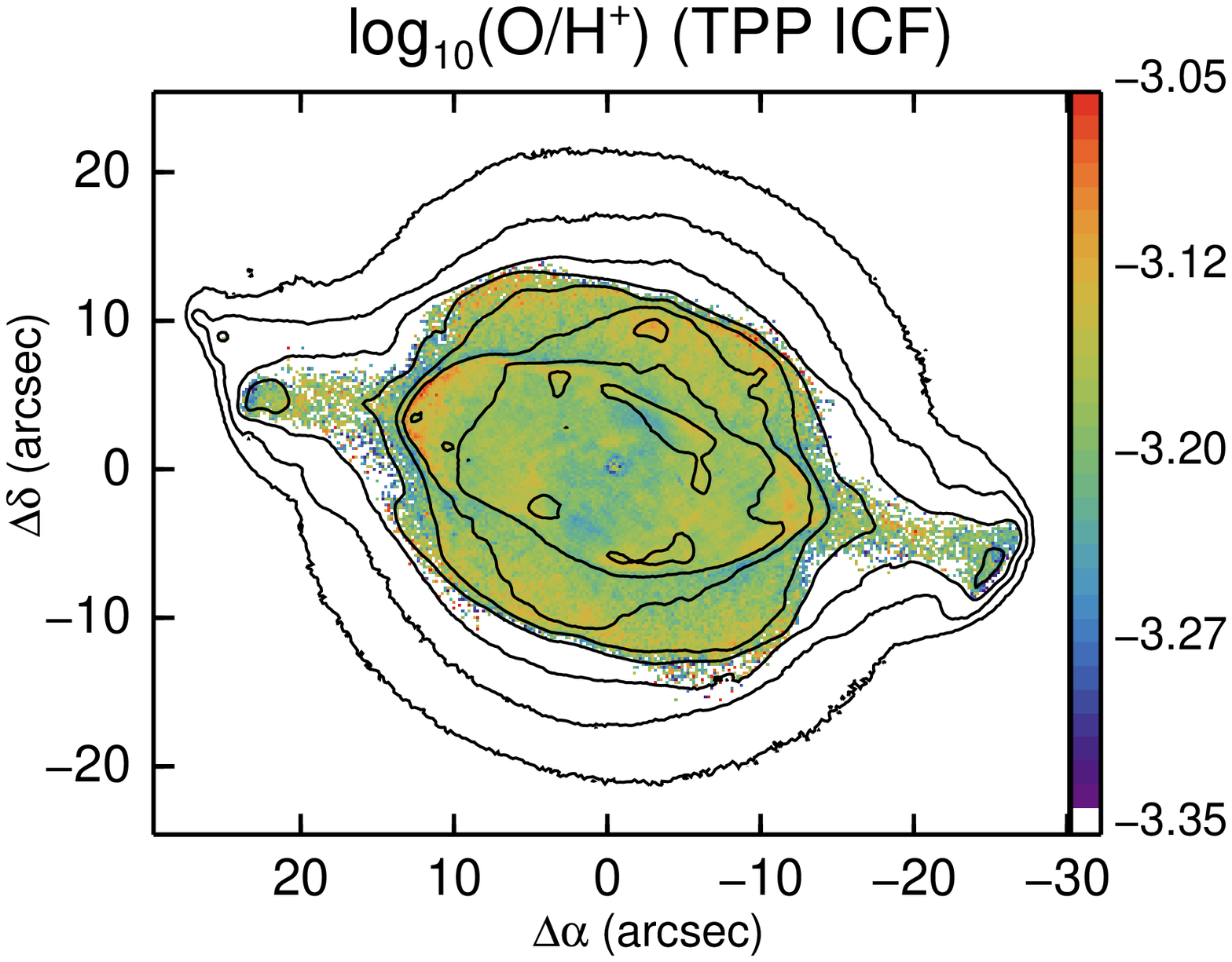}
\hspace{0.2truecm}
\includegraphics[width=0.50\textwidth,angle=0,clip]{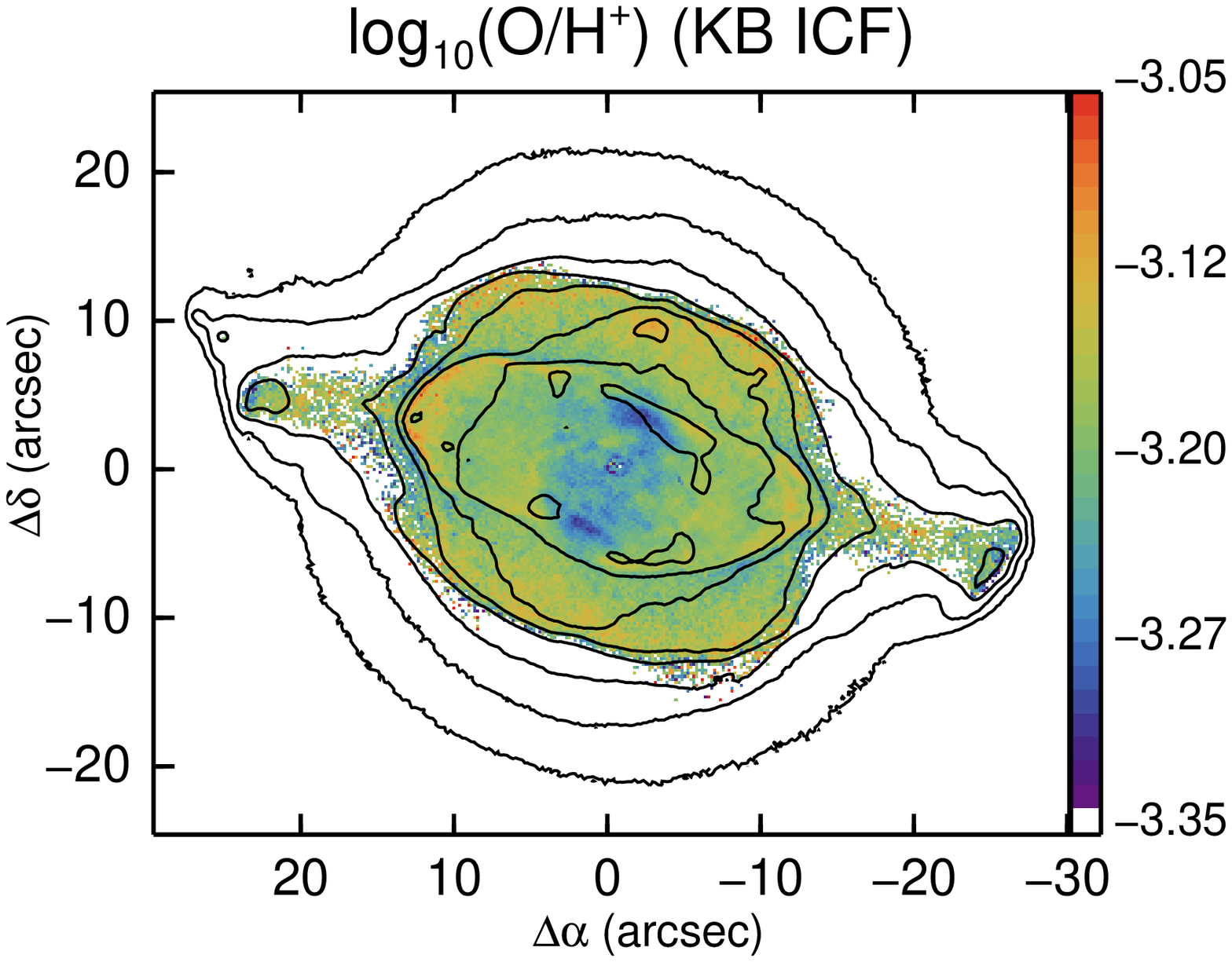}
\hspace{0.2truecm}
\includegraphics[width=0.50\textwidth,angle=0,clip]{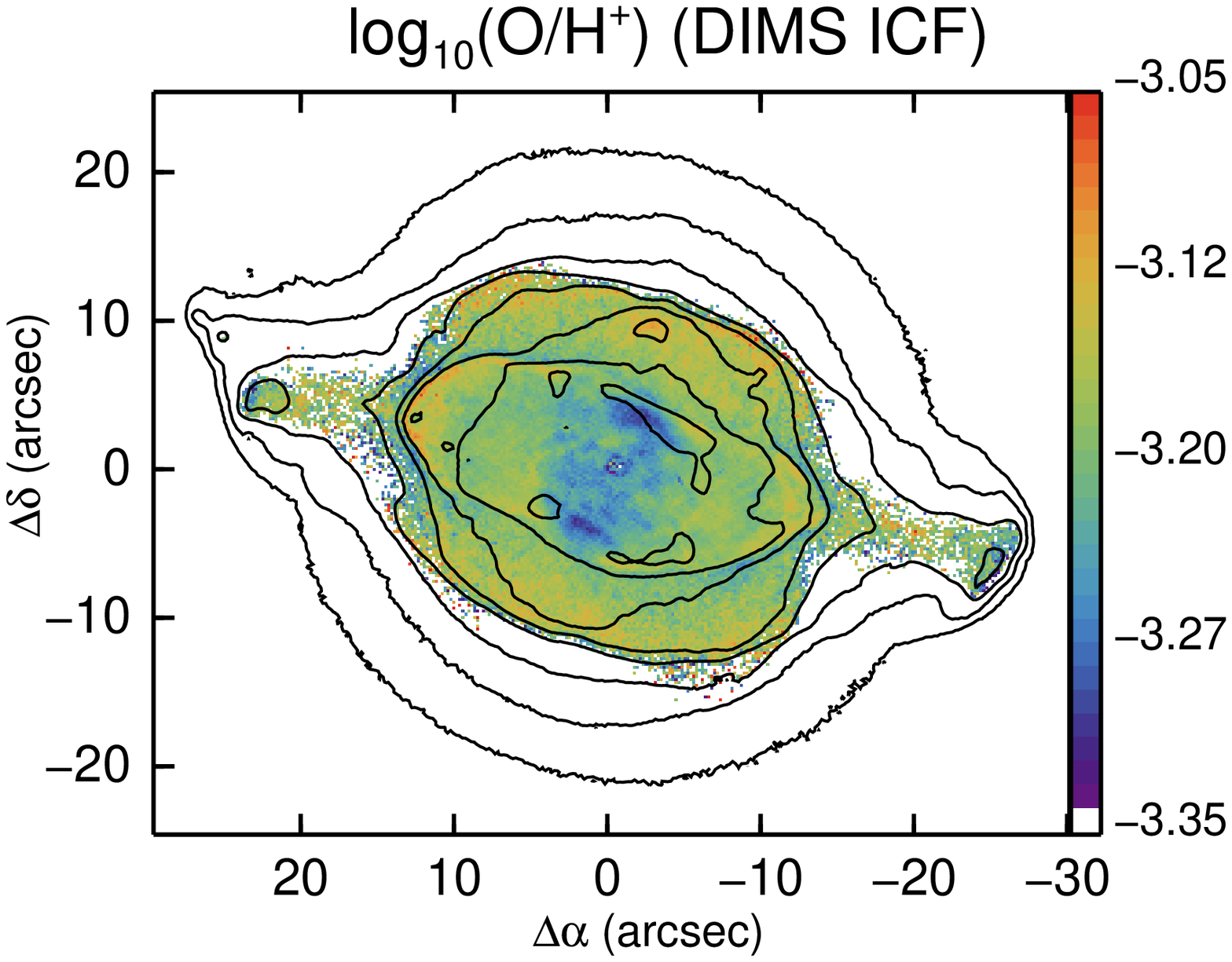}
}
\caption{Maps of total O/H$^{+}$ abundance for three ionization
correction factors adjusting for the presence of O$^{+++}$.
Left: the 'classical' scheme of \citet{TorresPeimbert1977};
centre: the corrected classical ICF advocated by \citet{KingBarlow1994};
right: the model-based approach of \citet{Delgado-Inglada2014}.
The log F(H$\beta$) surface brightness contours are as in Fig. 
\ref{fig:nemaps}.
}
\label{fig:Otot}
\end{figure*}

Figure \ref{fig:Otot} shows the comparison of the total O abundance 
from the ICF's of \citet{TorresPeimbert1977}, 
\citet{KingBarlow1994} and \citet{Delgado-Inglada2014}. It is clear
from these maps that the correction is not complete enough as a
deficit remains in the central regions where the He$^{++}$ is
strong. All three ICF's need a 20-30\% correction for the presence
of O$^{+++}$. However there are also traces of the inner rim
in the He abundance map, coincident with the peak of the wave in 
the dust extinction map shown in Paper I. This feature in
the region of the ionization front may indicate that the conditions 
in the He$^{++}$ zone, and possibly also in the He$^{+}$ zone, may 
not be well diagnosed by adopting the [\ion{S}{III}] $T_{\rm e}$ values.

Based on the detections of lines such as [\ion{K}{VI}], [\ion{Mn}{vi}] 
and [\ion{Fe}{vi}] by 
\citet{FangLiu2013} (although these lines are very weak), there may be
expected to be some ions with energies above 60 eV present in the
nebula, so correction for the presence of O$^{4+}$ must be considered. 
However, in principle, the O ICF schemes should include all ionic
stages above O$^{++}$. A weaker alternative proposition is that the
ICF is too low even where He$^{++}$ is not present; increasing the ICF
could raise the correction, including in the regions with
strong \ion{He}{II} emission, such that the discrepancy over the He$^{++}$ 
region is reduced.

\subsection{N/O abundance}

\begin{figure}
\resizebox{\hsize}{!}{
\includegraphics[width=0.50\textwidth,angle=0,clip]{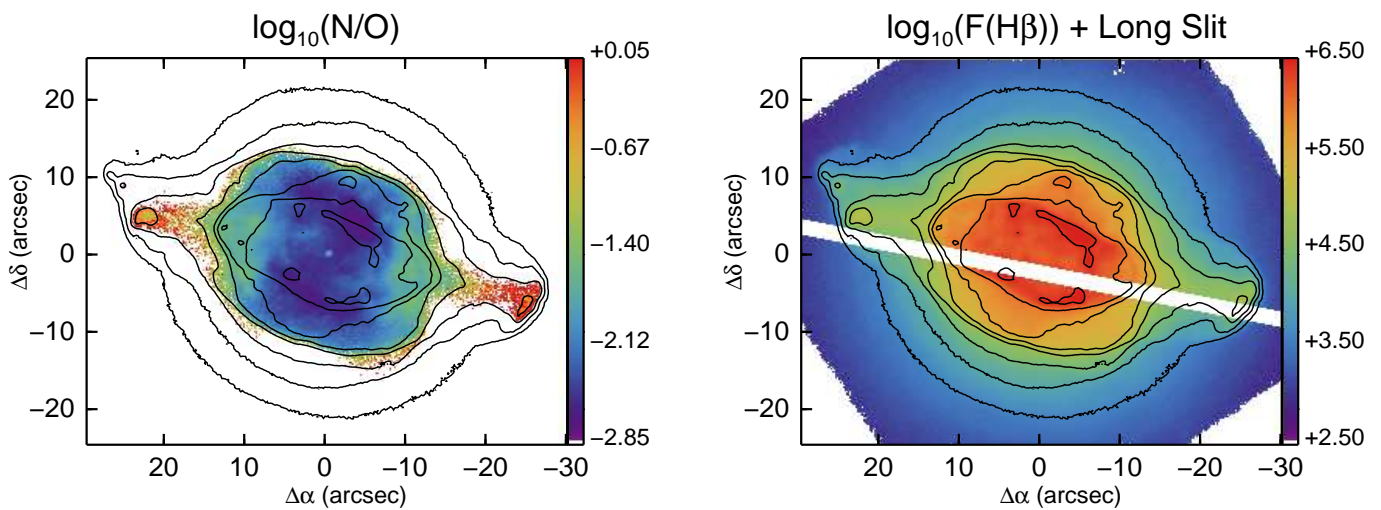}
}
\caption{Map of the N/O abundance based on the N$^{+}$/H$^{+}$ and
O$^{+}$/H$^{+}$ maps (Figs. \ref{fig:N+maps} and \ref{fig:Oabmaps}, 
respectively).
The log F(H$\beta$) surface brightness contours are as in Fig. 
\ref{fig:nemaps}. The mean value of signal-to-error is 13.
}
\label{fig:NOmap}
\end{figure}

The classical method of deriving the N/O abundance ratio is based on the
assumption that N$^{+}$ and O${^+}$, on account of the similarity of
their ionization energies, are co-located; thus a simple
ratio of their ionic emission can be employed to determine the 
total N/O \citep{Peimbert_Costero1969, Peimbert_TorresPeimbert1971, 
Kaler1979}; N/O is an important diagnostic of the dredge-up enrichment 
of the nebula by the central star. The resulting N/O 
map is shown in Fig. \ref{fig:NOmap}. The simple mean value 
of N/O is 0.132 but the values are far from constant and range over 
more than an order of magnitude to $>$ 1 at the extremities of knots K1 
and K4, as noted by many works. Although developed on long-slit or nebula 
integrated spectra, and compared with 1D photoionization models, it clear 
that, at least in the case of NGC~7009, this scheme for determining total 
N/O is a gross over-simplification, as demonstrated by 
\citet{Goncalves2006}. 

The ionization energies for O$^{+}$ 
and N$^{+}$ are 13.6 -- 35.1 and 14.5 -- 29.6 eV, respectively; thus the
range of N$^{+}$ ionization energies is entirely contained within that 
of O$^{+}$. However the physical conditions in NGC~7009, and perhaps 
also the central star far-UV energy distribution, disfavour these ions 
being co-extensive within the nebula volume and this method appears to 
be ill-suited to a spatially resolved ionization correction scheme. 
\citet{Goncalves2006} from 3D photoionization modelling point out that
the difference in ionization potentials make the spatially resolved 
ratio N$^{+}$/O${^+}$ very sensitive to the shape of the local 
radiation field. A relatively large fraction of N must be in the form 
of N$^{++}$ over a substantial area of the nebula, such
that correcting the amount of N$+$ by including ionization energies up
to 35.1 eV could lead to a flat N/O map, as suggested by the
N$^{++}$/H$^{+}$ ORL map (Fig. \ref{fig:N+maps}).   

\section{Integrated spectra}
\label{Intspec}
Integral field data provide an ideal basis for constructing 
summed spectra of regions defined by a variety of geometrical
(or other) criteria. These spectra have enhanced signal-to-noise 
based on the summation of spectra in many pixels and can lead to
the detection of lines much fainter than are accessible in the 
spaxel maps. In the case of NGC~7009, summed (spatially
integrated) spectra over its specific morphological features
can be formed: in addition to the obvious
ones of inner shell, outer shell and halo, summations over the ansae, 
high ionization poles on the minor axis (Fig. \ref{fig:He++map}), 
the cusp in the extinction map (Paper I) or the higher density 
lobes on the minor axis (Fig. \ref{fig:nemaps}) can be analysed,
for example.

\begin{figure}
\resizebox{\hsize}{!}{
\includegraphics[width=0.50\textwidth,angle=0,clip]{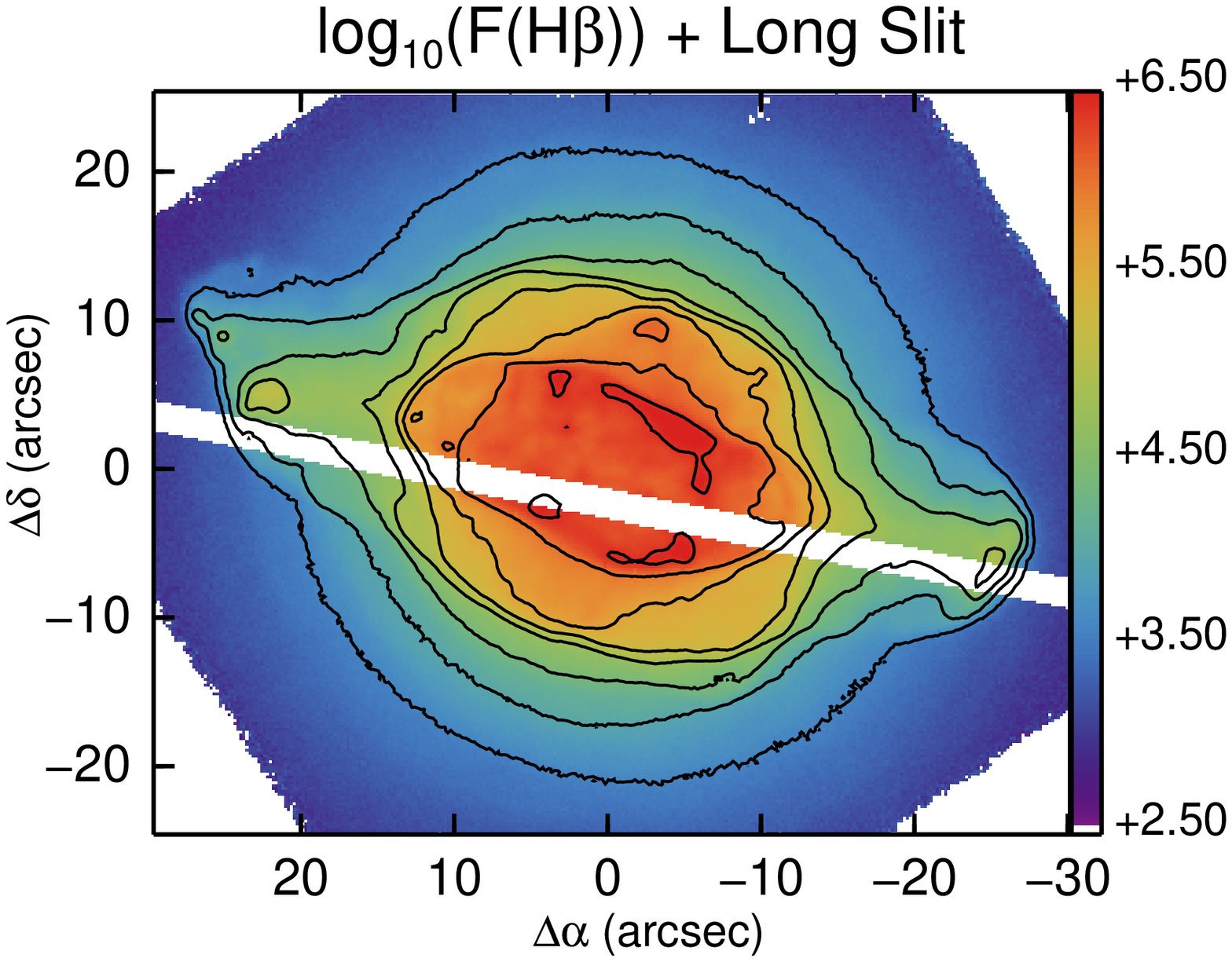}
}
\caption{The simulated footprint of the long slit (PA 79$^\circ$) 
used by \citet{FangLiu2011, FangLiu2013} in their deep spectroscopic 
study of NGC~7009 shown, in white, against the 10s H$\beta$ image. The 
slit samples 14\% of the total H$\beta$ flux. The log F(H$\beta$) 
surface brightness contours are as in Fig. \ref{fig:nemaps}.
}
\label{fig:Liuslit}
\end{figure}

Specific geometrical regions can be defined, such as 
slits and apertures that sample those from previous studies, to
determine how the derived physical quantities (and abundances) 
relate to the total (nebula summed) values. A specific example is the long-slit 
spectrum, 2$''$ wide, situated 2~--~3$''$ south of the central star 
oriented along the major axis (PA 79$^\circ$), in the deep 
spectroscopic study by \citet{FangLiu2011, FangLiu2013}.
This slit was numerically constructed and the total spectrum
was formed by all those spaxels within this slit. Figure
\ref{fig:Liuslit} shows the simulated slit on the H$\beta$ emission
line image. The slit samples 14.2\% of the total H$\beta$ flux 
(based on the H$\beta$ image) but since the \ion{He}{II} emission
is considerably extended along the ansae (Fig. \ref{fig:ratmaps}), 
fully 17.2\% of the total \ion{He}{II} emission is sampled by this slit,
presenting a slightly higher ionization view of the entire 
nebula. Given the diversity of data and instruments used in the 
\citet{FangLiu2011, FangLiu2013} study, the MUSE spectrum from this
simulated slit serves as a useful fiducial for this very
deep (and higher spectral resolution) spectrum. 

One of the 'bonuses' of IFU spectroscopy of an entire Galactic
planetary nebulae is the total spectrum integrated out to the 
halo, which can be compared to truly integrated spectra of 
more distant PNe in the Magellanic Clouds and nearby galaxies, 
for example. Summing all the spaxels of the 120s H$\beta$ cube 
over an area of 2340 square arcseconds (0.704 square arcminutes) 
results in an H$\beta$ flux of 1.469 $\times10^{-10}$ ergs cm$^{-2}$ 
s$^{-1}$. This can be compared with a literature photoelectric 
photometric H$\beta$ flux (-9.78 +/- 0.03 in log$_{10}$) of 
1.62 $\times$ 10$^{-10}$ ergs cm$^{-2}$ s$^{-1}$ 
\citep{Peimberts1971}. 

%

\subsection{1D photoionization model}
\label{Cloudy1D}
Although the maps show the 2D appearance of line ratios, diagnostics
and abundances, a simple 1D photoionization model is useful
for quantifying how the 2D discrepancies affect conclusions about
the structure of the nebula and the central star. Based on the
dereddened emission line images (Sect. \ref{CELTeNe}), an 
indicative basic set of line ratios for the integrated nebula 
was formed taking the H$\beta$ flux within the contour
of $7 \times 10^{-17}$ ergs cm$^{-2}$ s$^{-1}$ per spaxel
(surface brightness $2.8 \times 10^{-18}$ ergs cm$^{-2}$ s$^{-1}$
arcsec$^{-2}$ , enclosing a total flux of $2.08 \times 10^{-10}$ 
ergs cm$^{-2}$ s$^{-1}$). Table \ref{tab:integfluxes} lists the 
dereddened line fluxes relative to H$\beta$=1.0; the dereddened 
fluxes from \citet{FangLiu2011} are listed in col. 4 for comparison. 

As a starting basis, a spherical CLOUDY \citep[version 17.01, ][]{Ferland2017} 
model was constructed starting with the set of abundances from 
\citet{Sabbadin2004} and a black body stellar atmosphere. A 
satisfactory model matching the line ratios in
Table \ref{tab:integfluxes} could be found with a black body of 
95\,000 K, a constant density of 1600 cm$^{-3}$, and inner and outer
nebular radii of 0.0005 and 0.10 pc respectively; dust was not 
included in the model. Col. 5 of Table \ref{tab:integfluxes} 
lists the output CLOUDY relative fluxes; referred to as Model A. The 
footnote to the table lists the light element abundances; for 
elements not listed, abundances were taken from \citet{Sabbadin2004}. 

The stellar effective temperature is slightly higher than 
that determined by \citet{Sabbadin2004}, and considerably higher
than the value of 82\,000K found by \citet{Mendez1992} from a NLTE 
model atmosphere fit to the stellar lines profiles, but such a high value
is required to match the \ion{He}{II} / \ion{He}{I} line ratios. The 
constant H density was lower than the electron density values from
the integrated [\ion{S}{II}] 6716.4/6730.8\,\AA\ and 
[\ion{Cl}{III}]5517.7/5537.9\,\AA\ ratios of 3900 and 3300 
cm$^{-3}$, respectively, for an [\ion{S}{III}] electron temperature 
of 9900 K, but no filling factor was included. The outer radius was
restricted to ensure a density-bounded nebula to match the low 
ionization lines in particular.

A more elaborate model was constructed using an NLTE He-Ne model 
atmosphere from \citet{Rauch2003} (solar metallicity and
log g=5.0) and a spherical cylinder geometry with
elongation to match the major axis length. The abundances
of He, N, O and S were allowed to vary slightly (within 0.2 dex) 
of the values from \citet{Sabbadin2004} and \citet{FangLiu2013}. 
A stellar effective temperature of 98\,000K was used with a constant
H density of 1000 cm$^{-3}$, inner and outer radii of 0.0008 and
0.136 respectively, and a filling factor of 0.5; the abundances are 
listed in the footnote 
to Tab. \ref{tab:integfluxes}. The quality of the match to the emitted 
spectrum (Tab. \ref{tab:integfluxes}, Model B) is similar to Model A,
except that the higher ionization gas is hotter as indicated by the
stronger auroral [\ion{O}{III}]4363.2\,\AA\ and [\ion{S}{III}]6312.1\,\AA\ 
model line strengths. Model B could be used as a starting point for 
more elaborate 1D models to match the integrated spectrum and as the 
basis for 3D modelling.

\begin{table*}
\caption{Integrated NGC~7009 dereddened line ratios: observations 
and photoionization models}
\centering
\begin{tabular}{llllrr}
\hline\hline
Species  & $\lambda$(\AA)   & F(H$\beta$=1.0) & F(H$\beta$=1.0)  & Model           & Model \\
         &                  & MUSE            & Fang\&Liu        & F(H$\beta$=1.0) & F(H$\beta$=1.0) \\
         &                  &                 & (2011)           & \bf{A}         & \bf{B} \\ 
\hline
\ion{O}{II}   & 3726.0 &       & 0.137 &  0.162 & 0.151 \\
\ion{O}{III}  & 4363.2 &       & 0.073 &  0.081 & 0.139 \\
H$\beta$      & 4861.3 & 1.000 & 1.000 &  1.000 & 1.000 \\
\ion{O}{III}  & 4958.9 & 3.880 & 3.879 &  4.139 & 4.154 \\
\ion{He}{II}  & 5411.5 & 0.011 & 0.012 &  0.009 & 0.013 \\
\ion{Cl}{III} & 5517.7 & 0.005 & 0.005 &  0.007 & 0.006 \\ 
\ion{Cl}{III} & 5537.9 & 0.006 & 0.006 &  0.008 & 0.006 \\
\ion{O}{I}    & 6300.3 & 0.004 & 0.006 &  $<$0.001 & $<$0.001 \\
\ion{S}{III}  & 6312.1 & 0.014 & 0.014 &  0.014 & 0.022 \\
\ion{N}{II}   & 6583.5 & 0.147 & 0.156 &  0.164 & 0.126 \\
\ion{He}{I}   & 6678.2 & 0.040 & 0.037 &  0.036 & 0.039 \\
\ion{S}{II}   & 6716.4 & 0.014 & 0.014 &  0.017 & 0.024 \\
\ion{S}{II}   & 6730.8 & 0.023 & 0.023 &  0.023 & 0.028 \\
\ion{O}{II}   & 7320.0 & 0.013 & 0.012 &  0.007 & 0.007 \\
\ion{O}{II}   & 7330.2 & 0.011 & 0.011 &  0.006 & 0.006 \\
\ion{S}{III}  & 9068.6 & 0.252 & 0.231 &  0.227 & 0.236 \\
              &        &       &       &        &       \\
L(H$\beta$)   &        & 34.69 &       &  34.83 & 34.53 \\
\hline
\end{tabular}
\tablefoot{Model A log abundances: He -0.95; C -3.80; N -3.80; O -3.30; Ne -3.85
S -5.15; Cl -6.70. \\
Model B log abundances: He -0.90; C -3.79; N -4.00; O -3.55; Ne -3.80; 
S -5.25; Cl -7.00.
}
\label{tab:integfluxes}
\end{table*}


\section{Discussion}
\label{Discus}

A few aspects of the rich set of MUSE data presented here are discussed:
analysis of the spectra as opposed to the maps; comparisons among
the various $T_{\rm e}$ maps; the problem of abundance maps showing
features of the ionization structures and how ICF's might help to
correct this; the prospects for photoionization modelling of
these data.

\subsection{Limited IFU analysis}
The presentation and analysis has been restricted to mapping the 
emission lines and derived physical diagnostics at the native spatial
resolution of 0.2$''$ $\times$ 0.2$''$, which samples the measured image
quality of 0.5--0.6$''$ well. This analysis clearly misses many 
diagnostically crucial weaker lines. Increasing the signal-to-noise 
per spatial element through various approaches, such as simple 
$m \times n$ binning of spaxels, summation of spaxels across 
morphological or ionization features or Voronoi tesselated 
binning to specified S/N (Sect. \ref{Voronoi}), offer approaches to measure
weaker lines. In particular the ORL lines, with typical strengths
$<<$ 1\% of F($H\beta$) can be studied for how their properties
relate to the CEL lines in identical spatial elements. The MUSE
standard wavelength range provides access to the strongest \ion{N}{II}, 
\ion{N}{III}, \ion{C}{II} and \ion{C}{III} ORL lines but not to 
the strong \ion{O}{II} M1 2p$^{2}$ 3p 4D$_{o}$~--~2p$^{2}$3s 4P 
lines, which are however detectable with 
the MUSE extended mode (spectra to $\sim$4650\,\AA). Of course, 
obtaining deeper spectra provides a more direct approach, but 
given the modest total exposure of 1200s for the set of 120s cubes, 
this route cannot be feasibly be pushed more than a factor $\sim$10 
(and NGC~7009 is among the highest average surface brightness PN 
accessible with MUSE). Not insubstantial
are the problems associated with bright emission lines saturated in 
long exposures, which must be compensated by short unsaturated 
exposures, and can produce strong line wings around the brightest 
lines which hamper detection of surrounding faint lines.

Given the depth of the long slit spectrum data (detection of lines to 
10$^{-4}$ $F(H\beta$)) and detailed analysis of \citet{FangLiu2011, 
FangLiu2013}, it is paramount to relate the physical conditions 
in these long slit data to those of the global nebula, in order
to derive a truly nebula-integrated PN spectrum for comparison with
observations of compact Galactic PNe, unresolved extra-Galactic 
PNe and the few nearby PNe with integrated spectra from slit-scanning
observations (e.g. NGC~7027, \citet{Zhang2007}). A follow-on paper will 
present a study of this cube-integrated spectrum and comparison with 
the results from the long slit study of \citet{FangLiu2011, 
FangLiu2013}. 

\subsection{$T_{\rm e}$ diagnostics}
Apart from the strikingly different morphology which is displayed
in lines of differing ionization potential (Figs. \ref{fig:ratmaps}),
of most interest are the maps presenting the physical conditions,
principally $T_{\rm e}$ and $N_{\rm e}$ and extinction (the latter
presented in Paper I). Here features show up with distinct properties (Figs. 
\ref{fig:nemaps},\ref{fig:temaps}), such as the hotter inner shell (not 
unexpected on account of its strong \ion{He}{II} emission), the higher density 
and temperature polar knots and ansae, features associated with the
more filamentary knots K2 and K3 and the dust accumulation at the interface of
the inner and outer shells. Of particular interest are the maps derived 
from \ion{H}{I} and \ion{He}{I} (Figs. \ref{fig:He+Temaps}, \ref{fig:PJTemap}, 
\ref{fig:hinPasNemap}), which provide different views 
of the physical conditions prevailing in the bulk of the gas (Sect. 
\ref{TeNecomps}), in comparison with the CEL diagnostics. The
CEL and ORL density diagnostics present similar densities, with
their mean values only differing by $<$1000 cm$^{-3}$ (with [\ion{S}{II}] 
lowest to the highest values from the fit to the \ion{H}{I} 
Paschen lines). The CEL and ORL temperature diagnostics 
however differ strongly in value and morphology, with a range of 
3000 K in their mean values (in decreasing magnitude -- [\ion{S}{III}], 
\ion{H}{I} Paschen Jump to \ion{He}{I} $T_{\rm e}$). It will be 
important for the abundance discrepancy factor (ADF) problem to 
compare ORL and CEL ionic abundances spaxel-wise. This comparison can
however only be performed with MUSE for O, for which the extended 
spectral range is required.

Section \ref{t2mapping} presents, to our knowledge, the first map of
the temperature fluctuation parameter, $t^{2}$, calculated however 
from [\ion{S}{III}] $T_{\rm e}$ in comparison to the 'classical'
method using [\ion{O}{III}] $T_{\rm e}$ \citep{Peimbert1971}. 
The map shows the large scale structures with the inner shell 
framed by a rim of higher $t^{2}$ with values of $\sim$0.10.
The other well-known features -- knots K2 and K3 and the minor
axis polar knots -- notably show lower values, with $t^{2}$ 
approaching zero, particularly over knots K1 and K4,
where PJ $T_{\rm e}$ is comparable to [\ion{S}{III}] $T_{\rm e}$. Whilst
$t^{2}$ was formed from the difference of the [\ion{S}{III}] and PJ $T_{\rm e}$
and therefore typifies $t^{2}$ for H$^{+}$, it could just as well have been
formulated in terms of the \ion{He}{I} $T_{\rm e}$ (Sect. \ref{TeHe1}) and 
another CEL temperature diagnostic, such as [\ion{Ar}{III}]7135.8/5191.8\,\AA\ 
\citep{FangLiu2011}, giving the He$^{+}$ temperature fluctuations. 

\citet{Rubin2002} from HST WF/PC2 imaging in narrow band filters covering 
the [\ion{O}{III}]4363.2\AA\ and [\ion{O}{III}]5006.9\AA\ lines computed 
the spatial variation of $T_{\rm e}$, the mean $T_{\rm e}$ and 
fractional mean $T_{\rm e}$ variation across the nebula (called $t^{2}_{A}$). 
The resulting value of $t^{2}_{A}$ was very small, $\leq$0.01; much smaller 
than the value of the $t^{2}$ temperature fluctuation determined here 
(Sect. \ref{t2mapping}). From STIS long slit spectroscopy, a mean 
[\ion{O}{III}] $T_{\rm e}$ of 10140 K and $t^{2}_{A}$ 0.0035 were determined
\citep{Rubin2002}. Comparing the map of [\ion{O}{III}] 
$T_{\rm e}$ \citep[][, Fig. 2]{Rubin2002} with [\ion{S}{III}] 
$T_{\rm e}$ (Fig. \ref{fig:temaps}) shows similar aspects, such as
the elevations on the minor axis, the lower value over the outer
shell and even the depressions across the inner shell boundary; the 
[\ion{S}{III}] $T_{\rm e}$ map did not however extend to the outer 
halo where [\ion{O}{III}] shows elevated values of $T_{\rm e}$.
\citet{Rubin2002} calculated the mean $T_{\rm e}$ and 
$t^{2}_{A}$ over an elliptical area of 560 square arcseconds: 
over this area the mean [\ion{S}{III}] $T_{\rm e}$ and $t^{2}$ values
are 9130 K and 0.064 from the maps (Figs. \ref{fig:temaps} 
and \ref{fig:T0t2maps} respectively). $t^{2}_{A}$ from [\ion{S}{III}] 
$T_{\rm e}$ over the same area yields a comparable value of 0.0011
to that from [\ion{O}{III}] $T_{\rm e}$.

\subsection{Towards a flat He/H}
Unless the He abundance is indeed spatially structured, then the causes of
the non-uniform He/H$^{+}$ map (Fig. \ref{fig:Hetot}) must lie in the
physical conditions applied in calculating the He$^{++}$ and He$^{+}$ /H$^{+}$
ratios. Both $T_{\rm e}$ and $N_{\rm e}$ in the He$^{++}$ are not well
constrained and the [\ion{S}{III}] and [\ion{Cl}{III}] derived 
values were applied; these ions however sample a lower ionization 
medium (IP $<$ 55 eV), while He$^{++}$ exists at energies above 54 eV. 
The [\ion{Ar}{IV}]4711.4/4740.2\,\AA\ 
ratio (just outside the MUSE standard wavelength coverage) samples 
$N_{\rm e}$ at energies between 40.7 and 59.8 eV and \citet{FangLiu2011} 
determined an integrated $N_{\rm e}$ of 4890 cm$^{-3}$, only slightly higher 
than the lower ionization species (see their table 4). 

\citet{FangLiu2011} determined $T_{\rm e}$ of 11000 K for the He$^{++}$ 
zone from the 5694\,\AA\ free-bound Pfund series jump. In comparison, the
1D photoionization models of Sect. \ref{Cloudy1D} predict area
averaged $T_{\rm e}$(He$^{++}$) of 12800 K (Model A) and 15600 K (Model B). 
Adopting the \citet{FangLiu2011} $T_{\rm e}$(He$^{++}$) value to determine 
He$^{++}$/H$^{+}$ makes an insignificant difference 
to the values in the map and does not alter the lack of 
flatness of the resulting total He map. Even increasing $T_{\rm e}$
to the upper edge of the error bound quoted by \citet{FangLiu2011},
i.e., 13000 K, does not change this situation since both He$^{++}$ 
and H$^{+}$ have very similar dependence of emissivity on $T_{\rm e}$.
The 1D photoionization models (Tab. \ref{tab:integfluxes}) predict 
that the contribution of He$^{++}$ to total He/H$^{+}$ is 23\% 
(Model A) and 24\% (Model B), so the He$^{++}$/H$^{+}$ contribution is 
a modest fraction of the total. 

Higher ionization CEL $T_{\rm e}$ diagnostic line ratios, such 
as [\ion{Ar}{V}], or [\ion{Fe}{VI}] and [\ion{Fe}{vii}], would be 
required to probe the temperature in the He$^{++}$ region; lines of
these species are present in the tabulation of \citet{FangLiu2011} but 
enough well-detected lines to determine $T_{\rm e}$ are not available. 
Perhaps a dedicated spectroscopic search covering the regions with 
strongest \ion{He}{II} could result in useful $N_{\rm e}$ and $T_{\rm e}$ 
diagnostics for He$^{++}$. If very hot gas ($T_{\rm e} >$ 15000 K) is
present in the He$^{++}$ region, then CEL cooling through the observed
O, N and S emission is not efficient, so CEL $T_{\rm e}$ (and $N_{\rm e}$) 
diagnostics do not probe such hot gas effectively.

The results on the flatness of the He/H image imply that, in order to 
accurately determine abundances, it is important to use 
the value of $T_{\rm e}$ appropriate to the emissivity for all the 
ionic species, and H$^{+}$. This sounds obvious, but the enhanced
He in the inner shell, which results by applying $T_{\rm e}$ derived 
from the ratio of the \ion{He}{I} lines to the He$^{+}$/H$^{+}$ map,
shows that this He$^{+}$ temperature is not everywhere appropriate
to producing a flat He/H result. The CLOUDY models 
(Sect. \ref{Cloudy1D}), by comparison, predict electron temperatures 
in the He$^{+}$ zone of 9800 and 11800 K (Models A and B respectively), 
much higher than determined in Sect. \ref{TeHe1}.   
One possible approach to explore would
be to use the assumption of flatness of the He/H map to determine 
$T_{\rm e}$ as a variable in the He$^{++}$ and He$^{+}$ emission 
spaxels. Crude versions of this approach, such as scaling the [\ion{S}{III}] 
$T_{\rm e}$ map by a constant factor, were applied, but it did not prove 
possible to improve the flatness of the resulting He/H map beyond that
shown in Fig. \ref{fig:Hetot}.    
 
In computing the fraction
of He$^{++}$ v. H$^{+}$ in the inner shell, the fraction of H$^{+}$ emitting 
in this zone also required; the latter is not available from a line-of-sight 
integrated map. Imaging with high enough velocity resolution offers a
first approach to observational determination of the fraction of H$^{+}$
in the He$^{++}$ region, by apportioning the emission on the basis of
the radial velocity (or perhaps from the line widths), as explored
by \citet{Barlow2006} from slit spectroscopy. MUSE with its low
resolution is not suitable for such a study, but FLAMES with higher
spectral resolution modes (although a much smaller IFU) offers one
approach \citep[c.f.,][]{Tsamis2008} as do high fidelity slit scanning 
and Fourier transform spectrometry. 

\subsection{ICFs}
The peaks and boundaries in the map of He/H$^{+}$ (Fig. \ref{fig:Hetot}) 
directly imply that the O/H$^{+}$ map will not be flat since the O ICF 
schemes divide by some ratio of He/He$^{+}$. Whilst flatness can be 
viewed as an ideal in a real nebula, without prior information that 
O abundance (usually considered not to be significantly enriched by the 
dredge-up episodes, c.f. \citet{Richer2007}) is spatially altering, the 
deviations from flatness at the positions
of the boundaries in the ionization structures (inner/outer shell)
suggest flaws in determining the physical conditions (extinction,
temperature and density) or the ICF, or both. While the three ICF methods 
for O abundance yield very similar mean values (differing by 2.5\%), the 
flatness of the O maps is highest for the \citet{TorresPeimbert1977} ICF 
and lowest for the \citet{Delgado-Inglada2014} one. With comparable MUSE data
for other PNe, widely applicable solutions for spatial ICF's can be 
investigated across a range of ionization conditions.

\subsection{Potential for 3D modelling}
The emission line and diagnostic maps presented provide 
observational inputs for 3-D photoionization modelling, where 
the emission along all lines of sight across the whole nebula defined by
the spaxels can be matched with the MUSE observations. As made clear
in this presentation and the crudeness of 1D photoionization modelling 
(Sect. \ref{Cloudy1D}), IFU data of the depth and spatial coverage 
provided by MUSE demand the development of 3D models for 
explaining the interaction between emergent emission lines, 
diagnostics of extinction, $N_{\rm e}$, $T_{\rm e}$ and the ionic
abundances. Codes such as MOCASSIN \citep{Ercolano2003} and the 
multi-1D code, CLOUDY\_3D \citep{Morisset2011}, offer 
the desired approach. \citet{Goncalves2006} already performed 
MOCASSIN modelling of NGC~7009 to match long slit and selected HST 
emission line images in a study of the outer knots K1 and K4.
Inclusion of ultra-violet, infrared, mm/sub-mm and 
radio data, both resolved and unresolved, can lead to a comprehensive 
photoionization model considering the neutral and ionized gas, dust and 
molecular constituents, as performed by \citet{Otsuka2017} for NGC~6781.


\section{Conclusions}
\label{Concl}
The first detailed study of a planetary nebula with the MUSE
spectrograph has been presented. The observations demonstrate
the huge potential of this instrument for advancing optical
spectroscopic studies of extended emission nebulae. Even with 
very short exposures, the high surface brightness of NGC~7009 
allows a wealth of data to be exploited. The excellent image quality
during the Science Verification observations enables quantitative 
spectroscopy at spatial resolution approaching that of HST. 

The MUSE data on NGC~7009 have been presented
in their reduced state as sets of flux maps, distinguished by sufficient
signal-to-noise to be displayed and analysed in their native 
(0.2$''$ $\times$ 0.2$''$) spaxel form. A limited amount of analysis of
these unbinned maps has been presented, such as observed line ratios
(Sect. \ref{Emismaps}), $N_{\rm e}$ and $T_{\rm e}$ maps (Sect. 
\ref{NeTemaps}), ionic and total abundances (Sects. \ref{Ionmaps} and 
\ref{Abundmaps} respectively). The only exceptions to this approach were 
the presentation of Voronoi tesselated images in H$\alpha$ and H$\beta$, 
in order to illustrate the potential for studies of the lower surface brightness 
halo (Sect. \ref{Voronoi}), and tabulation of an area-integrated spectrum for
comparison to photoionization models (Sect. \ref{Cloudy1D} ). These results
by no means saturate the richness of the data cube.

In addition to images in a variety of emission lines from
neutral species, such as [\ion{O}{I}], to the high ionization 
potential species at good spatial resolution ([\ion{Mn}{V}], 51.2 
$<$ ionization energy $<$ 72.4 eV), line ratios have been used to
derive ORL and CEL diagnostics of reddening, electron temperature
and density. For the first time, maps of the differences in
CEL and ORL $T_{\rm e}$ have been derived, and correspondingly
a map of $t^{2}$ between a CEL and ORL temperature; they show 
considerable detail not limited to the differing ionization
zones. 

From the spatial coverage of physical conditions,
derived parameters have been mapped, such as ionization
fractions of some light elements. For He, the total
abundance across the nebula could be determined; within 2.3\%
the He/H abundance is flat, but there are worrying signatures of
the ionization structure still present, with remnants of the 
strongest He$^{++}$ regions and the demarcation between inner and
outer shells remaining. Data of the quality shown have the potential 
to provide general solutions for accurate abundance determination
and derivation of ICF's. But MUSE observations of a broader 
sample of PNe are required to develop such general conclusions 
that are applicable to nebulae of a range of ionization level, and 
abundance.


\begin{acknowledgements}
We sincerely thank the whole MUSE Science Verification team for achieving 
such a high level of data quality during this phase of the instrument 
testing. Peter Scicluna and Dave Stock are gratefully thanked for their 
contributions to the SV proposal. We thank Peter Storey for advice on
possible recombination contribution to the [\ion{S}{III}] 6312.1\,\AA\
emission, important for $T_{\rm e}$ estimation.
\newline
AMI acknowledges support from the Spanish MINECO through project 
AYA2015-68217-P. M.L.L.-F. was supported by CNPq, Conselho Nacional 
de Desenvolvimento Cient\'{I}fico e Tecnol\'{o}gico - Brazil, process 
number 248503/2013-8.
\newline
We are very grateful to the referee for a deep review and for a 
multitude of suggestions, many of which were included into the revised 
version.
\end{acknowledgements}


\begin{appendix}
\section{Custom definition of Paschen Jump and calibration with $T_{\rm e}$}
\label{Appendix}

\begin{figure}
\resizebox{\hsize}{!}{
\includegraphics[width=0.50\textwidth,angle=-90,clip]{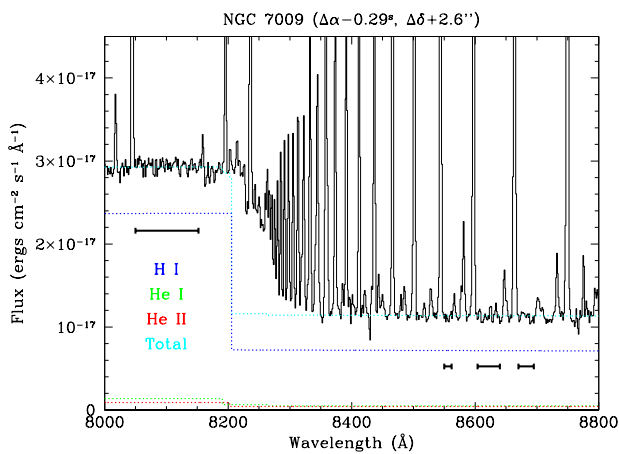}
}
\caption{The dereddened spectrum of a single spaxel over the inner shell 
(offsets from position of central star indicated) around the wavelength of 
the \ion{H}{I} Paschen Jump, showing the four continuum sections used to 
measure the magnitude of the jump. The calculated \ion{H}{I}, \ion{He}{I} 
and \ion{He}{II} nebular continua are shown by coloured dotted lines. 
The spaxel has an extinction, c, of 0.186, an He$^{+}$/H$^{+}$ value of 
0.094, an He$^{++}$/H$^{+}$ of 0.014, [\ion{Cl}{III}] $N_{\rm e}$ 7520 cm$^{-3}$ 
and a PJ $T_{\rm e}$ of 7790 K. A continuum of 3.3$\times$ 10$^{-18}$ 
ergs cm$^{-2}$ s$^{-1}$ \AA$^{-1}$ had to be added to the summed 
theoretical continua of \ion{H}{I}, \ion{He}{I} and \ion{He}{II} in order 
match the total dereddened continuum flux. Given the displacement from the 
central star (of 5.0$''$), it is unclear if this is scattered continuum 
from the central star (either instrumental or intrinsic to the nebula), 
or instrument scattered light or residual sky background; the simplest 
option of a flat continuum was therefore chosen.
}
\label{AppFig:ObsSpec}
\end{figure}

The map of electron temperature derived from the Paschen jump described in 
Section \ref{TePasJum} and shown in Fig. \ref{fig:PJTemap} was derived using 
a custom method to measure the magnitude of the Paschen jump with respect to 
the P11 line tuned to MUSE spectra. This method was developed with the aim
to produce a robust map of Paschen Jump $T_{\rm e}$ over the many thousands 
of spaxel spectra where the jump could be reliably measured, without 
individually fitting the continuum jump as a function of $T_{\rm e}$ and 
$N_{\rm e}$ at all spaxels. Figure \ref{AppFig:ObsSpec}
shows a single spaxel spectrum from the 120s MUSE cube over the bright
inner shell (offsets $\Delta \alpha$ = -4.2$''$, $\Delta \delta$ = +2.6$''$ 
from the central star) in the region of the Paschen jump. 

Continuum windows were selected based on the line-free sections
of this typical spectrum, with the total continuum on the low and high 
wavelength sides of the jump (referred to simply as 'blue' and 'red') 
not necessarily single consecutive sections. To the low wavelength side of 
the jump there is a line-free continuum section around 8050~--~8150\,\AA\
and this was selected; this window has the advantage that it contains no
strong telluric absorption lines (there is a strong molecular 
absorption centred at around 8230\,\AA, extending to $\sim$8180\,\AA). 

At the MUSE (measured) resolution of about 2.8\,\AA, the higher Paschen lines 
are strongly blended (Fig. \ref{AppFig:ObsSpec}) and so it proves challenging
to define line-free continuum windows for estimation of the true nebula
continuum to the long wavelength side of the jump. After some trials, 
three windows were chosen, 8550~--~8562, 8604~--~8640 and 8670~--~8695\,\AA, 
totalling 73\,\AA\ width; all but the bluest window suffer negligible telluric 
absorption. Although these windows are some hundreds of \,\AA\ 
separated from the Paschen Jump, the theoretical nebular continuum can be 
calculated in these same windows to derive an expression for the definition
of Paschen Jump with $T_{\rm e}$.

The continuum in a planetary nebula arises from four processes:
\begin{itemize}
\item bound-free (b--f) emission from captures of 
free electrons by H$^{+}$, He$^{+}$ and He$^{++}$ (b--f emission from 
atoms heavier than He is neglected as it makes a negligible contribution);
\item free-free (f--f), or Bremsstrahlung, emission from electrons being 
scattered by the nuclear charge of H$^{+}$, He$^{+}$ and He$^{++}$;
\item two-photon (2--$\nu$) gaseous continuum by radiative decay
from the $2^{2}S$~--~$1^{2}S$ levels of H$^{+}$ and He$^{++}$, which is 
forbidden by direct transition, and, much weaker, two photon decay from the
$1s2s ^{1} - 1s2s ^{3}S$ states to the ground state in He I \citep{Drake1969};
\item scattered continuum of non-gaseous origin, such as starlight scattered by
dust (and electrons) and thermal emission of dust within the nebula.
\end{itemize}

The first three effects can be computed using well-known atomic recombination 
theory whilst the latter is object and geometry specific and is not amenable
to direct computation. Since a map of the Paschen jump is produced, deviations 
in $T_{\rm e}$ over the map can indicate non-gaseous scattering processes. A case
in point is the central star: the $T_{\rm e}$ map in Fig. \ref{fig:PJTemap} shows
very low values ($<$ 100 K) over the image of the central star since no attempt was
made to subtract the stellar continuum from the total (stellar + nebular) continuum,
the jump is thus dominated by the strong blue stellar continuum. 

\subsection{Computation of gaseous continuum emission}
\citet{Schirmer2016} provides a concise summary for computation of the nebular
continuum together with a $c$ routine to calculate the output spectrum (although
this routine was not used for the computations described). Given its very minor
contribution, the computation of \ion{He}{I} 2--$\nu$ was neglected.    

\noindent{\bf b--f}\\
The bound free continuum for H$^{+}$, He$^{+}$ and He$^{++}$ has been calculated 
by \citet{Ercolano2006} and coefficients are tabulated for an extensive set of 
transitions and for temperatures $100 \leq T \leq 10^{5}$ K. The
interpolation scheme for $T_{\rm e}$ and frequency was used to calculate the b--f
emission on a fine wavelength scale across the region of the Paschen jump. 

\noindent{\bf f--f}\\
The free-free emissivity of a hydrogenic ion of nuclear charge $Z$ at a
temperature $T$ as a function 
of frequency ($\nu$) is given by \citet{BrownMathews1970} as:
\begin{multline}
  j_{\rm ff}(\nu) = \frac{1}{4\pi}\,N_X\,N_e\; \frac{32\,Z^2 e^4 h}{3 m_e^2 c^3}
  \left(\frac{\pi h \nu_0}{3k_BT}\right)^{1/2}\;\times\\
       {\rm exp}(-h \nu/k_B T)\;g_{\rm ff}(Z,T,\nu)
\end{multline}
where $N_X$ and $N_e$ are the ionic and electron number densities, $h \nu_0$ is
the ionization energy of hydrogen and the other symbols have their usual
meaning ($e$ the electron charge, $c$ the speed of light, $m_e$ the mass of the
electron, $h$ the Planck constant and $k_B$ the Boltzmann constant). $g_{\rm ff}$
is the thermally averaged mean free-free Gaunt factor which holds a quantum
mechanical correction to the classical electron scattering formula. The 
non-relativistic mean free-free Gaunt factor has been recently presented by
\citet{vanHoof2014} and tabulated as a function of scaled ionic energy and 
frequency ($\gamma^2$, $u$). The polynomial interpolation routine $interpolate.F$
provided by \citet{vanHoof2014} was employed to calculate the Gaunt factor 
as a function of frequency and nuclear energy. There are very slight
differences in the new computation of the mean free-free Gaunt factor from 
\citet{vanHoof2014} and the earlier tabulation by \citet{Hummer1988} using the 
subroutine provided in \citet{StoreyHummer1991} for interpolation of
the analytic approximation to $g_{\rm ff}(\gamma^2, u)$: the differences amount 
to up to 0.2\% for \ion{H}{I} and 0.3\% for \ion{He}{II} over the wavelength range around
the Paschen Jump and the range of $T_{\rm e}$ encountered in PNe. 

\noindent{\bf 2-photon}\\
The two-photon continuum of \ion{He}{I} and \ion{He}{II} peak at half 
the Lyman-$\alpha$ frequency (for \ion{H}{I} at a wavelength of 1420\,\AA\ 
in a flux per \AA\ spectrum; 608\,\AA\ for \ion{He}{II}). The emissivity can be 
written as:
\begin{equation}
\gamma_{2q}(\nu)=\alpha^{\rm eff}_{2^2{\rm S}}(T,N_e)\,g(\nu)\,
\left(1+\frac{\sum_X N_X\,q^X}{A_{2q}}\right)^{-1}
\end{equation}
following \citet{Schirmer2016}. Here, $g(\nu)$ is the frequency dependence 
of the two-photon emission presented and fitted by \citet{Nussbaumer1984} 
and $\alpha^{\rm eff}_{2^2{\rm S}}$ is the effective recombination 
coefficient of the upper ($2^{2}$ S) level. The evaluation of 
$\alpha^{\rm eff}_{2^2{\rm S}}$ is given by interpolation as a function of 
$T_{\rm e}$ and $N_{\rm e}$ in tables 4 (for \ion{H}{I}) and 5 (for 
\ion{He}{II}) of \citet{Schirmer2016} 
\citep[reproduced from][]{HummerStorey1987}.
$g(\nu)$ expresses the frequency dependence of the 2-photon spectrum in 
terms of the transition probability of the two photon transition of 
$A_{2q}$ = 8.2249 s$^{-1}$ given by \citet{Nussbaumer1984}. The term in brackets 
expresses the effect of de-population of the $2^{2} {\rm S}$ level by means of 
angular momentum changing collisions, which can reduce the 2--$\nu$ flux at typical 
nebular densities; the collisional transition rate coefficient, $q^{X}(T,N{\rm e})$ 
has been computed by \citet{PengellySeaton}, but see \citet{Schirmer2016} for 
a discussion. These corrections are implemented for \ion{H}{I} and 
\ion{He}{II} following \citet{Schirmer2016}. 


\subsection{Fitting the Paschen Jump {\it v.} $T_{\rm e}$}
\begin{figure}
\resizebox{\hsize}{!}{
\includegraphics[width=0.50\textwidth,angle=-90,clip]{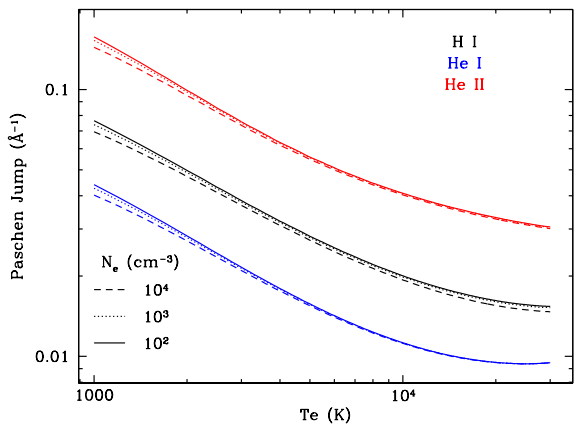}
}
\caption{The variation of the jump around 8200\,\AA\ for \ion{H}{I}, \ion{He}{I} 
and \ion{He}{II} as defined in the text with $T_{\rm e}$ for several values of
$N_{\rm e}$ ($10^2$, $10^3$ and $10^4$ cm$^{-3}$) is shown in a 
$log - log$ plot.
}
\label{AppFig:PJTe}
\end{figure}

The flux of the \ion{H}{I}, \ion{He}{I} and \ion{He}{II} continua were 
calculated on a fine wavelength grid around the Paschen Jump and the 
mean flux in the window to the blue of the jump and the three windows to 
the red computed for a range of $T_{\rm e}$ and at several $N_{\rm e}$ 
values. The emissivity of the \ion{H}{I} P11 line for ratioing the 
Paschen Jump flux by that of an \ion{H}{I} line 
was provided by the tables of \citet{StoreyHummer1995} for Case B. The 
variation of the defined jumps (blue - red continuum, employing means 
weighted by the number of pixels) ratioed by the P11 flux was then plotted 
for \ion{H}{I}, \ion{He}{I} and \ion{He}{II} $v.$ $T_{\rm e}$ at several $N_{\rm e}$. Even in 
a log PJ (\AA$^{-1}$) $v.$ log $T_{\rm e}$ plot, PJ can be multi-valued, 
particularly for the \ion{He}{I} jump $v$ $T_{\rm e}$ at high $T_{\rm e}$ (see Fig. 
\ref{AppFig:PJTe}), so it was decided to fit the variation in defined 
$T_{\rm e}$ ranges by simple fits following the form $T_{\rm e} 
\propto PJ^{-n}$ adopted by \citet{FangLiu2011}. However differing from 
\citet{FangLiu2011}, the fit of jump $v$ $T_{\rm e}$ was made independently
for \ion{H}{I}, \ion{He}{I} and \ion{He}{II}; the total (observed) jump 
was then formed as the sum of the three jumps for \ion{H}{I}, \ion{He}{I} 
and \ion{He}{II}. This is more rigorous
of the measurement situation and allows the possibility of different 
values of $T_{\rm e}$ for the He$^{+}$ and He$^{++}$ emitting plasma 
to be included. 

Table \ref{AppTab:PJTefit} lists the fits over three $T_{\rm e}$ ranges 
chosen to retain an error on the fits of $^{<}_{\sim}$ 300 K. These fits 
were then employed to produce the PJ $T_{\rm e}$ map in Fig. \ref{fig:PJTemap} 
using the measured [\ion{Cl}{III}] electron density to determine the fit of jump 
$v$ $T_{\rm e}$ to be applied within narrow ranges. Initially the 
[\ion{S}{III}] $T_{\rm e}$ was employed to choose the fits range, then the 
resulting PJ $T_{\rm e}$ map was used to position the fits; this process was 
iterated twice until the deviation on the $T_{\rm e}$ values was 
$^{<}_\sim$100 K, typically less than the fit errors. Estimates of errors 
on the PJ temperature were determined by Monte Carlo propagation of the error 
on the Paschen jump and the fit errors in K on the PJ $v$ $T_{\rm e}$ 
relations (Fig. \ref{AppFig:PJTe}). Figure \ref{AppFig:ObsSpec} shows an example
of the nebular continuum fit to an individual spaxel spectrum over the bright 
rim.

\begin{table*}
\caption{Fits to $log_{10} PJ = a - b \times log_{10} T_{\rm e}$ for \ion{H}{I}, \ion{He}{I} and \ion{H}{II} }
\centering
\begin{tabular}{llrrrr}
\hline\hline
Start $T_{\rm e}$ & End $T_{\rm e}$ & \ion{H}{I} a, b & \ion{He}{I} a,b & \ion{He}{II} a,b & RMS (\ion{H}{I}, \ion{He}{I}, \ion{He}{II}) \\
          (K)      & (K)              &          &          &           &  $T_{\rm e}$  (K) \\
\hline
\multicolumn{1}{c}{$N_{\rm e}$ = 100 cm$^{-3}$} \\ 
 1000 &  7000 &   0.73948, 0.61868 &  0.56095, 0.63906 &   1.11139, 0.63940 & 95.4, 108.4, 160.5 \\
 7000 & 15000 &  -0.08653, 0.40215 & -0.48580, 0.36487 &   0.08158, 0.36703 & 319.3, 247.6, 185.5 \\
15000 & 22000 &  -0.79633, 0.23120 & -1.40631, 0.14320 &  -0.38012, 0.25564 & 148.6, 299.6, 65.5 \\

\multicolumn{1}{c}{$N_{\rm e}$ = 1000 cm$^{-3}$} \\
 1000 &  7000 &   0.69099, 0.60674 &  0.50618, 0.62471 &   1.05870, 0.62579 & 76.1,100.7, 157.5 \\
 7000 & 15000 &  -0.09157, 0.40200 & -0.49917, 0.36175 &   0.07152, 0.36500 & 168.4, 239.9, 139.5 \\
15000 & 22000 &  -0.73123, 0.24765 & -1.32643, 0.16217 &  -0.35185, 0.26271 & 243.2, 465.0, 99.3 \\

\multicolumn{1}{c}{$N_{\rm e}$ = 10000 cm$^{-3}$}  \\
 1000 &  7000 &   0.60974, 0.58724 &  0.40644, 0.59878 &   0.96336, 0.60137 & 201.8, 127.6, 198.3 \\
 7000 & 15000 &  -0.06881, 0.41044 & -0.52410, 0.35599 &   0.06060, 0.36332 & 160.3, 242.0, 131.3 \\
15000 & 22000 &  -0.77695, 0.23999 & -1.43933, 0.13568 &  -0.39114, 0.25441 & 151.4, 325.9, 70.9 \\
\hline
\end{tabular}
\label{AppTab:PJTefit}
\end{table*}

\end{appendix}


%
\bibliographystyle{aa} 
\bibliography{Walsh33445.bib} 

\begin{thebibliography}{104}
\expandafter\ifx\csname natexlab\endcsname\relax\def\natexlab#1{#1}\fi

\bibitem[{{Aleman} {et~al.}(2014){Aleman}, {Ueta}, {Ladjal}, {Exter},
  {Kastner}, {Montez}, {Tielens}, {Chu}, {Izumiura}, {McDonald}, {Sahai},
  {Si{\'o}dmiak}, {Szczerba}, {van Hoof}, {Villaver}, {Vlemmings},
  {Wittkowski}, \& {Zijlstra}}]{Aleman2014}
{Aleman}, I., {Ueta}, T., {Ladjal}, D., {et~al.} 2014, \aap, 566, A79

\bibitem[{{Ali} {et~al.}(2015){Ali}, {Amer}, {Dopita}, {Vogt}, \&
  {Basurah}}]{Ali2015}
{Ali}, A., {Amer}, M.~A., {Dopita}, M.~A., {Vogt}, F.~P.~A., \& {Basurah},
  H.~M. 2015, \aap, 583, A83

\bibitem[{{Ali} \& {Dopita}(2017)}]{Ali2017}
{Ali}, A. \& {Dopita}, M.~A. 2017, \pasa, 34, e036

\bibitem[{{Ali} {et~al.}(2016){Ali}, {Dopita}, {Basurah}, {Amer}, {Alsulami},
  \& {Alruhaili}}]{Ali2016}
{Ali}, A., {Dopita}, M.~A., {Basurah}, H.~M., {et~al.} 2016, \mnras, 462, 1393

\bibitem[{{Bacon} {et~al.}(2010){Bacon}, {Accardo}, {Adjali}, {Anwand},
  {Bauer}, {Biswas}, {Blaizot}, {Boudon}, {Brau-Nogue}, {Brinchmann},
  {Caillier}, {Capoani}, {Carollo}, {Contini}, {Couderc}, {Daguis{\'e}},
  {Deiries}, {Delabre}, {Dreizler}, {Dubois}, {Dupieux}, {Dupuy}, {Emsellem},
  {Fechner}, {Fleischmann}, {Fran{\c c}ois}, {Gallou}, {Gharsa}, {Glindemann},
  {Gojak}, {Guiderdoni}, {Hansali}, {Hahn}, {Jarno}, {Kelz}, {Koehler},
  {Kosmalski}, {Laurent}, {Le Floch}, {Lilly}, {Lizon}, {Loupias}, {Manescau},
  {Monstein}, {Nicklas}, {Olaya}, {Pares}, {Pasquini}, {P{\'e}contal-Rousset},
  {Pell{\'o}}, {Petit}, {Popow}, {Reiss}, {Remillieux}, {Renault}, {Roth},
  {Rupprecht}, {Serre}, {Schaye}, {Soucail}, {Steinmetz}, {Streicher}, {Stuik},
  {Valentin}, {Vernet}, {Weilbacher}, {Wisotzki}, \& {Yerle}}]{Bacon2010}
{Bacon}, R., {Accardo}, M., {Adjali}, L., {et~al.} 2010, in \procspie, Vol.
  7735, Ground-based and Airborne Instrumentation for Astronomy III, 773508

\bibitem[{{Balick} {et~al.}(1998){Balick}, {Alexander}, {Hajian}, {Terzian},
  {Perinotto}, \& {Patriarchi}}]{Balick1998}
{Balick}, B., {Alexander}, J., {Hajian}, A.~R., {et~al.} 1998, \aj, 116, 360

\bibitem[{{Balick} {et~al.}(1994){Balick}, {Perinotto}, {Maccioni}, {Terzian},
  \& {Hajian}}]{Balick1994}
{Balick}, B., {Perinotto}, M., {Maccioni}, A., {Terzian}, Y., \& {Hajian}, A.
  1994, \apj, 424, 800

\bibitem[{{Barlow} {et~al.}(2006){Barlow}, {Hales}, {Storey}, {Liu}, {Tsamis},
  \& {Aderin}}]{Barlow2006}
{Barlow}, M.~J., {Hales}, A.~S., {Storey}, P.~J., {et~al.} 2006, in IAU
  Symposium, Vol. 234, Planetary Nebulae in our Galaxy and Beyond, ed. M.~J.
  {Barlow} \& R.~H. {M{\'e}ndez}, 367--368

\bibitem[{{Benjamin} {et~al.}(1999){Benjamin}, {Skillman}, \&
  {Smits}}]{Benjamin1999}
{Benjamin}, R.~A., {Skillman}, E.~D., \& {Smits}, D.~P. 1999, \apj, 514, 307

\bibitem[{{Bohigas} {et~al.}(1994){Bohigas}, {Lopez}, \&
  {Aguilar}}]{Bohigas1994}
{Bohigas}, J., {Lopez}, J.~A., \& {Aguilar}, L. 1994, \aap, 291, 595

\bibitem[{{Brown} \& {Mathews}(1970)}]{BrownMathews1970}
{Brown}, R.~L. \& {Mathews}, W.~G. 1970, \apj, 160, 939

\bibitem[{{Cappellari} \& {Copin}(2003)}]{Capellari2003}
{Cappellari}, M. \& {Copin}, Y. 2003, \mnras, 342, 345

\bibitem[{{Corradi} {et~al.}(2004){Corradi}, {S{\'a}nchez-Bl{\'a}zquez},
  {Mellema}, {Gianmanco}, \& {Schwarz}}]{Corradi2004}
{Corradi}, R.~L.~M., {S{\'a}nchez-Bl{\'a}zquez}, P., {Mellema}, G.,
  {Gianmanco}, C., \& {Schwarz}, H.~E. 2004, \aap, 417, 637

\bibitem[{{Delgado-Inglada} {et~al.}(2014){Delgado-Inglada}, {Morisset}, \&
  {Stasi{\'n}ska}}]{Delgado-Inglada2014}
{Delgado-Inglada}, G., {Morisset}, C., \& {Stasi{\'n}ska}, G. 2014, \mnras,
  440, 536

\bibitem[{{Dopita} \& {Meatheringham}(1990)}]{Dopita1990}
{Dopita}, M.~A. \& {Meatheringham}, S.~J. 1990, \apj, 357, 140

\bibitem[{{Drake}(1969)}]{Drake1969}
{Drake}, G.~W.~F. 1969, \apj, 158, 1199

\bibitem[{{Dufour} {et~al.}(2015){Dufour}, {Kwitter}, {Shaw}, {Henry},
  {Balick}, \& {Corradi}}]{Dufour2015}
{Dufour}, R.~J., {Kwitter}, K.~B., {Shaw}, R.~A., {et~al.} 2015, \apj, 803, 23

\bibitem[{{Ercolano} {et~al.}(2003){Ercolano}, {Barlow}, {Storey}, \&
  {Liu}}]{Ercolano2003}
{Ercolano}, B., {Barlow}, M.~J., {Storey}, P.~J., \& {Liu}, X.-W. 2003, \mnras,
  340, 1136

\bibitem[{{Ercolano} \& {Storey}(2006{\natexlab{a}})}]{ErcolanoStorey2006}
{Ercolano}, B. \& {Storey}, P.~J. 2006{\natexlab{a}}, \mnras, 372, 1875

\bibitem[{{Ercolano} \& {Storey}(2006{\natexlab{b}})}]{Ercolano2006}
{Ercolano}, B. \& {Storey}, P.~J. 2006{\natexlab{b}}, \mnras, 372, 1875

\bibitem[{{Etxaluze} {et~al.}(2014){Etxaluze}, {Cernicharo}, {Goicoechea}, {van
  Hoof}, {Swinyard}, {Barlow}, {van de Steene}, {Groenewegen}, {Kerschbaum},
  {Lim}, {Lique}, {Matsuura}, {Pearson}, {Polehampton}, {Royer}, \&
  {Ueta}}]{Etxaluze2014}
{Etxaluze}, M., {Cernicharo}, J., {Goicoechea}, J.~R., {et~al.} 2014, \aap,
  566, A78

\bibitem[{{Fang} \& {Liu}(2011)}]{FangLiu2011}
{Fang}, X. \& {Liu}, X.-W. 2011, \mnras, 415, 181

\bibitem[{{Fang} \& {Liu}(2013)}]{FangLiu2013}
{Fang}, X. \& {Liu}, X.-W. 2013, \mnras, 429, 2791

\bibitem[{{Ferland} {et~al.}(2017){Ferland}, {Chatzikos}, {Guzm{\'a}n},
  {Lykins}, {van Hoof}, {Williams}, {Abel}, {Badnell}, {Keenan}, {Porter}, \&
  {Stancil}}]{Ferland2017}
{Ferland}, G.~J., {Chatzikos}, M., {Guzm{\'a}n}, F., {et~al.} 2017, \rmxaa, 53,
  385

\bibitem[{{Freeman} {et~al.}(2014){Freeman}, {Montez}, {Kastner}, {Balick},
  {Frew}, {Jones}, {Miszalski}, {Sahai}, {Blackman}, {Chu}, {De Marco},
  {Frank}, {Guerrero}, {Lopez}, {Zijlstra}, {Bujarrabal}, {Corradi},
  {Nordhaus}, {Parker}, {Sandin}, {Sch{\"o}nberner}, {Soker}, {Sokoloski},
  {Steffen}, {Toal{\'a}}, {Ueta}, \& {Villaver}}]{Freeman2014}
{Freeman}, M., {Montez}, Jr., R., {Kastner}, J.~H., {et~al.} 2014, \apj, 794,
  99

\bibitem[{{Fried}(1966)}]{Fried1966}
{Fried}, D.~L. 1966, Journal of the Optical Society of America (1917-1983), 56,
  1372

\bibitem[{{Gon{\c c}alves} {et~al.}(2003){Gon{\c c}alves}, {Corradi},
  {Mampaso}, \& {Perinotto}}]{Goncalves2003}
{Gon{\c c}alves}, D.~R., {Corradi}, R.~L.~M., {Mampaso}, A., \& {Perinotto}, M.
  2003, \apj, 597, 975

\bibitem[{{Gon{\c c}alves} {et~al.}(2006){Gon{\c c}alves}, {Ercolano},
  {Carnero}, {Mampaso}, \& {Corradi}}]{Goncalves2006}
{Gon{\c c}alves}, D.~R., {Ercolano}, B., {Carnero}, A., {Mampaso}, A., \&
  {Corradi}, R.~L.~M. 2006, \mnras, 365, 1039

\bibitem[{{Gray} {et~al.}(2012){Gray}, {Matsuura}, \& {Zijlstra}}]{Gray2012}
{Gray}, M.~D., {Matsuura}, M., \& {Zijlstra}, A.~A. 2012, \mnras, 422, 955

\bibitem[{{Guerrero} {et~al.}(2000){Guerrero}, {Chu}, \&
  {Gruendl}}]{Guerrero2000}
{Guerrero}, M.~A., {Chu}, Y.-H., \& {Gruendl}, R.~A. 2000, \apjs, 129, 295

\bibitem[{{Howarth}(1983)}]{Howarth1983}
{Howarth}, I.~D. 1983, \mnras, 203, 301

\bibitem[{{Hummer}(1988)}]{Hummer1988}
{Hummer}, D.~G. 1988, \apj, 327, 477

\bibitem[{{Hummer} \& {Storey}(1987)}]{HummerStorey1987}
{Hummer}, D.~G. \& {Storey}, P.~J. 1987, \mnras, 224, 801

\bibitem[{{Jacoby} \& {Ford}(1983)}]{JacobyFord1983}
{Jacoby}, G.~H. \& {Ford}, H.~C. 1983, \apj, 266, 298

\bibitem[{{James} \& {Roos}(1975)}]{James1975}
{James}, F. \& {Roos}, M. 1975, Comput.Phys.Commun., 10, 343

\bibitem[{{Kaler}(1979)}]{Kaler1979}
{Kaler}, J.~B. 1979, \apj, 228, 163

\bibitem[{{Karakas} \& {Lattanzio}(2014)}]{KarakasLatt2014}
{Karakas}, A.~I. \& {Lattanzio}, J.~C. 2014, \pasa, 31, e030

\bibitem[{{Kastner} {et~al.}(2012){Kastner}, {Montez}, {Balick}, {Frew},
  {Miszalski}, {Sahai}, {Blackman}, {Chu}, {De Marco}, {Frank}, {Guerrero},
  {Lopez}, {Rapson}, {Zijlstra}, {Behar}, {Bujarrabal}, {Corradi}, {Nordhaus},
  {Parker}, {Sandin}, {Sch{\"o}nberner}, {Soker}, {Sokoloski}, {Steffen},
  {Ueta}, \& {Villaver}}]{Kastner2012}
{Kastner}, J.~H., {Montez}, Jr., R., {Balick}, B., {et~al.} 2012, \aj, 144, 58

\bibitem[{{Kingsburgh} \& {Barlow}(1994)}]{KingBarlow1994}
{Kingsburgh}, R.~L. \& {Barlow}, M.~J. 1994, \mnras, 271, 257

\bibitem[{{Lame} \& {Pogge}(1996)}]{Lame1996}
{Lame}, N.~J. \& {Pogge}, R.~W. 1996, \aj, 111, 2320

\bibitem[{{Leroy} {et~al.}(1986){Leroy}, {Le Borgne}, \& {Arnaud}}]{Leroy1986}
{Leroy}, J.~L., {Le Borgne}, J.~F., \& {Arnaud}, J. 1986, \aap, 160, 171

\bibitem[{{Liu} {et~al.}(1995{\natexlab{a}}){Liu}, {Barlow}, {Danziger}, \&
  {Clegg}}]{Liuetal1995}
{Liu}, X.-W., {Barlow}, M.~J., {Danziger}, I.~J., \& {Clegg}, R.~E.~S.
  1995{\natexlab{a}}, \mnras, 273, 47

\bibitem[{{Liu} {et~al.}(2006){Liu}, {Barlow}, {Zhang}, {Bastin}, \&
  {Storey}}]{Liu2006}
{Liu}, X.-W., {Barlow}, M.~J., {Zhang}, Y., {Bastin}, R.~J., \& {Storey}, P.~J.
  2006, \mnras, 368, 1959

\bibitem[{{Liu} \& {Danziger}(1993{\natexlab{a}})}]{Liu1993}
{Liu}, X.-W. \& {Danziger}, J. 1993{\natexlab{a}}, \mnras, 263, 256

\bibitem[{{Liu} \& {Danziger}(1993{\natexlab{b}})}]{LiuDanziger1993}
{Liu}, X.-W. \& {Danziger}, J. 1993{\natexlab{b}}, \mnras, 263, 256

\bibitem[{{Liu} {et~al.}(1995{\natexlab{b}}){Liu}, {Storey}, {Barlow}, \&
  {Clegg}}]{Liu1995}
{Liu}, X.-W., {Storey}, P.~J., {Barlow}, M.~J., \& {Clegg}, R.~E.~S.
  1995{\natexlab{b}}, \mnras, 272, 369

\bibitem[{{Liu} {et~al.}(2000){Liu}, {Storey}, {Barlow}, {Danziger}, {Cohen},
  \& {Bryce}}]{Liu2000}
{Liu}, X.-W., {Storey}, P.~J., {Barlow}, M.~J., {et~al.} 2000, \mnras, 312, 585

\bibitem[{{Luridiana} {et~al.}(2015){Luridiana}, {Morisset}, \&
  {Shaw}}]{Luridiana2015}
{Luridiana}, V., {Morisset}, C., \& {Shaw}, R.~A. 2015, \aap, 573, A42

\bibitem[{{Martin} {et~al.}(2016){Martin}, {Prunet}, \& {Drissen}}]{Martin2016}
{Martin}, T.~B., {Prunet}, S., \& {Drissen}, L. 2016, \mnras, 463, 4223

\bibitem[{{Matsuura} {et~al.}(2007){Matsuura}, {Speck}, {Smith}, {Zijlstra},
  {Viti}, {Lowe}, {Redman}, {Wareing}, \& {Lagadec}}]{Matsuura2007}
{Matsuura}, M., {Speck}, A.~K., {Smith}, M.~D., {et~al.} 2007, \mnras, 382,
  1447

\bibitem[{{Meaburn} \& {Walsh}(1981)}]{Meaburn1981}
{Meaburn}, J. \& {Walsh}, J.~R. 1981, \apss, 78, 473

\bibitem[{{Mendez} {et~al.}(1992){Mendez}, {Kudritzki}, \&
  {Herrero}}]{Mendez1992}
{Mendez}, R.~H., {Kudritzki}, R.~P., \& {Herrero}, A. 1992, \aap, 260, 329

\bibitem[{{Miller Bertolami}(2016)}]{MillerBerto2016}
{Miller Bertolami}, M.~M. 2016, \aap, 588, A25

\bibitem[{{Monreal-Ibero} {et~al.}(2005){Monreal-Ibero}, {Roth},
  {Sch{\"o}nberner}, {Steffen}, \& {B{\"o}hm}}]{MonrealIbero2005}
{Monreal-Ibero}, A., {Roth}, M.~M., {Sch{\"o}nberner}, D., {Steffen}, M., \&
  {B{\"o}hm}, P. 2005, \apjl, 628, L139

\bibitem[{{Monreal-Ibero} {et~al.}(2006){Monreal-Ibero}, {Roth},
  {Sch{\"o}nberner}, {Steffen}, \& {B{\"o}hm}}]{MonrealIbero2006}
{Monreal-Ibero}, A., {Roth}, M.~M., {Sch{\"o}nberner}, D., {Steffen}, M., \&
  {B{\"o}hm}, P. 2006, \nar, 50, 426

\bibitem[{{Monreal-Ibero} {et~al.}(2013){Monreal-Ibero}, {Walsh},
  {Westmoquette}, \& {V{\'{\i}}lchez}}]{MonrealIbero2013}
{Monreal-Ibero}, A., {Walsh}, J.~R., {Westmoquette}, M.~S., \&
  {V{\'{\i}}lchez}, J.~M. 2013, \aap, 553, A57

\bibitem[{{Monteiro} {et~al.}(2013){Monteiro}, {Gon{\c c}alves},
  {Leal-Ferreira}, \& {Corradi}}]{Monteiro2013}
{Monteiro}, H., {Gon{\c c}alves}, D.~R., {Leal-Ferreira}, M.~L., \& {Corradi},
  R.~L.~M. 2013, \aap, 560, A102

\bibitem[{{Montez} {et~al.}(2015){Montez}, {Kastner}, {Balick}, {Behar},
  {Blackman}, {Bujarrabal}, {Chu}, {Corradi}, {De Marco}, {Frank}, {Freeman},
  {Frew}, {Guerrero}, {Jones}, {Lopez}, {Miszalski}, {Nordhaus}, {Parker},
  {Sahai}, {Sandin}, {Schonberner}, {Soker}, {Sokoloski}, {Steffen},
  {Toal{\'a}}, {Ueta}, {Villaver}, \& {Zijlstra}}]{Montez2015}
{Montez}, Jr., R., {Kastner}, J.~H., {Balick}, B., {et~al.} 2015, \apj, 800, 8

\bibitem[{{Moreno-Corral} {et~al.}(1998){Moreno-Corral}, {de La Fuente}, \&
  {Guti{\'e}rrez}}]{Moreno-Corral1998}
{Moreno-Corral}, M., {de La Fuente}, E., \& {Guti{\'e}rrez}, F. 1998, \rmxaa,
  34, 117

\bibitem[{{Morisset}(2011)}]{Morisset2011}
{Morisset}, C. 2011, {Cloudy\_3D: Quick Pseudo-3D Photoionization Code},
  Astrophysics Source Code Library

\bibitem[{{Morisset}(2017)}]{Morisset2017}
{Morisset}, C. 2017, in IAU Symposium, Vol. 323, Planetary Nebulae:
  Multi-Wavelength Probes of Stellar and Galactic Evolution, ed. X.~{Liu},
  L.~{Stanghellini}, \& A.~{Karakas}, 43--50

\bibitem[{{Noll} {et~al.}(2012){Noll}, {Kausch}, {Barden}, {Jones}, {Szyszka},
  {Kimeswenger}, \& {Vinther}}]{Noll2012}
{Noll}, S., {Kausch}, W., {Barden}, M., {et~al.} 2012, \aap, 543, A92

\bibitem[{{Nussbaumer} \& {Schmutz}(1984)}]{Nussbaumer1984}
{Nussbaumer}, H. \& {Schmutz}, W. 1984, \aap, 138, 495

\bibitem[{{Otsuka} {et~al.}(2017){Otsuka}, {Ueta}, {van Hoof}, {Sahai},
  {Aleman}, {Zijlstra}, {Chu}, {Villaver}, {Leal-Ferreira}, {Kastner},
  {Szczerba}, \& {Exter}}]{Otsuka2017}
{Otsuka}, M., {Ueta}, T., {van Hoof}, P.~A.~M., {et~al.} 2017, \apjs, 231, 22

\bibitem[{{Peimbert}(1967)}]{Peimbert1967}
{Peimbert}, M. 1967, \apj, 150, 825

\bibitem[{{Peimbert}(1971)}]{Peimbert1971}
{Peimbert}, M. 1971, Boletin de los Observatorios Tonantzintla y Tacubaya, 6,
  29

\bibitem[{{Peimbert} \& {Costero}(1969)}]{Peimbert_Costero1969}
{Peimbert}, M. \& {Costero}, R. 1969, Boletin de los Observatorios Tonantzintla
  y Tacubaya, 5, 3

\bibitem[{{Peimbert} \& {Torres-Peimbert}(1971{\natexlab{a}})}]{Peimberts1971}
{Peimbert}, M. \& {Torres-Peimbert}, S. 1971{\natexlab{a}}, Boletin de los
  Observatorios Tonantzintla y Tacubaya, 6, 21

\bibitem[{{Peimbert} \&
  {Torres-Peimbert}(1971{\natexlab{b}})}]{Peimbert_TorresPeimbert1971}
{Peimbert}, M. \& {Torres-Peimbert}, S. 1971{\natexlab{b}}, \apj, 168, 413

\bibitem[{{Pengelly} \& {Seaton}(1964)}]{PengellySeaton}
{Pengelly}, R.~M. \& {Seaton}, M.~J. 1964, \mnras, 127, 165

\bibitem[{{Persi} {et~al.}(1999){Persi}, {Cesarsky}, {Marenzi}, {Preite-Mar
  tinez}, {Rouan}, {Siebenmorgen}, {Lacombe}, \& {Tiphene}}]{Persi1999}
{Persi}, P., {Cesarsky}, D., {Marenzi}, A.~R., {et~al.} 1999, \aap, 351, 201

\bibitem[{{Phillips} {et~al.}(2010){Phillips}, {Cuesta}, \&
  {Ramos-Larios}}]{Phillips2010}
{Phillips}, J.~P., {Cuesta}, L.~C., \& {Ramos-Larios}, G. 2010, \mnras, 409,
  881

\bibitem[{{Pitrou} {et~al.}(2018){Pitrou}, {Coc}, {Uzan}, \&
  {Vangioni}}]{Pitrou2018}
{Pitrou}, C., {Coc}, A., {Uzan}, J.-P., \& {Vangioni}, E. 2018, ArXiv e-prints
  [\eprint[arXiv]{1801.08023}]

\bibitem[{{Porter} {et~al.}(2012){Porter}, {Ferland}, {Storey}, \&
  {Detisch}}]{Porter2012}
{Porter}, R.~L., {Ferland}, G.~J., {Storey}, P.~J., \& {Detisch}, M.~J. 2012,
  \mnras, 425, L28

\bibitem[{{Porter} {et~al.}(2013){Porter}, {Ferland}, {Storey}, \&
  {Detisch}}]{Porter2013}
{Porter}, R.~L., {Ferland}, G.~J., {Storey}, P.~J., \& {Detisch}, M.~J. 2013,
  \mnras, 433, L89

\bibitem[{{Rauch}(2003)}]{Rauch2003}
{Rauch}, T. 2003, \aap, 403, 709

\bibitem[{{Reay} {et~al.}(1983){Reay}, {Atherton}, \& {Taylor}}]{Reay1983a}
{Reay}, N.~K., {Atherton}, P.~D., \& {Taylor}, K. 1983, \mnras, 203, 1079

\bibitem[{{Reay} {et~al.}(1984){Reay}, {Atherton}, \& {Taylor}}]{Reay1984}
{Reay}, N.~K., {Atherton}, P.~D., \& {Taylor}, K. 1984, \mnras, 206, 71

\bibitem[{{Richard} {et~al.}(2017){Richard}, {Bacon}, \&
  {Vernet}}]{Richard2017}
{Richard}, J., {Bacon}, R., \& {Vernet}, J. 2017, {MUSE User Manual. Issue 8.1}
  (ESO)

\bibitem[{{Richer}(2007)}]{Richer2007}
{Richer}, M.~G. 2007, Journal of Korean Astronomical Society, 40, 183

\bibitem[{{Richer} {et~al.}(1991){Richer}, {McCall}, \& {Martin}}]{Richer1991}
{Richer}, M.~G., {McCall}, M.~L., \& {Martin}, P.~G. 1991, \apj, 377, 210

\bibitem[{{Robbins}(1968)}]{Robbins}
{Robbins}, R.~R. 1968, \apj, 151, 511

\bibitem[{{Rubin}(1986)}]{Rubin1986}
{Rubin}, R.~H. 1986, \apj, 309, 334

\bibitem[{{Rubin} {et~al.}(2002){Rubin}, {Bhatt}, {Dufour}, {Buckalew},
  {Barlow}, {Liu}, {Storey}, {Balick}, {Ferland}, {Harrington}, \&
  {Martin}}]{Rubin2002}
{Rubin}, R.~H., {Bhatt}, N.~J., {Dufour}, R.~J., {et~al.} 2002, \mnras, 334,
  777

\bibitem[{{Sabbadin} {et~al.}(2004){Sabbadin}, {Turatto}, {Cappellaro},
  {Benetti}, \& {Ragazzoni}}]{Sabbadin2004}
{Sabbadin}, F., {Turatto}, M., {Cappellaro}, E., {Benetti}, S., \& {Ragazzoni},
  R. 2004, \aap, 416, 955

\bibitem[{{Sandin}(2014)}]{Sandin2014}
{Sandin}, C. 2014, \aap, 567, A97

\bibitem[{{Sandin} {et~al.}(2008){Sandin}, {Sch{\"o}nberner}, {Roth},
  {Steffen}, {B{\"o}hm}, \& {Monreal-Ibero}}]{Sandin2008}
{Sandin}, C., {Sch{\"o}nberner}, D., {Roth}, M.~M., {et~al.} 2008, \aap, 486,
  545

\bibitem[{{Schirmer}(2016)}]{Schirmer2016}
{Schirmer}, M. 2016, \pasp, 128, 114001

\bibitem[{{Sch{\"o}nberner} {et~al.}(2014){Sch{\"o}nberner}, {Jacob},
  {Lehmann}, {Hildebrandt}, {Steffen}, {Zwanzig}, {Sandin}, \&
  {Corradi}}]{Schoenberner2014}
{Sch{\"o}nberner}, D., {Jacob}, R., {Lehmann}, H., {et~al.} 2014, Astronomische
  Nachrichten, 335, 378

\bibitem[{{Seaton}(1979)}]{Seaton1979}
{Seaton}, M.~J. 1979, \mnras, 187, 73P

\bibitem[{{Storey} \& {Hummer}(1991)}]{StoreyHummer1991}
{Storey}, P.~J. \& {Hummer}, D.~G. 1991, Comp. Phys. Comm, 66, 129

\bibitem[{{Storey} \& {Hummer}(1995)}]{StoreyHummer1995}
{Storey}, P.~J. \& {Hummer}, D.~G. 1995, \mnras, 272, 41

\bibitem[{{Taylor} \& {Atherton}(1980)}]{Taylor1980}
{Taylor}, K. \& {Atherton}, P.~D. 1980, \mnras, 191, 675

\bibitem[{{Torres-Peimbert} \& {Peimbert}(1977)}]{TorresPeimbert1977}
{Torres-Peimbert}, S. \& {Peimbert}, M. 1977, \rmxaa, 2, 181

\bibitem[{{Tsamis} {et~al.}(2008){Tsamis}, {Walsh}, {P{\'e}quignot}, {Barlow},
  {Danziger}, \& {Liu}}]{Tsamis2008}
{Tsamis}, Y.~G., {Walsh}, J.~R., {P{\'e}quignot}, D., {et~al.} 2008, \mnras,
  386, 22

\bibitem[{{Ueta} {et~al.}(2014){Ueta}, {Ladjal}, {Exter}, {Otsuka}, {Szczerba},
  {Si{\'o}dmiak}, {Aleman}, {van Hoof}, {Kastner}, {Montez}, {McDonald},
  {Wittkowski}, {Sandin}, {Ramstedt}, {De Marco}, {Villaver}, {Chu},
  {Vlemmings}, {Izumiura}, {Sahai}, {Lopez}, {Balick}, {Zijlstra}, {Tielens},
  {Rattray}, {Behar}, {Blackman}, {Hebden}, {Hora}, {Murakawa}, {Nordhaus},
  {Nordon}, \& {Yamamura}}]{Ueta2014}
{Ueta}, T., {Ladjal}, D., {Exter}, K.~M., {et~al.} 2014, \aap, 565, A36

\bibitem[{{van Hoof} {et~al.}(2014){van Hoof}, {Williams}, {Volk}, {Chatzikos},
  {Ferland}, {Lykins}, {Porter}, \& {Wang}}]{vanHoof2014}
{van Hoof}, P.~A.~M., {Williams}, R.~J.~R., {Volk}, K., {et~al.} 2014, \mnras,
  444, 420

\bibitem[{{Walsh} \& {Clegg}(1994)}]{Walsh1994}
{Walsh}, J.~R. \& {Clegg}, R.~E.~S. 1994, \mnras, 268, L41

\bibitem[{{Walsh} {et~al.}(2016){Walsh}, {Monreal-Ibero}, {Barlow}, {Ueta},
  {Wesson}, \& {Zijlstra}}]{Walsh2016}
{Walsh}, J.~R., {Monreal-Ibero}, A., {Barlow}, M.~J., {et~al.} 2016, \aap, 588,
  A106

\bibitem[{{Weilbacher} {et~al.}(2015){Weilbacher}, {Monreal-Ibero},
  {Kollatschny}, {Ginsburg}, {McLeod}, {Kamann}, {Sandin}, {Palsa}, {Wisotzki},
  {Bacon}, {Selman}, {Brinchmann}, {Caruana}, {Kelz}, {Martinsson},
  {P{\'e}contal-Rousset}, {Richard}, \& {Wendt}}]{Weilbacher2015}
{Weilbacher}, P.~M., {Monreal-Ibero}, A., {Kollatschny}, W., {et~al.} 2015,
  \aap, 582, A114

\bibitem[{{Weilbacher} {et~al.}(2014){Weilbacher}, {Streicher}, {Urrutia},
  {P{\'e}contal-Rousset}, {Jarno}, \& {Bacon}}]{Weilbacher2014}
{Weilbacher}, P.~M., {Streicher}, O., {Urrutia}, T., {et~al.} 2014, in
  Astronomical Society of the Pacific Conference Series, Vol. 485, Astronomical
  Data Analysis Software and Systems XXIII, ed. N.~{Manset} \& P.~{Forshay},
  451

\bibitem[{{Zhang} {et~al.}(2005){Zhang}, {Liu}, {Liu}, \& {Rubin}}]{Zhang2005}
{Zhang}, Y., {Liu}, X.-W., {Liu}, Y., \& {Rubin}, R.~H. 2005, \mnras, 358, 457

\bibitem[{{Zhang} {et~al.}(2007){Zhang}, {Liu}, {Luo}, {P{\'e}quignot}, \&
  {Barlow}}]{Zhang2007}
{Zhang}, Y., {Liu}, X.-W., {Luo}, S.-G., {P{\'e}quignot}, D., \& {Barlow},
  M.~J. 2007, \aap, 472, 555

\bibitem[{{Zhang} {et~al.}(2004){Zhang}, {Liu}, {Wesson}, {Storey}, {Liu}, \&
  {Danziger}}]{Zhang2004}
{Zhang}, Y., {Liu}, X.-W., {Wesson}, R., {et~al.} 2004, \mnras, 351, 935

\end{thebibliography}
%


\end{document}